\documentclass[%
reprint,twocolumn,
amsmath,amssymb,
aps,
pra,
]{revtex4-2}
\usepackage{graphicx}% Include Figure~files
\usepackage{dcolumn}% Align table columns on decimal point
\usepackage{bm}% bold math
\usepackage{float}
\usepackage{enumerate}
\usepackage{algorithm}
\usepackage{algpseudocode}
\usepackage{algcompatible}

\usepackage{subfigure}
\usepackage{booktabs}
\usepackage[cmyk]{xcolor}
\usepackage{multirow}
\usepackage{makecell}
\usepackage{color}
\usepackage{amsmath,amssymb}
\usepackage{hyperref}
\usepackage[normalem]{ulem}

\hypersetup{colorlinks=true,citecolor=blue,linkcolor=blue,filecolor=blue,urlcolor=blue,breaklinks=true}
\newcommand{\nc}{\newcommand}
\nc{\rnote}[1]{\marginpar{\small\raggedright #1}}
\nc{\txtr}[1]{\textcolor{red}{#1}}
\nc{\txtb}[1]{\textcolor{blue}{#1}}
\nc{\txtg}[1]{\textcolor[RGB]{56,168,68}{#1}}
\nc{\txtbr}[1]{\textcolor{brown}{#1}}

\nc{\rr}[1]{\rnote{\txtr{#1}}}
\nc{\rg}[1]{\rnote{\txtg{#1}}}
\nc{\rb}[1]{\rnote{\txtb{#1}}}

\nc{\bra}[1]{\langle#1|}
\nc{\ket}[1]{|#1\rangle}
\nc{\op}[1]{\operatorname{#1}}
\nc{\ketbra}[2]{|#1\rangle\!\langle#2|}
\nc{\braket}[2]{\langle#1|#2\rangle}
\nc{\tr}{\op{tr}}

\begin{document}
	% \fontsize{12pt}{14pt}\selectfont
	%\preprint{APS/123-QED}
	
	\title{Robust Control of High-dimensional Quantum Systems against \\ Coherent and Incoherent Errors}% Force line breaks with \\
	
	\author{Yidian Fan}
	\affiliation{%
		Department of Automation, Tsinghua University, Beijing, 100084, China}%
	\author{Re-Bing Wu}
	\thanks{E-mail: rbwu@tsinghua.edu.cn}
	\affiliation{
		Department of Automation, Tsinghua University, Beijing, 100084, China}%
	
	\date{\today}% It is always \today, today,
	\begin{abstract}
        Toward scalable quantum computing, the control of quantum systems needs to be robust against both coherent errors induced by parametric uncertainties and incoherent errors induced by environmental decoherence. This poses significant challenges for high-dimensional systems due to the computational intensity involved in the control design process. In this paper, we propose a systematic framework to improve the design efficiency. By employing the Taylor series expansion of uncertain parameters, the problem of robust control for an uncertain quantum system is reformulated as the optimal control problem of an augmented deterministic system. The Suzuki-Trotter expansion is then applied to accelerate the calculation of the system dynamics. Numerical simulations of quantum state preparation and quantum gate synthesis demonstrate that the proposed algorithm can successfully identify robust solutions in multi-qubit systems. The enhanced efficiency effectively extends the feasibility of high-order robust control for realistic high-dimensional quantum systems with multiple error sources. 
	\end{abstract}

	\maketitle
	\section{Introduction}
    In the hardware implementation of quantum algorithms, control protocols for the realization of high-fidelity quantum gates are the building blocks of quantum circuits~\cite{krantz2019quantum, preskill2018quantum, nielsen2010quantum}. When the model of a quantum system is precisely known, control solutions can be analytically or numerically designed~\cite{khaneja2005optimal}. However, under most circumstances the precise model is unavailable due to uncertain factors in experimental settings, e.g., incorrectly identified system parameters~\cite{koswara2021quantum}, pulse distortions~\cite{wu2018data,rol2020time}, unwanted cross-talks~\cite{zhao2022quantum}. These factors can severely degrade the control performance and introduce coherent errors into quantum gates, which calls for the enhancement of control robustness to the corruption of uncertainties~\cite{weidner2024robust}.
	
    The dynamics of the uncertain quantum system can be envisioned as following some unknown model among a collection of deterministic models. The goal of robust control design is to find a universal control solution that performs effectively across all deterministic quantum systems in this collection. In the literature, there are two classes of approaches. The first approach is to optimize the average control performance over a collection of uncertainty parameter samples. The average performance can be evaluated over a carefully chosen set of representative samples~\cite{chen2014sampling}, but the required size of the sample set rapidly grows with the number of uncertain parameters~\cite{wu2019learning}. Alternatively, stochastic gradient-based algorithms can be applied, which significantly reduces the computational cost of performance evaluation, at the price of slow convergence of the control optimization~\cite{wu2019learning, turinici2019stochasticgradient, ge2020robust}. 
	
    The second approach is to transform the original nondeterministic system model into an augmented deterministic system by expressing the quantum state as a polynomial expansion of the uncertain parameters~\cite{li2006control,van2017robust, li2022moment, cao2024robust,poggi2024universally}. This treatment allows the analysis of the dynamics' sensitivity to potential uncertainties~\cite{propson2022robust,weidner2024robust} and facilitates the application of efficient deterministic optimization methods for control updates~\cite{liu1989limited,byrd1995limited}. A similar concept has also been explored in the context of Dyson and Magnus expansions~\cite{green2013arbitrary,koswara2014robustness,  haas2019engineering, koswara2021robust, puzzuoli2023algorithms,liu2024robust, shao2024multiple,chen2025robust}. The sampling complexity can be substantially reduced by truncating the series expansion to the first few terms, which suffice to account for the dominant contributions. While the achieved level of robustness is inherently determined by the truncation order, the effective dimensionality of the augmented system can be extremely high when dealing with multiple uncertain parameters or high-order robustness.
    
    % In order to scale up quantum processors to handle complex computational tasks, e.g., quantum error correction~\cite{fowler2012surface, michael2016new, acharya2024quantum}, the ability to simulate and optimize a moderate size of interacting qubits~\cite{leung2017speedup, shi2019optimized,klimov2020snake,piveteau2023circuit} or the extended Hilbert space of quantum computing elements~\cite{heeres2017implementing, hu2019quantum, ni2023beating} is becoming pivotal.
    
    In addition to system uncertainties, incoherent errors caused by interactions with the environment can also be detrimental to the performance of control strategies that are optimized solely for closed quantum systems~\cite{schulte2011optimal}. Control solutions designed under the evolution governed by the master equation, which simulates environmental effects through incorporating Lindblad operators into the Schrodinger equation, have demonstrated improved tolerance to the incoherent errors~\cite{schulte2011optimal, floether2012robust, goerz2014optimal, koch2016controlling, boutin2017resonator, gautier2025optimal}. However, modeling the open system dynamics of a $d$-dimensional closed system requires a $d^2$-dimensional representation in the form of a density matrix~\cite{breuer2002theory,jing2022one}, resulting in a quadratic increase in dimensionality. Consequently, the dimensionality challenge becomes even more severe when formulating an augmented system aimed at suppressing both coherent and incoherent errors~\cite{haas2019engineering, puzzuoli2023algorithms, liu2024robust}.

    To alleviate computational burden of simulating open system dynamics, the method proposed in~\cite{abdelhafez2019gradient} employs the quantum trajectories technique that propagates an ensemble of pure states instead of a full density matrix~\cite{dum1992monte,daley2014quantum}. The ensemble average over these trajectories reproduces the solution of the master equation. The work in~\cite{chen2025robust} approximates the quantum state to first order with respect to both system uncertainties and incoherent effects, and minimizes the average control error over the uncertainty distribution. Both methods can reduce the computational complexity of designing optimal control for open quantum systems, bringing it close to that of closed-system scenarios. However, optimizing the average control performance does not provide flexibility to suppress sensitivity to specific uncertainty directions or perturbative orders~\cite{dalgaard2022dynamical}. In contrast, the  augmented-system formulation allows explicit control over such sensitivity. Numerical results further support its effectiveness in identifying more robust solutions~\cite{propson2022robust}. Therefore, developing an efficient method that reduces the computational overhead of evolving the augmented system is crucial for practical control optimization.

    The method in~\cite{bhole2018practical,jensen2021approximate,jensen2021achieving} utilizes the Suzuki-Trotter expansion to approximate the dynamics of closed system evolution, upon which the control optimization is performed. The results demonstrate a substantial improvement in computational efficiency while largely preserving the optimality of the solution. This trade-off between efficiency and performance provides a promising direction for scalable control design, particularly when the Hilbert space dimension is so large that solving the exact system dynamics becomes computationally demanding or even intractable. As quantum processors continue to scale up~\cite{fowler2012surface, michael2016new, acharya2024quantum}, the ability to efficiently simulate and optimize the control of moderately sized interacting multi-qubit systems~\cite{leung2017speedup, shi2019optimized, klimov2020snake, piveteau2023circuit} or quantum computing elements with extended Hilbert spaces~\cite{heeres2017implementing, hu2019quantum, ni2023beating} is becoming increasingly critical. These systems often pose significant challenges for conventional full-density-matrix simulations. Consequently, efficient approximation techniques that retain high-fidelity control performance without prohibitive computational cost are essential to enable the practical deployment of quantum control strategies in near-term and future quantum technologies.
    
    In this paper, we propose a systematic framework for designing robust control pulses to implement gate operations in open quantum systems with multiple uncertain parameters. Specifically, we formulate this task as a set of state transfer problems and reduce the number of required instances from the conventional $d^2$ to $d+1$~\cite{goerz2014optimal}, significantly lowering the computational burden. Furthermore, we extend the Suzuki-Trotter expansion approach in~\cite{bhole2018practical,jensen2021approximate,jensen2021achieving} to approximate the dynamics of the augmented system governed by the master equation, enabling robust control optimization that accounts for both coherent and incoherent errors throughout the evolution. While the computational complexity of simulating the system dynamics is substantially reduced, the resulting optimized control solutions exhibit improved robustness against uncertainties and decoherence.
	
    The remainder of this work is organized as follows. Section~\ref{sec:uncertain_dynamics} introduces both an ODE solver and the S-T expansion for computing the dynamics of uncertain quantum systems. Section~\ref{sec:robust_control_design} presents a gradient-based optimization method for robust quantum state preparation and gate synthesis. Section~\ref{sec:simulation} demonstrates the effectiveness of the proposed algorithms through simulations on a spin-chain model. Finally, conclusions are provided in Sec.~\ref{sec:conclude_discuss}.
	
    \section{Dynamics of open quantum systems with uncertainties}
    \label{sec:uncertain_dynamics}
    In this section, we utilize the Taylor expansion to analyze the impacts of uncertainties presented in open quantum systems. The resulting expansion terms form a high-dimensional augmented system, whose dynamics can be computed using standard ODE solvers. To further lower the computational overhead, we propose leveraging the Suzuki-Trotter expansion to efficiently approximate the evolution of the system while maintaining sufficient accuracy for quantum control design.
	
	\subsection{Model of uncertain open quantum systems}
	\label{sec:uncertain_sys}
	Consider the manipulation of a $d$-dimensional quantum system dynamics modeled by the system Hamiltonian
	\begin{equation}
		H_S(t) = H_0 +  \sum_{c=1}^{n_c}u_c(t)H_c.
	\end{equation} 
	Here, $H_0$ and $H_1,\cdots,H_c$ are the drift and the control Hamiltonians, and the real number valued functions $u_1(t),\cdots,u_c(t)$ are the associated control pulses. The parametric uncertainties in realistic systems are described by the following Hamiltonian: 
	\begin{equation}\label{eq:uncertain_ham}
		H(t)=H_S(t) + \sum_{j=1}^{m} \epsilon_j E_j,
	\end{equation}
    where the effects of uncertain factors, such as system parameter shifts, unwanted crosstalk between qubits, control distortion, can be modeled as additive perturbation terms $E_j$ with unknown strengths $\epsilon_j$ for $j=1,\cdots,m$. In addition, the interaction of the system with the environment further introduces incoherent dynamics (e.g., dissipation and decoherence), which can be modeled by the following Lindblad master equation involving uncertain parameters in~(\ref{eq:uncertain_ham}): %TODO:need citations
	\begin{equation}\label{eq:open_system_dynamics}
		\dot{\rho}(t) = \left[\mathcal{L}(t) + \sum_{j=1}^{m}\epsilon_j\mathcal{E}_j\right] \rho(t),
	\end{equation}
	where $\rho(t)\in \mathbb{C}^{d\times d}$ is the density matrix, $\mathcal{L}(t)$ is the Lindblad superoperator acting on $\rho(t)$~\cite{breuer2002theory}, and $\mathcal{E}_j$ is the superoperator associated with the $j$-th uncertainty. Their representations are expressed as:
	\begin{equation}\label{eq:quantum_channel}
		\begin{aligned}
			\mathcal{L}(t)\rho&:= -i [H_S(t), \rho]+\sum_{i=1}^{n_d}\gamma_i(c_i \rho c_i^{\dagger} - \frac{1}{2}c_i^{\dagger} c_i \rho - \frac{1}{2}\rho c_i^{\dagger} c_i),\\
			\mathcal{E}_j\rho&:= -i [E_j, \rho],
		\end{aligned}
	\end{equation}
	where $c_1,\cdots,c_{n_d}$ are Lindblad operators representing the system's coupling to the environment, $\gamma_1,\cdots,\gamma_{n_d}$ are the corresponding coupling strengths, and $[A, B]:=AB-BA$ is the commutator.
	
    In practice, the uncertainty parameters are usually small, and thus we can expand the state $\rho(t)$ into a Taylor series as follows~\cite{van2017robust,cao2024robust}:
	\begin{equation}\label{eq:taylor}
		\rho(t) = \sum_{p_1,\cdots, p_{m}\geq 0}\epsilon_1^{p_1}\cdots \epsilon_{m}^{p_{m}}\rho_{p_1,\cdots,p_m}(t),
	\end{equation}
	where each term $\epsilon_1^{p_1}\cdots \epsilon_m^{p_m}$ captures the parametric dependence at a given perturbative order, and the corresponding expansion coefficient $\rho_{p_1,\cdots,p_m}(t)$ describes the system's response at that order. Replacing Eq.~(\ref{eq:taylor}) into Eq.~(\ref{eq:open_system_dynamics}) and comparing terms of equal order in $\epsilon_1^{p_1}\cdots\epsilon_m^{p_m}$, we find that the zero-th order coefficient $\rho_0(t)$ obeys the standard, noise-free Lindblad master equation:
	\begin{equation}\label{eq:lindblad}
		\dot{\rho}_0(t)  = \mathcal{L}(t) \rho_0(t),
	\end{equation}
	with the initial condition $\rho_0(0)=\rho(0)$. The higher-order terms follow the recurrence relation:
	\begin{equation}\label{eq:recurrence}
		\begin{aligned}
			\dot{\rho}_{p_1,\cdots,p_m}(t) & =  \mathcal{L}(t)\rho_{p_1,\cdots,p_m}(t) \\
            &\quad +\sum_{\substack{j=1 \\ p_j \geq 1}}^{m}\mathcal{E}_j\rho_{p_1,\cdots,p_j-1,\cdots,p_m}(t),
		\end{aligned}
	\end{equation}
    with the initial condition $\rho_{p_1,\cdots,p_m}(0)=0$. It becomes clear from Eq.~(\ref{eq:taylor}) that $\rho_0(T)$ represents the ideal final state evolved under the nominal master equation, while the terms $\rho_{p_1,\cdots,p_m}(T)$ capture varying degrees of deviations arising from the effects of system uncertainties. 

    As evaluating infinite terms in the expansion is usually impractical, we truncate the series to include only the dominant terms $\rho_{p_1,\cdots,p_m}$ up to the $n$th-order, where $0\leq p_1+\cdots+p_m\leq n$. The total number of these truncated terms is~\cite{puzzuoli2023algorithms} 
    \begin{equation}\label{eq:truncate_terms_N}
        N = \binom{m+n}{m}=\frac{(m+n)!}{m!n!}.
    \end{equation}
    Based on the recurrence relation~(\ref{eq:recurrence}), these terms can be organized into an augmented system and computed simultaneously. This formalism was proposed in Ref.~\cite{puzzuoli2023algorithms}, and is analogous to the block-matrix approach~\cite{van1978computing,carbonell2008computing} used in quantum control for gradient evaluation~\cite{goodwin2015auxiliary} and robustness enhancement~\cite{haas2019engineering,shao2024multiple,liu2024robust}. In the following, we generalize the framework from Ref.~\cite{puzzuoli2023algorithms} to address the dynamics of uncertain open quantum systems, represented in a superoperator form as outlined in Eqs.~(\ref{eq:lindblad}) and (\ref{eq:recurrence}).
	
    We stack all the truncated terms into a column in descending order to form an \textit{augmented state} $\vec{\rho}$, where each expansion order $(p_1,\cdots,p_m)$ is interpreted as a base-$(n+1)$ representation. For instance, when $m=2$ and $n=2$, the truncated terms are then rearranged as
    \begin{equation}\label{eq:m_2_n_2}
        \vec{\rho}(t)=\begin{bmatrix}\rho_{20}(t)\\ \rho_{11}(t)\\ \rho_{10}(t)\\ \rho_{02}(t)\\ \rho_{01}(t)\\ \rho_{00}(t)\end{bmatrix}
    \end{equation}
	following the descending order of their corresponding expansion orders $(p_1,p_2)$ in a base-3 representation. We denote the $k$-th sub-block in $\vec{\rho}(t)$ as $\rho_{[k]}(t)$. For this example, $\rho_{[1]}(t)=\rho_{20}(t)$, $\rho_{[2]}(t)=\rho_{11}(t)$, and so on. Utilizing the recurrence relationship from Eq.~(\ref{eq:recurrence}), we construct an augmented deterministic system for $\vec{\rho}(t)$ as:
	\begin{equation}\label{eq:augment_sys}
		\dot{\vec{\rho}}(t)=\mathbf{L}(t){\vec{\rho}}(t)+\sum_{j=1}^m\mathbf{E}_j{\vec{\rho}}(t),
	\end{equation}
	where 
	\begin{equation}\label{eq:L_and_E}
		\begin{aligned}
			\mathbf{L}(t)=\mathbb{I}_{N}\otimes \mathcal{L}(t),~~
			\mathbf{E}_j=\mathbb{R}_j\otimes \mathcal{E}_j.
		\end{aligned}
	\end{equation}
    Here, $\mathbb{I}_N$ denotes the $ N $-dimensional identity matrix, and the $N\times N$ matrix $\mathbb{R}_j$ is constructed according to Eq.~(\ref{eq:recurrence}): For each $1\leq k\leq N$, 
    let the order of $\rho_{[k]}(t)$ be represented as $(p_1,\cdots,p_j,\cdots,p_m)$. If there exists $\rho_{[l]}(t)$ with the order $(p_1,\cdots,p_j-1,\cdots,p_m)$, then $(\mathbb{R}_j)_{kl} = 1$. All other entries of $\mathbb{R}_j$ are zero. The concise formulation~(\ref{eq:augment_sys}) is convenient for adopting the S-T expansion, as will be detailed in Sec.~\ref{subsec:suzuki_trotter}.
 
	\subsection{Computational complexity for simulating the dynamics}
	\label{subsec:computational_complexity}
    In the simulations, the control pulses $u_c(t)$, for $c=1,\cdots,n_c$, are chosen to be piecewise constant over $[0, T]$. The control variables are the amplitudes $\{u_c(t_k)\}$ defined at every discretized time interval $\left[t_k, t_{k+1}\right)$, where $\Delta t=t_{k+1}-t_k=T/N_T$ and $k=0,\cdots,N_T-1$. Let the evolution over the $k$-th time step can be represented as 
    \begin{equation}\label{eq:S_k}
        \mathbf{S}_k = \exp\left\{\left[\mathbf{L}(t_k)+\sum_{j=1}^m\mathbf{E}_j\right]\Delta t\right\},
    \end{equation}
    then the final state is given by
	\begin{equation}
		\label{eq:exact_dynamics}
		\vec{\rho}(T) = \mathbf{S}_{N_T-1}\cdots \mathbf{S}_1\mathbf{S}_0\vec{\rho}(0).
	\end{equation} 
	A standard approach for deriving matrix representations of the superoperator $\mathbf{S}_k$ is to \emph{vectorize} $\vec{\rho}(t)$ into
    \begin{equation}\label{eq:vec_rho}
        \vec{\rho}_{\rm vec}(t)=\begin{bmatrix}\op{vec}\left[\rho_{n_1,\cdots,n_m}(t)\right]\\ \vdots \\ \op{vec}\left[\rho_0(t)\right]\end{bmatrix}\in\mathbb{C}^{Nd^2}
    \end{equation}
    where $\op{vec}\left[\rho_{p_1,\cdot,p_m}(t)\right]\in \mathbb{C}^{d^2}$ is a vector constructed by stacking all columns of $\rho_{p_1,\cdots,p_m}(t)$~\cite{schulte2011optimal,johansson2013qutip}. Therefore, the effective dimension of the augmented system is
    \begin{equation}\label{eq:effective_dim}
        d_{\rm aug}=Nd^2 = \frac{(m+n)!}{m!n!}d^2,
    \end{equation}
    which can be extremely high even for quantum systems of moderate size. The superoperators $\mathcal{L}(t)$ and $\mathcal{E}_j$ can then be represented as $d^2\times d^2$ matrices under properly chosen basis of $\rho(t)$~\cite{havel2003robust}. Consequently, $\mathbf{L}(t)$ and $\mathbf{E}_j$ become $Nd^2 \times Nd^2$ matrices~\cite{schulte2011optimal,floether2012robust,goerz2014optimal}. See Appendix~\ref{app:superoperator} for their detailed expressions. The term $\mathbf{S}_k$ defined in Eq.~(\ref{eq:S_k}) can then be evaluated by matrix exponentiation, which typically involves multiple matrix multiplications of size $Nd^2\times Nd^2$~\cite{higham2005scaling,al2010new}. The computational complexity of multiplying two dense matrices of dimensions $(n_a, n_b)$ and $(n_b, n_c)$ is taken as $\mathcal{O}(n_an_bn_c)$~\cite{boutin2017resonator}. In this context, without accounting for potential speedups due to matrix sparsity, each computation of $\mathbf{S}_k$ entails operations of complexity $\mathcal{O}(Nd^2\times Nd^2\times Nd^2) = \mathcal{O}(N^3d^6)$~\cite{boutin2017resonator,chen2025robust}. This computational cost becomes increasingly prohibitive as the robustness order or the system dimension grows. 
	
    To speed up the computation, we may  use high-performance ODE solvers that are customized for calculating evolution involved in quantum system dynamics~\cite{johansson2013qutip,virtanen2020scipy, kramer2018quantumoptics,puzzuoli2023qiskit,puzzuoli2023algorithms}. For example, QuTiP~\cite{johansson2013qutip} incorporates a Fortran-based solver~\cite{hindmarsh1983odepack} that can employ a linear multi-step method~\cite{Hairer1993} to solve the ODE described by Eq.~(\ref{eq:augment_sys}). The numerical integration depends on the product of the supermatrices with the vector $\vec{\rho}_{\rm vec}(t)$, reducing the complexity to $\mathcal{O}(Nd^2\times Nd^2\times 1)=\mathcal{O}(N^2d^4)$ for propagating a single initial state $\vec{\rho}_{\rm vec}(0)$. Moreover, this solver well exploits the sparsity of typical quantum system Hamiltonians to further accelerate the computation, as illustrated in Sec.~\ref{sec:simulation}.
	
    The ODE solver can only compute the evolution of one state trajectory. To determine the entire process for quantum gate design, the evolution of a complete set of initial states is needed. It is evident from Eq.~(\ref{eq:open_system_dynamics}) that the $d^2$ basis states spanning the Liouville space of density matrices provide a sufficient choice~\cite{schulte2011optimal,floether2012robust}, resulting in a total complexity of $\mathcal{O}(N^2d^4\times d^2)=\mathcal{O}(N^2d^6)$ per optimization iteration. This computational cost becomes unbearable when $d$ is large. To address this issue, Sec.~\ref{subsec:suzuki_trotter} introduces the S-T expansion that offers additional speedup for solving the dynamics~(\ref{eq:augment_sys}). Furthermore, Sec.~\ref{subsec:robust_control_objectives} proves that the number of basis states required for robust gate design can be reduced to $d+1$.

	\subsection{Suzuki-Trotter expansion}
	\label{subsec:suzuki_trotter}
    The Suzuki-Trotter expansion (S-T expansion)~\cite{suzuki1991general,childs2021theory} is a widely applied method for approximating the evolution of quantum systems~\cite{bhole2018practical,jensen2021achieving, jensen2021approximate,han2021experimental}, offering a practical balance between computational efficiency and numerical accuracy.  In this section, we generalize the S–T expansion approach proposed in Refs.~\cite{bhole2018practical,jensen2021approximate}, which was originally developed for control optimization of closed systems, to the augmented system~(\ref{eq:augment_sys}) formulated for robust control design of open quantum systems. As shown in the subsequent analysis, the computational complexity of evolving the augmented state using the adapted method scales approximately as $\mathcal{O}(Nd^3)$, demonstrating a significant reduction compared to the $\mathcal{O}(N^2d^4)$ complexity required by ODE solvers.
	
	Following the idea of Ref.~\cite{jensen2021approximate}, we first consider the case where there is only one diagonal control Hamiltonian $H_c$. The more general cases will be discussed in Appendix~\ref{app:multi_control}. The Lindblad evolution $\mathcal{L}(t)$ described in Eq.~(\ref{eq:quantum_channel}) can be reformulated as~\cite{daley2014quantum}%todo: need to consider the multi-control case.
	\begin{equation}
		\begin{aligned}
			&\mathcal{L}(t) \rho(t) \\
			& = -i\left[H_{\rm eff}\rho(t)-\rho(t) H_{\rm eff}^\dagger\right] -iu_c(t)[H_c, \rho(t)] \\
            &\hspace{1em} + \sum_{i=1}^{n_d} \gamma_ic_i\rho(t) c_i^\dagger
		\end{aligned}
	\end{equation}
	where $H_{\rm eff} = H_0 - \frac{i}{2}\sum_{i=1}^{n_d}\gamma_ic_i^\dagger c_i$ is the effective drift Hamiltonian. We define the associated superoperators as follows:
    \begin{equation}
        \begin{aligned}
        \mathcal{H}_{\rm eff}\rho(t)&:=-i\left[H_{\rm eff}\rho(t)-\rho(t) H_{\rm eff}^\dagger\right],\\
        \mathcal{H}_{c}\rho(t)&:=-i\left[H_{c}, \rho(t)\right],\\
        \mathcal{C}\rho(t)&:=\sum_{i=1}^{n_d} \gamma_ic_i\rho(t) c_i^\dagger,
    \end{aligned}
    \end{equation}
    so that the Lindblad evolution can be written as $\mathcal{L}(t)=\mathcal{H}_{\rm eff}+u_c(t)\mathcal{H}_c+\mathcal{C}$. Denote the Kronecker product of $\mathbb{I}_N$ with the superoperator $\mathcal{H}_{\rm eff}$, $\mathcal{H}_c$ and $\mathcal{C}$ by $\mathbf{H}_{\rm eff}$, $\mathbf{H}_{c}$ and $\mathbf{C}$, respectively. Following the definition~(\ref{eq:S_k}), the evolution $\mathbf{S}_k$ can be expressed as
    \begin{equation}
        \mathbf{S}_k=\exp\left\{\left[\mathbf{H}_{\rm eff}+u_c(t_k)\mathbf{H}_c+\mathbf{C}+\sum_{j=1}^m\mathbf{E}_j\right]\Delta t\right\}.
    \end{equation}
    
    The first-order S-T expansion for approximating $\mathbf{S}_k$ is given by~\cite{childs2021theory}
	\begin{equation}
		\begin{aligned}
		\mathbf{S}_k = e^{\mathbf{H}_{\rm eff}\Delta t}e^{u_c(t_k)\mathbf{H}_{c}\Delta t}e^{\mathbf{C}\Delta t}\left(\prod_{j=1}^{m}e^{\mathbf{E}_j\Delta t}\right) + \mathcal{O}(\Delta t^2),
		\end{aligned}
	\end{equation}
	where $\prod_{j=1}^{m}e^{\mathbf{E}_j\Delta t}=e^{\mathbf{E}_m\Delta t}\cdots e^{\mathbf{E}_2\Delta t}e^{\mathbf{E}_1\Delta t}$. The second-order expansion increases the approximation precision by re-ordering the perturbation terms as~\cite{bhole2018practical, childs2021theory, jensen2021approximate}:
	\begin{equation}\label{eq:2order_trotter}
		\begin{aligned}
			\mathbf{S}_k		=\hat{\mathbf{S}}_k+\mathcal{O}(\Delta t^3),
		\end{aligned}
	\end{equation}
	where
	\begin{equation}\label{eq:trotter_s}
		\begin{aligned}
			\hat{\mathbf{S}}_k &= 
			\left( \prod_{j=m}^{1} e^{\mathbf{E}_j\frac{\Delta t}{2}} \right) 
			e^{\mathbf{C}\frac{\Delta t}{2}} 
			e^{u_c(t_k)\mathbf{H}_{c}\frac{\Delta t}{2}} \\
			&\quad \times e^{\mathbf{H}_{\rm eff}\Delta t} 
			e^{u_c(t_k)\mathbf{H}_{c}\frac{\Delta t}{2}} 
			e^{\mathbf{C}\frac{\Delta t}{2}} 
			\left( \prod_{j=1}^{m} e^{\mathbf{E}_j\frac{\Delta t}{2}} \right).
		\end{aligned}
	\end{equation}
	Using the approximated propagators~(\ref{eq:trotter_s}), one can approximate $\vec{\rho}(T)$ as~\cite{han2021experimental}:
	\begin{equation}\label{eq:rho_approx}
		\vec{\rho}(T) = \hat{\vec{\rho}}(T)+\mathcal{O}(\Delta t^2),
	\end{equation}
	where
	\begin{equation}\label{eq:rho_approx_2}
		\hat{\vec{\rho}}(T) = \hat{\mathbf{S}}_{N_T-1}\cdots \hat{\mathbf{S}}_1\hat{\mathbf{S}}_0 \vec{\rho}(0),
	\end{equation}
	and the error decreases quadratically with a finer sampling step $\Delta t$.
	
    To illustrate the computational benefits of the S-T expansion in robust control design, we now detail the decomposition of each term in Eq.~(\ref{eq:trotter_s}) and analyze their respective computational properties.
	
    (i) The calculation of the propagators $e^{\mathbf{H}_{\rm eff}\Delta t}$ and $e^{u_c(t_k)\mathbf{H}_{c}\frac{\Delta t}{2}}$ scales linearly with $N$. The operations of these propagators on an augmented state $\vec{\rho}$ with $N$ sub-blocks can be expressed as:
    \begin{equation}\label{eq:Ueff}
        \begin{aligned}
            e^{\mathbf{H}_{\rm eff}\Delta t} \vec{\rho}  & =\begin{bmatrix}
                U_{\rm eff}\rho_{[1]} U_{\rm eff}^\dagger \\
                \vdots \\
                U_{\rm eff}\rho_{[N]} U_{\rm eff}^\dagger
            \end{bmatrix},\\
            e^{u_c(t_k)\mathbf{H}_{c}\frac{\Delta t}{2}} \vec{\rho} 
            & =\begin{bmatrix}
                U_{c}(t_k)\rho_{[1]} U_{c}^\dagger(t_k) \\
                \vdots \\
                U_{c}(t_k)\rho_{[N]} U_{c}^\dagger(t_k)
            \end{bmatrix} ,
        \end{aligned}
    \end{equation}
    where $U_{\rm eff}=e^{-iH_{\rm eff}\Delta t}$, $U_c(t_k)=e^{-iu_c(t_k)H_c\frac{\Delta t}{2}}$. Since $U_{\rm eff}$ is time-independent, it only needs to be computed once, with a cost of $\mathcal{O}(d^3)$, and can be saved for repeated use. Under proper basis, the control Hamiltonian $H_c$ can be treated as diagonal, enabling $U_c(t_k)$ to be calculated with minimal effort as the element-wise exponential of its diagonal entries, resulting in a complexity of $\mathcal{O}(d)$. From Eq.~(\ref{eq:Ueff}), it becomes evident that the computational complexity scales as $\mathcal{O}(N d^3)$.
	
    (ii) The calculation of $e^{\mathbf{E}_j\frac{\Delta t}{2}}$ can be simplified by the nilpotent property of $\mathbb{R}_j$ defined in Eq.~(\ref{eq:L_and_E}), i.e., $\mathbb{R}_j^\alpha=0$ for $\alpha>n$. This implies that $\mathbf{E}_j$ is also nilpotent:
	\begin{equation}
		\begin{aligned}
			\mathbf{E}_j^\alpha = \mathbb{R}_j^\alpha\otimes \mathcal{E}_j^\alpha = 0,~~\forall \alpha>n.
		\end{aligned}
	\end{equation}
	The proof is provided in Appendix~\ref{app:nilpotency}. Therefore, $e^{\mathbf{E}_j\frac{\Delta t}{2}}$ in Eq.~(\ref{eq:trotter_s}) includes only a finite number of terms:
	\begin{equation}\label{eq:expE}
		e^{\mathbf{E}_j\frac{\Delta t}{2}}=I+\frac{\Delta t}{2}\mathbf{E}_j+\cdots + \frac{1}{n!}\left(\frac{\Delta t}{2}\mathbf{E}_j\right)^{n},
	\end{equation}
    since all higher-order Taylor terms vanish. By successively applying $\mathbf{E}_j$ for $n$ iterations, we can construct $e^{\mathbf{E}_j\frac{\Delta t}{2}}\vec{\rho}$ following the formulation~(\ref{eq:expE}). We prove that the computational complexity of the process is approximately $\mathcal{O}(Nd^3)$ when the robustness order $n$ is not large. The detailed algorithm for computing Eq.~(\ref{eq:expE}), along with its complexity analysis, is presented in Appendix~\ref{app:nilpotency}.
	
    (iii) Since modern devices for quantum information applications can maintain low decoherence rates $\gamma_i$, the propagator $e^{\mathbf{C}\frac{\Delta t}{2}}$ can be truncated to the second order as: 
    \begin{equation}\label{eq:C_rho}
        e^{\mathbf{C}\frac{\Delta t}{2}} \vec{\rho} = \vec{\rho} + \frac{\Delta t}{2}\mathbf{C} \vec{\rho} + \frac{\Delta t^2}{8}\mathbf{C}^2 \vec{\rho} + \mathcal{O}(\gamma^3\Delta t^3),
    \end{equation}
    where 
    \begin{equation}\label{eq:collapse}
		\mathbf{C}\vec{\rho}= 
            \begin{bmatrix}
                \sum_{i=1}^{n_d}\gamma_i c_i\rho_{[1]}c_i^\dagger \\
                \vdots \\
                \sum_{i=1}^{n_d}\gamma_i c_i\rho_{[N]}c_i^\dagger
            \end{bmatrix}.
	\end{equation}
    The approximation error introduced by Eq.~(\ref{eq:C_rho}) is of the same order in $\Delta t$ as the S-T approximation error in Eq.~(\ref{eq:2order_trotter}), thus the overall approximation accuracy described in Eq.~(\ref{eq:trotter_s}) remains valid. The computational cost of this approximation also scales as $\mathcal{O}(Nd^3)$. It is worth noting that, in practice, matrix sparsity of the Lindblad operators can also be utilized to accelerate the computation of Eq.~(\ref{eq:collapse}), especially when the number of Lindblad operators $n_d$ is sizable. For further details, see Appendix~\ref{app:superoperator}.
	
    The S-T expansion described by Eq.~(\ref{eq:trotter_s}) divides the overall evolution operator into multiple operators, each of which can be evaluated with reduced computational complexity. Table~\ref{tab:complexity} compares the computational complexity of different approaches. The S-T expansion offers potential speedup when the Hilbert dimension $d$ or the robustness order $n$ is modestly large, achieved by tolerating a controlled approximation error of order $\mathcal{O}(\Delta t^2)$ in the resulting quantum state, as shown in Eq.~(\ref{eq:rho_approx}). We evaluate the approximation performance with parameters typical of modern arbitrary waveform generators (AWGs) in Sec.~\ref{sec:simulation}, showing that the second-order S-T expansion is accurate enough for robust control design. 

	\begin{table}[h]
		\centering
		\caption{Computational complexity of different methods for evolving the augmented state in terms of the system dimension $d$ and the number of truncation terms $N$ (see Eq.~(\ref{eq:truncate_terms_N}) for the expression of $N$). Here, we do not account for additional speedups that may arise from matrix sparsity.}
		\begin{tabular}{l| c}
			\toprule
			~Method & ~Computational Complexity~ \\
			\midrule
			~Direct Matrix Exponential~ & \( \mathcal{O}(N^3 d^6) \) \\
			~ODE Solver & \( \mathcal{O}(N^2 d^4) \) \\
			~Suzuki-Trotter & \( \mathcal{O}(N d^3) \) \\
			\bottomrule
		\end{tabular}
		\label{tab:complexity}
	\end{table}

	\section{Robust control design for open quantum systems}
	\label{sec:robust_control_design}
	Based on the above augmented-system formulation, this section will propose robust control objective functions for quantum state preparation and quantum gate synthesis. Subsequently, a gradient-based algorithm is introduced to perform robust control optimization.
	
	\subsection{Robust control objective for state preparation}
	\label{subsec:robust_control_objectives}
	The state preparation task aims to steer the system's state toward a target state $\rho_{\rm targ}$, usually accomplished by optimizing the control to maximize the fidelity defined by the overlap function:
	\begin{equation}\label{eq:state_overlap}
F\left[\rho(T)\right]:=\tr\left[\rho(T)\rho_{\rm targ}\right]
	\end{equation}
    between $\rho_{\rm targ}$ and the final state $\rho(T)$. An optimal control is said to be robust if it can maintain high fidelity in the presence of weak uncertainties. According to Eq.~(\ref{eq:taylor}), this requirement translates to realizing $\rho_0(T)=\rho_{\rm targ}$, with all higher-order terms equal to zero. Since in practice the uncertainties are usually weak, only the first few dominant higher-order terms are to be mitigated. In this regard, a control is said to be $n$th-order robust if
	\begin{equation}\label{eq:robustness_condition}
		\rho_0(T)=\rho_{\rm targ},~~ \rho_{p_1,\cdots,p_m}(T)=0
	\end{equation}
	for all $0<p_1+\cdots+p_m\leq n$. By evolving the augmented system (\ref{eq:augment_sys}), we can solve all the truncated terms at the final time $T$. Consequently, an $n$th-order robust control should achieve the augmented state transfer:
	\begin{equation}\label{eq:aug_state_transfer}
		\vec{\rho}(0)=\begin{bmatrix}
			0 \\
			\vdots \\
			0 \\
			\rho(0)
		\end{bmatrix}\to
		\vec{\rho}(T)=\begin{bmatrix}
			0 \\
			\vdots \\
			0 \\
			\rho_{\rm targ}
		\end{bmatrix},
	\end{equation}
    which can be sought by maximizing the following robust control objective function~\cite{propson2022robust,puzzuoli2023algorithms,weidner2024robust,shao2024multiple,liu2024robust}:
	\begin{equation}\label{eq:robust_objective_state}
		\begin{aligned}
			J\left[\vec{\rho}(T)\right] =  F\left[\rho_{0}(T)\right] - \frac{1}{2}\sum_{p_1,\cdots,p_m}\lambda_{p_1,\cdots,p_m}\|\rho_{p_1,\cdots,p_m}(T)\|^2_F,
		\end{aligned}
	\end{equation}
	where $\|\cdot\|_F$ is the Frobenious norm, and $\lambda_{p_1,\cdots,p_m}>0$ are weights used to balance the objectives of reaching the target state and mitigating higher-order perturbation terms. 
    
    Note that the objective function defined in Eq.~(\ref{eq:robust_objective_state}) used in this work is different with the commonly used average performance~\cite{wu2019learning, ge2021risk, dalgaard2022dynamical,chen2025robust}, defined as
	\begin{equation}\label{eq:average_gate_performance}
		F_{\rm avg} = \langle F\rangle_{\epsilon\sim\mathcal{P}} = \mathbb{E}_{\epsilon\sim\mathcal{P}}\left\{F\left[\rho(T)\right]\right\},
	\end{equation}
	where $\mathbb{E}_{\epsilon\sim\mathcal{P}}$ denotes the expectation value over the distribution $\mathcal{P}$ of uncertain parameters $\epsilon$. Upon certain assumptions, $F_{\rm avg}$ can be approximated as
	\begin{equation}\label{eq:average_control_objective}
		F_{\rm avg}\approx \tilde{J} = \tr\left[\rho_{\rm targ} \rho_0(T)\right] + \sum_{i=1}^{n_d}\sigma_{i}^2 \tr\left[\rho_{\rm targ} \rho_{ii}(T)\right], 
	\end{equation}
	where $\sigma_i^2$ is the variance of $\epsilon_i$, and $\rho_{ii}(t)$ is the expansion coefficient associated with the term $\epsilon_i^2$ described in Eq.~(\ref{eq:taylor}) (see Appendix~\ref{app:average_gate_performance} for details). The formulation of $\tilde{J}$ agrees with the result in Ref.~\cite{dalgaard2022dynamical}, where a similar perturbative treatment of system uncertainty yields a second-order correction to fidelity. It also resembles the objective function optimized in Ref.~\cite{chen2025robust}, with some subtle differences in implementation details. Notably, $\tilde{J}$ is fully determined once the noise distribution is specified. In contrast, the objective function $J$ offers greater flexibility in balancing robustness and nominal performance, as the weights $\lambda_{p_1\cdots,p_m}$ can be tuned to selectively emphasize or suppress the sensitivity to specific uncertainty directions or perturbative orders. This tunability allows the optimization to be tailored to the specific demands of a given application. In our simulations, by appropriately choosing values for $\lambda_{p_1,\cdots,p_m}$, optimizing $J$ yields control fields with improved robustness compared to those obtained by optimizing $\tilde{J}$, as illustrated in Fig.~\ref{fig:gs_uniform_sampling}. A more comprehensive evaluation of the effectiveness of the two objective functions is left for future work.

    \subsection{Robust control objective for gate synthesis}
	The quantum gate synthesis task aims to realize a high-fidelity implementation of a target gate $U_{\rm targ}$, which is an essential building block for large-scale quantum computation and error correction~\cite{barends2014superconducting}. In the closed-system dynamics, this task can be accomplished by transforming a set of states $\{\ket{\psi_i}\}_{i=1,\cdots,d}$, which form the basis of the $d$-dimensional Hilbert space, into their corresponding target states $\{U_{\rm targ}\ket{\psi_i}\}_{i=1,\cdots,d}$~\cite{palao2003optimal}. Existing studies directly extend this concept to the design of robust quantum gates in open quantum systems through optimizing $d^2$ robust state preparation tasks (i.e., the optimization of $J$ defined in Eq.~(\ref{eq:robust_objective_state})), which span the basis of the Liouville space~\cite{floether2012robust,haas2019engineering,puzzuoli2023algorithms,liu2024robust}. However, the quadratic scaling of these subroutines with the system dimension results in substantial computational costs.
	
    Thanks to the unitarity of quantum gates, $U_{\rm targ}$ is perfectly realized if and only if three carefully chosen initial density matrices $\{\rho^{(i)}(0)\}_{i=1,2,3}$ are steered to $\{U_{\rm targ}\rho^{(i)}(0)U_{\rm targ}^\dagger\}_{i=1,2,3}$~\cite{reich2013minimum}. The optimization demonstrates faster convergence by using a selected set of $d+1$ states $\{\rho^{(i)}(0)\}_{i=1,\cdots,d+1}$~\cite{reich2013minimum, goerz2014optimal}, and the quantum gate synthesis problem can be formulated as:
    \begin{equation}\label{eq:quantum_gate_synthesis}
		\max_{u(t)} \sum_{i=1,\cdots,d+1}w_iF\left[\rho^{(i)}(T)\right],
    \end{equation}
    where $w_i>0$ are weights satisfying $\sum_i w_i=1$.	The selection of $\{\rho^{(i)}(0)\}_{i=1,\cdots,d+1}$ is summarized in Appendix~\ref{app:density_matrix}, with the detailed proof found in Ref.~\cite{reich2013minimum, goerz2014optimal}. Similarly, an optimal control for gate synthesis is said to be $n$th-order robust if $\rho^{(i)}_0(T)=U_{\rm targ}\rho^{(i)}(0)U_{\rm targ}^\dagger$ and
	\begin{equation}
	\rho^{(i)}_{p_1,\cdots,p_m}(T)=0
	\end{equation}
	for all $i=1,\cdots,d+1$ and $0< p_1+\cdots+p_m\leq n$. Therefore, a robust gate can be synthesized by maximizing the objective function:
	\begin{equation}\label{eq:robust_gate_synthesis}
		\max_{u(t)} \sum_{i=1,\cdots,d+1}w_iJ\left[\vec{\rho}^{(i)}(T)\right].
	\end{equation}
	The result in Ref.~\cite{goerz2014optimal} is effectively extended to robust control design, significantly reducing the number of robust state preparation tasks from $d^2$ to $(d+1)$. Furthermore, the optimization of the function~(\ref{eq:robust_gate_synthesis}) facilitates parallelized computing, as discussed in Sec.~\ref{subsec:gradient_optimization}.
	
	\subsection{Gradient-based optimization}
	\label{subsec:gradient_optimization}
	Quantum optimal control design has been shown to exhibit a trap-free landscape under the assumptions of sufficient control resources~\cite{rabitz2004quantum, rabitz2006optimal, rabitz2005landscape}, suggesting that local gradient-based optimization algorithms can achieve optimal control solutions. The GRAPE algorithm~\cite{khaneja2005optimal}, as a renowned gradient-based optimization method for quantum control design, operates effectively when the control is piecewise constant, as assumed in Sec.~\ref{subsec:computational_complexity}. In this work, we employ GRAPE to maximize the objective function $J\left[\vec{\rho}(t)\right]$ defined in Eq.~(\ref{eq:robust_objective_state}). The main procedure is summarized as follows:
	
	Step 1: Propagate the dynamics~(\ref{eq:augment_sys}) forward, as described in Eq.~(\ref{eq:exact_dynamics}), and store the intermediate states $\vec{\rho}(t_k)=\mathbf{S}_{k-1}\cdots \mathbf{S}_1\mathbf{S}_0 \vec{\rho}(0)$ for $k=1,\cdots,N_T$.
	
	Step 2: Propagate the dynamics~(\ref{eq:augment_sys}) backward as:
	\begin{equation}\label{eq:backpropagate}
		\vec{O}(t_{k+1}) = \mathbf{S}^\dagger_{k+1}\cdots \mathbf{S}^\dagger_{N_T-2}\mathbf{S}^\dagger_{N_T-1}\vec{O}(T),
	\end{equation} 
    for $k=0,\cdots,N_T-1$. Here, $\mathbf{S}^\dagger_k$ is the adjoint of the propagation $\mathbf{S}_k$ (see Appendix~\ref{app:superoperator} for its detailed expression), and $\vec{O}(T)$ is the co-state defined as~\cite{goerz2022quantum}
	\begin{equation} %todo: check its correctness
		\begin{aligned}
			\vec{O}(T) = \nabla_{\vec{\rho}(T)}J\left[\vec{\rho}(T)\right] = 
            \begin{bmatrix}
                -\lambda_{n_1,\cdots,n_m}\rho_{n_1,\cdots,n_m}(T)\\
                \vdots\\
                -\lambda_{0,\cdots,0,1}\rho_{0,\cdots,0,1}(T)\\
                \rho_{\rm targ}
            \end{bmatrix}.
		\end{aligned}
	\end{equation}
	
	Step 3: Compute the gradient $\frac{\partial J}{\partial u_c(t_k)}$ with respect to each control variable $u_c(t_k)$ as
    \begin{equation}
        \frac{\partial J}{\partial u_c(t_k)} = \tr\left[\vec{O}^\dagger(t_{k+1})\frac{\partial \mathbf{S}_k}{\partial u_c(t_k)}\rho(t_k)\right].
    \end{equation}
	
	Step 4: Update the control field along the gradient-ascent direction:
	\begin{equation}\label{eq:gradient_update}
		u_c(t_k)\leftarrow u_c(t_k)+\alpha\frac{\partial J}{\partial u_c(t_k)},
	\end{equation}
    for $c=1,\cdots,n_c$ and $k=0,\cdots,N_T-1$. Here, $\alpha>0$ is a small learning rate. In practice, quasi-Newton optimizers, like L-BFGS-B~\cite{byrd1995limited}, are efficient~\cite{goerz2022quantum}. These optimizers utilize previously calculated gradients to approximate the Hessian, facilitating the identification of an improved local search direction for faster convergence.
	
	Step 5: Repeat Steps 1-4 until $u_c(t)$ converges.
	
	In Secs.~\ref{subsec:computational_complexity} and \ref{subsec:suzuki_trotter}, we introduce an ODE solver and the S-T expansion method for computing the forward propagation in Step 1. For clarity, we refer to the algorithm that uses the ODE solver as simply GRAPE, and the one using the S-T expansion as ST-GRAPE. The backward propagation outlined in Step 2 involves the adjoint propagator $\mathbf{L}^\dagger(t_{\alpha})$ and $\mathbf{E}^\dagger_j$, with their expressions and implementations provided in Appendix~\ref{app:superoperator}. This backward evolution has the same computational complexity as the forward mode presented in Table 1. 
	
	The evaluation of the gradient $\frac{\partial J}{\partial u_c(t_k)}$ in Step 3 differs between the two methods. In GRAPE, the gradient can be approximated to the first order as~\cite{khaneja2005optimal, jensen2021approximate}
	\begin{equation}\label{eq:gradient_J}
        \begin{aligned}
            \frac{\partial J}{\partial u_c(t_k)} & \approx \tr\left[\vec{O}^\dagger(t_{k+1})(\mathbf{H}_{c}\Delta t)\mathbf{S}_k\vec{\rho}(t_{k})\right]\\
            & =\tr\left[\vec{O}^\dagger(t_{k+1})(\mathbf{H}_{c}\Delta t)\vec{\rho}(t_{k+1})\right],
        \end{aligned}
	\end{equation}
	which is valid for small $\Delta t$. While higher-order approximations can be derived, the associated computational cost grows significantly with the system dimension~\cite{jensen2021approximate}. This scaling is problematic for the system modeled by Eq.~(\ref{eq:augment_sys}), where the state space typically exhibits a large dimensionality of $Nd^2$. Therefore, we follow the standard protocol~\cite{khaneja2005optimal} that employs the first-order approximation of the gradient for optimization. 
	
	In ST-GRAPE, the objective function $J\left[\vec{\rho}(T)\right]$ is approximated by $\hat{J}:=J\left[\hat{\vec{\rho}}(T)\right]$. The propagator in Eq.~(\ref{eq:trotter_s}) manipulated by $u_c(t_k)$ is $e^{u_c(t_k)\mathbf{H}_{c}\frac{\Delta t}{2}}$, whose derivative at $u_c(t_k)$ is 
	\begin{equation}\label{eq:st_gradient}
		\frac{\partial e^{u_c(t_k)\mathbf{H}_{c}\frac{\Delta t}{2}}}{\partial u_c(t_k)}=\frac{\Delta t}{2}\mathbf{H}_{c}e^{u_c(t_k)\mathbf{H}_{c}\frac{\Delta t}{2}} 
	\end{equation}
	Notably, this gradient~(\ref{eq:st_gradient}) is exact, and the expression for $\frac{\partial \hat{J}}{\partial u(t_k)}$ can be computed as
	\begin{equation}\label{eq:gradient_hatJ}
		\frac{\partial \hat{J}}{\partial u(t_k)} = \tr\left[\hat{\vec{O}}^\dagger(t_{k+1})\frac{\partial \hat{\mathbf{S}}(t_k)}{\partial u(t_k)}\hat{\vec{\rho}}(t_k)\right],
	\end{equation}
	where $\frac{\partial \hat{\mathbf{S}}(t_k)}{\partial u(t_k)}$ can be obtained by plugging the result from Eq.~(\ref{eq:st_gradient}) into the formulation~(\ref{eq:trotter_s}), following the rule of differentiation~\cite{jensen2021approximate}. 
	
\begin{figure}
		\centering
		\includegraphics[width=\columnwidth]{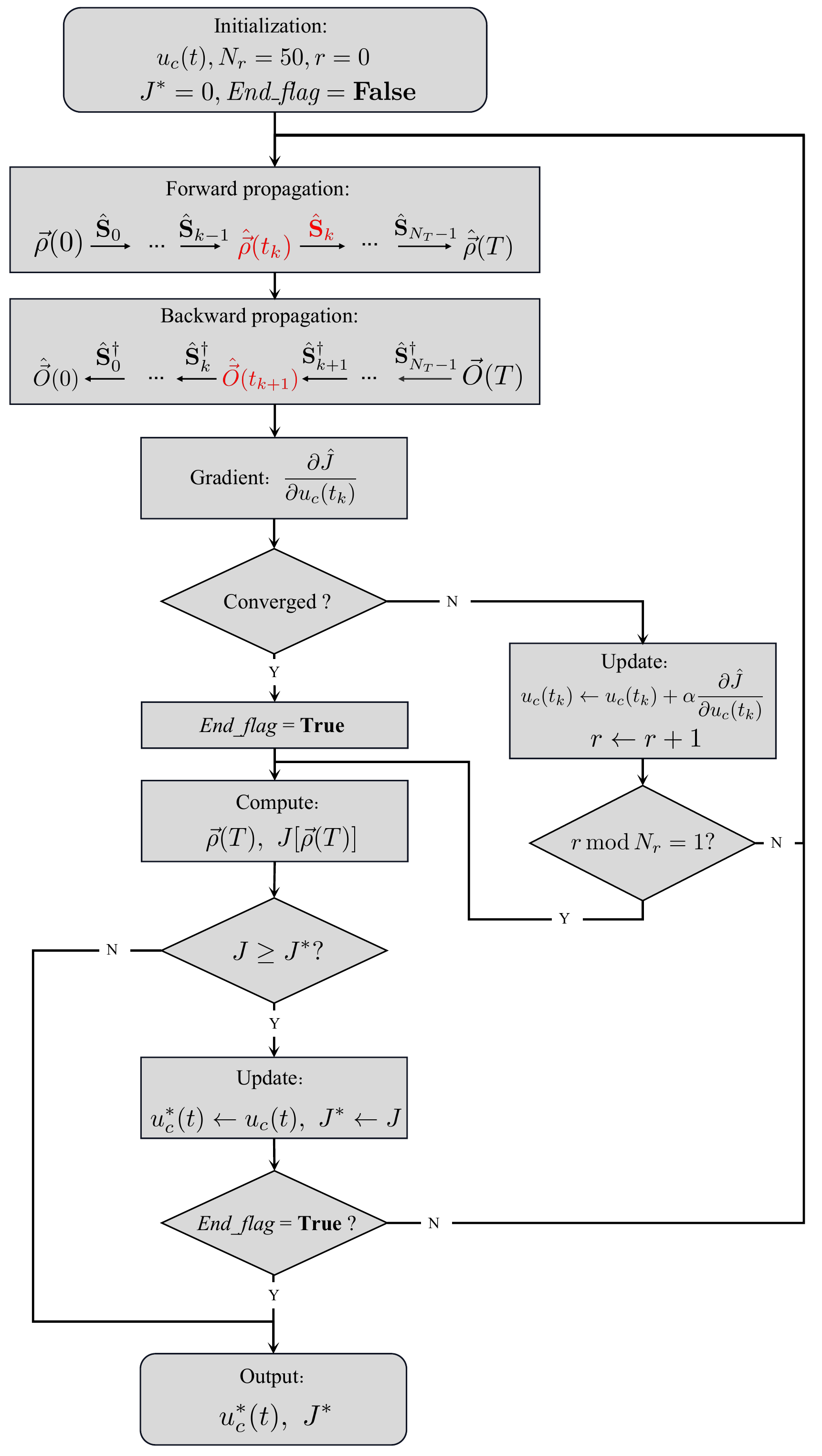}
		\caption{Flow chart of the ST-GRAPE algorithm. Here, $r$ and $N_r$ denote the current optimization iteration and the interval at which the true objective function $J$ is evaluated, respectively. The value $N_r=50$ is set in our simulations, indicating that we compute the true objective function $J$ every $50$ iterations. The terms $u_c^*(t)$ and $J^*$ are the optimal control solution and the corresponding function value. The notations highlighted in red indicate the terms involved in computing the gradient $\frac{\partial \hat{J}}{\partial u_c(t_k)}$, as described in Eq.~(\ref{eq:gradient_hatJ}). }
		\label{fig:flow_chart}
	\end{figure}

	According to Eq.~(\ref{eq:rho_approx}), the discrepancy between the Trotterized objective function $\hat{J}$ and the exact objective function $J$ is of the order $\mathcal{O}(\Delta t^2)$, which may not be negligible when the optimization of $\hat{J}$ approaches an optimum that is sensitive to variations in control parameters. To avoid being trapped by the potential "false" optima, we periodically examine the true value of $J$ during the optimization process (every 50 iterations in our simulations). When $J$ shows a decrease, we terminate the optimization because the S-T approximation has led to errors that cannot be ignored. The optimized result from the most recent evaluation is then adopted as the final solution. This strategy effectively balances computational efficiency with algorithmic performance, as shown in Sec.~\ref{sec:simulation}. Figure~\ref{fig:flow_chart} illustrates the flowchart of the ST-GRAPE algorithm.
	
    Although $\hat{J}$ serves as an estimation of the true objective function $J$, the exactness of the gradient~(\ref{eq:gradient_hatJ}) benefits quasi-Newton optimizers according to Ref.~\cite{jensen2021approximate}. In algorithms like BFGS and L-BFGS-B, the gradient approximation error can accumulate over successive iterations, potentially leading to a coarse estimation of the Hessian and an increased risk of premature termination of the optimization. While in the context of open quantum systems, decoherence imposes a fundamental limit on the achievable fidelity~\cite{jensen2021approximate}. Nevertheless, ST-GRAPE remains a powerful tool for achieving robust controls under these constraints.
	
	The computational complexities of the gradient evaluations in Eqs.~(\ref{eq:gradient_J}) and (\ref{eq:gradient_hatJ}) are $\mathcal{O}(N^2d^4)$ and $\mathcal{O}(Nd^3)$, respectively, which align with their costs of propagating the dynamics in Steps~1 and 2. These algorithms focus on optimizing $J\left[\vec{\rho}(T)\right]$ in the state preparation task. Similarly, a robust quantum gate can be synthesized by optimizing the weighted summation of $J\left[\vec{\rho}^{(i)}(T)\right]$ as defined in Eq.~(\ref{eq:robust_gate_synthesis}). 
	Since the evolution of each $\vec{\rho}^{(i)}(t)$ within a single iteration is independent, the gate optimization can be readily parallelized across multiple cores~\cite{johansson2013qutip,lambert2024qutip}. In this scheme, each core computes the gradient for its assigned $i$-th subroutine, and the results are summed according to the weights $\{w_i\}_{i=1,\cdots,d+1}$ at the end of every iteration to update the control.

	\section{Simulation results}
	\label{sec:simulation}
	In this section, we numerically demonstrate the effectiveness of the algorithms proposed in Sec.~\ref{subsec:gradient_optimization} for robust quantum control design. We consider a spin-chain of $N_q$ qubits with the following Hamiltonian:
	\begin{equation}\label{eq:spin_chain}
		\begin{aligned}
			H(t) & = J_{XY}\sum_{i=1}^{N_q-1} (\sigma^x_i\sigma^x_{i+1}+\sigma^y_i\sigma^y_{i+1}) \\
			& \quad + \sum_{i=1}^{N_q}\left[ u_{i,x}(t)\sigma^x_i + u_{i,y}(t)\sigma^y_i\right],
		\end{aligned}
	\end{equation}
	where $\sigma_i^{\alpha}$ with $\alpha\in\{x,y,z\}$ are the Pauli matrices of the $i$-th qubit, and $J_{XY}/2\pi=30$~MHz are the $XY$ interaction strengths between nearby qubits. The decoherence effects are modeled by Lindblad operators associated with the $i$-th qubit:
	\begin{equation}
		c_{i,1}=\sigma_i^{-},~~c_{i,2} = \sigma_i^+ \sigma_i^-,
	\end{equation}
	with $\sigma_i^\pm=\sigma_i^x\pm \sigma_i^y$. The rates $\gamma_{i,1}=\frac{1}{T_{i,1}}$, $\gamma_{i,2}=\frac{1}{T_{i,2}}$, $i=1,\cdots,N_q$. Here, we choose $T_{i,1}=T_{i,2}=30~\mu$s, which is a conservative setting compared with state-of-the-art experimental conditions~\cite{acharya2024quantum, li2024realization}. Each control $u_{i,\alpha}(t)$ is evenly discretized into $N_T$ pieces, each with duration $\Delta t = 0.5$~ns, leading to a total control pulse duration $T = N_T\Delta t$. The control amplitude is restricted to $u_c(t_k)/2\pi\in [-100,~100]$~MHz for $k=0,\cdots,N_T-1$. The quasi-Newton optimizer L-BFGS-B is used to update the controls. All numerical tests in this section are performed on an Intel(R) Core(TM) i9-14900KF CPU (24 cores).

	\subsection{Analysis of complexity and accuracy}
	\label{subsec:complexity}
	Both GRAPE and ST-GRAPE have analytical expressions for the gradient, which can be solved efficiently. As a result, the computational cost of the two algorithms is dominated by the forward and backward propagation of the system dynamics. We investigate how the computational time for propagation scales with the Hilbert dimension $d=2^{N_q}$ of the spin-chain model~(\ref{eq:spin_chain}). The quantum state is initialized to $\rho(0)=\frac{1}{d}\mathbf{1}_d$, where $\mathbf{1}_d$ denotes a $d\times d$ all-one matrix, serving as a representative example. The total control duration is set to $T= 10$~ns with $N_T=20$. The considered uncertainty terms are 
	\begin{equation}\label{eq:uncertain_E12}
		E_1 = \sigma^x_{1},~~E_2 = \sigma_{N_q}^x,
	\end{equation}
    which represents residual couplings with other devices. The robustness order is chosen as $n=1$, resulting in an augmented system of dimension $d_{\rm aug}=\binom{2+1}{2}d^2=3\times 4^{N_q}$. 
	
\begin{figure}
		\centering
		\includegraphics[width=\columnwidth]{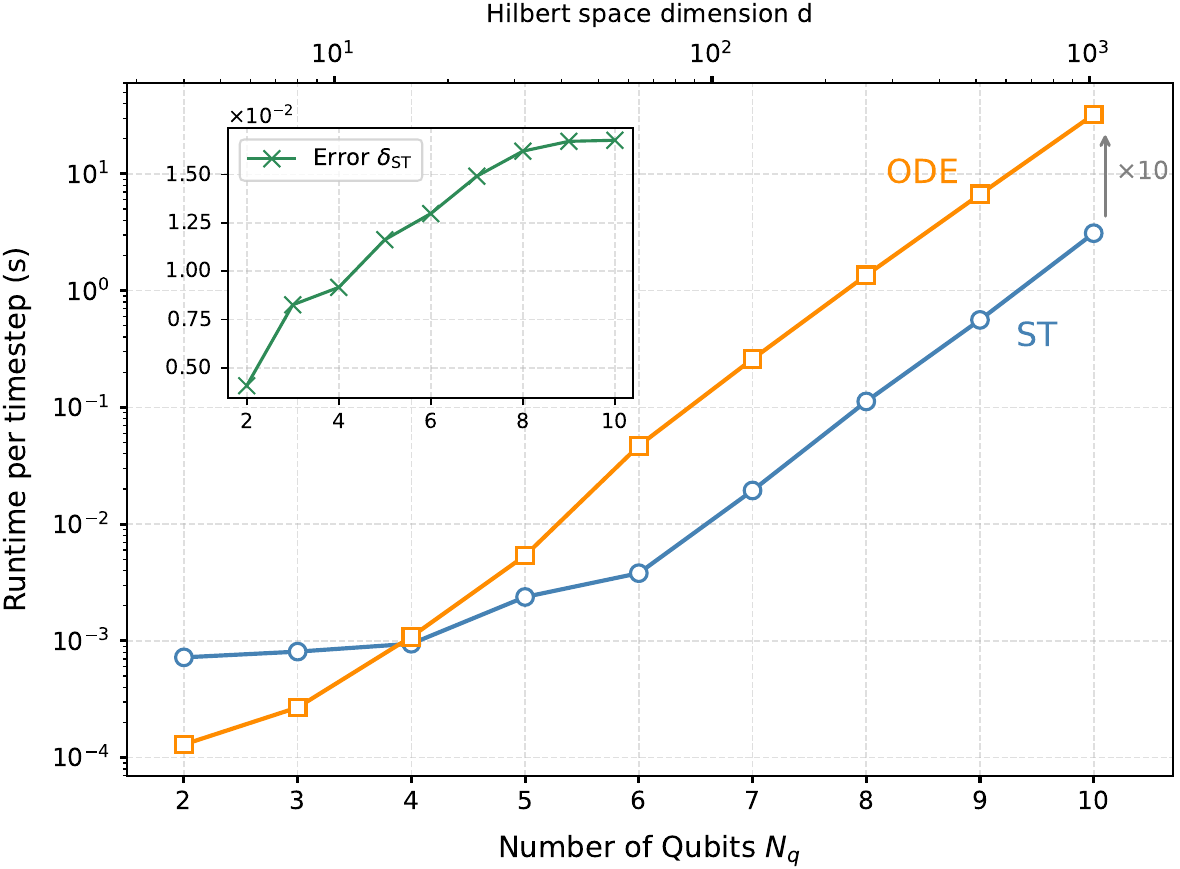}
		\caption{Benchmarking the ODE solver and the S-T expansion in propagating the augmented system dynamics. The yellow squares and the blue dots correspond to the runtime of the ODE solver (provided by QuTiP~\cite{johansson2013qutip,lambert2024qutip}) and the S-T expansion, respectively, for solving the evolution~(\ref{eq:augment_sys}) over one timestep $\Delta t=0.5$~ns. The comparison indicates a $10\times$ acceleration when the number of qubits is sufficiently large. The inset demonstrates the average approximation errors $\delta_{\rm ST}$ calculated from these simulations.}
		\label{fig:complexity_d}
	\end{figure}

	Figure~\ref{fig:complexity_d} plots the average runtime for forward propagation over one time step as a function of the number of qubits $N_q = 2,\cdots,10$. For each system configuration, the results are averaged over ten random controls. The acceleration introduced by the S-T expansion becomes significant for $N_q\geq 4$ ($d_{\rm aug}\geq 768$), achieving approximately $10\times$ speedup for $N_q\geq 6$ ($d_{\rm aug}\gtrsim 1.2\times 10^4$). The slopes of two curves are fitted, revealing a scaling of $d^{2.36}$ for the ODE solver and $d^{1.56}$ for the S-T expansion. The results differ with $d^4$ and $d^3$ listed in Tab.~\ref{tab:complexity}, potentially due to the sparsity of the Hamiltonian. Nonetheless, the S-T expansion still exhibits greater efficiency in practice when the Hilbert space is moderately large. We also test the computational cost with respect to the truncation order $N$. Both approaches demonstrate a sublinear scaling, possibly attributed to the sparsity as well (results not shown).
	
	In this example, we define the metric
	\begin{equation}
		\delta_{\rm ST} = \frac{\|\vec{\rho}(T)-\hat{\vec{\rho}}(T)\|_F}{\|\vec{\rho}(T)\|_F}
	\end{equation}
	to quantify the accuracy of the S-T expansion, which measures the relative error between the exact augmented state $\vec{\rho}(T)$, computed by the ODE solver, and its approximation $\hat{\vec{\rho}}(T)$, obtained from the S-T expansion. Here, the Frobenius norm $\|\cdot\|_F$ provides a measure of the distance between two states. As shown in the inset of Fig.~\ref{fig:complexity_d},
	the values of $\delta_{\rm ST}$ grow with the system dimension but remain below $2\%$ for $N_q\leq 10$. The relative discrepancy between the function values of $J$ and $\hat{J}$ is likely of the same order, thereby certifying the feasibility of optimizing the Trotterized objective $\hat{J}$.
	
	\subsection{Example 1: Multi-qubit Hadamard transform}
	\label{subsec:hadamard}
	To examine the performance of the algorithms in large-scale quantum control design, we begin by considering the task of realizing the Hadamard transform~\cite{leung2017speedup}:
	\begin{equation}
		|0\rangle^{\otimes N_q} \xrightarrow{{\rm H}^{\otimes N_q}} \frac{1}{\sqrt{2^{N_q}}} \sum_{x=0}^{2^{N_q}-1}|x\rangle,
	\end{equation}
	where H is the Hadamard gate. The uncertain factors are $E_1$ and $E_2$ specified in Eq.~(\ref{eq:uncertain_E12}). The robustness order of $n=1$ and a pulse duration of $T=40$~ns with $N_T=80$ are set for control design. As an illustrative example, consider the spin-chain model~(\ref{eq:spin_chain}) with $N_q=6$ qubits. The dimension of the corresponding augmented system is $d_{\rm aug} = \binom{1+2}{1} (2^6)^2 \approx 1.2\times 10^4$, which presents a challenge for control optimization. 

	\begin{figure*}
		\centering
		\includegraphics[width=0.95\textwidth]{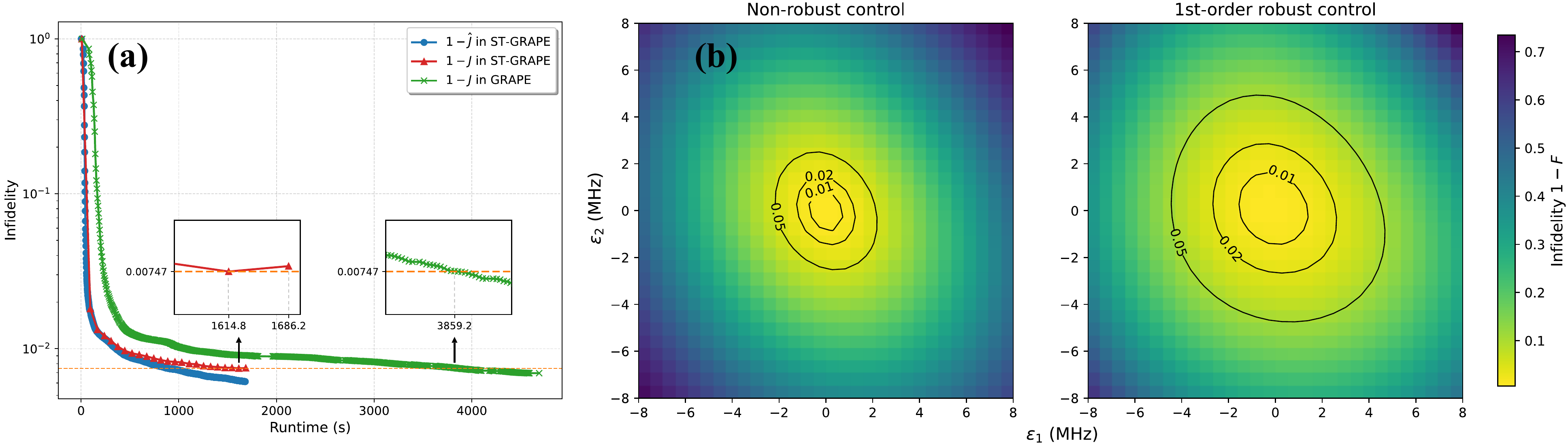}
		\caption{Robust control optimization for the 6-qubit Hadamard transform. (a) Infidelity of controls optimized by GRAPE and ST-GRAPE as a function of the runtime. The blue and red lines represent the approximated infidelity $1-\hat{J}$ and the true infidelity $1-J$ (recorded every $50$ iterations), respectively, during the optimization process of ST-GRAPE. Both algorithms successfully reduce the infidelity $1-J$ to comparably low values, with further optimization potentially limited by the finite coherence time of the system.  (b) State preparation fidelity $F$ versus the two uncertain parameters $(\epsilon_1,\epsilon_2)$ for a non-robust control solution and the first-order robust control solution obtained by ST-GRAPE, as shown in panel (a).}
		\label{fig:learning_curves}
	\end{figure*}

	Starting from the same randomly chosen initial guess on the control, we apply both GRAPE and ST-GRAPE to find robust control solutions. Defining the infidelity as $1-J$ (or $1-\hat{J}$), Figure~\ref{fig:learning_curves}(a) compares the training curves of GRAPE and ST-GRAPE during the optimization process. It takes 1614.8~s for ST-GRAPE to reach objective value of $J=0.9925$, after which the infidelity starts to increase. In contrast, it takes 3859.2~s for GRAPE to achieve the same fidelity, and 4689.2~s to reach a slightly higher performance of $J = 0.9930$ in the end. Therefore, ST-GRAPE demonstrates superior optimization efficiency due to the computational speedup provided by the S-T expansion for the given system size, as shown in Fig.~\ref{fig:complexity_d}. 
	
	During the implementation of ST-GRAPE, the gap between $\hat{J}$ and $J$ is apparent in the logarithm plot, as discussed in Sec.~\ref{subsec:gradient_optimization}. Nevertheless, the optimization can still proceed to a satisfactory precision in this example. To evaluate the enhanced robustness of control solutions, we employ the standard GRAPE algorithm to obtain an optimal solution from the same initial control under the ideal Lindblad master equation (i.e., the robustness order is set as $n=0$) as a comparison. Figure~\ref{fig:learning_curves}(b) shows the fidelity $F$ of the non-robust solution and the robust solution optimized by ST-GRAPE, plotted against the uncertain parameters $(\epsilon_1,\epsilon_2)$. The robustness can be quantified by the areas enclosed by three representative level sets of 0.99, 0.98, and 0.95. These areas of the robust solution are $4.17\times$, $4.28\times$, and $3.93\times$ those of the non-robust control, respectively. Therefore, the robust control maintains a high fidelity over a broader domain of uncertain parameters $(\epsilon_1, \epsilon_2)$, validating the effectiveness of the algorithm.
	
	\begin{figure}
		\centering
		\includegraphics[width=\columnwidth]{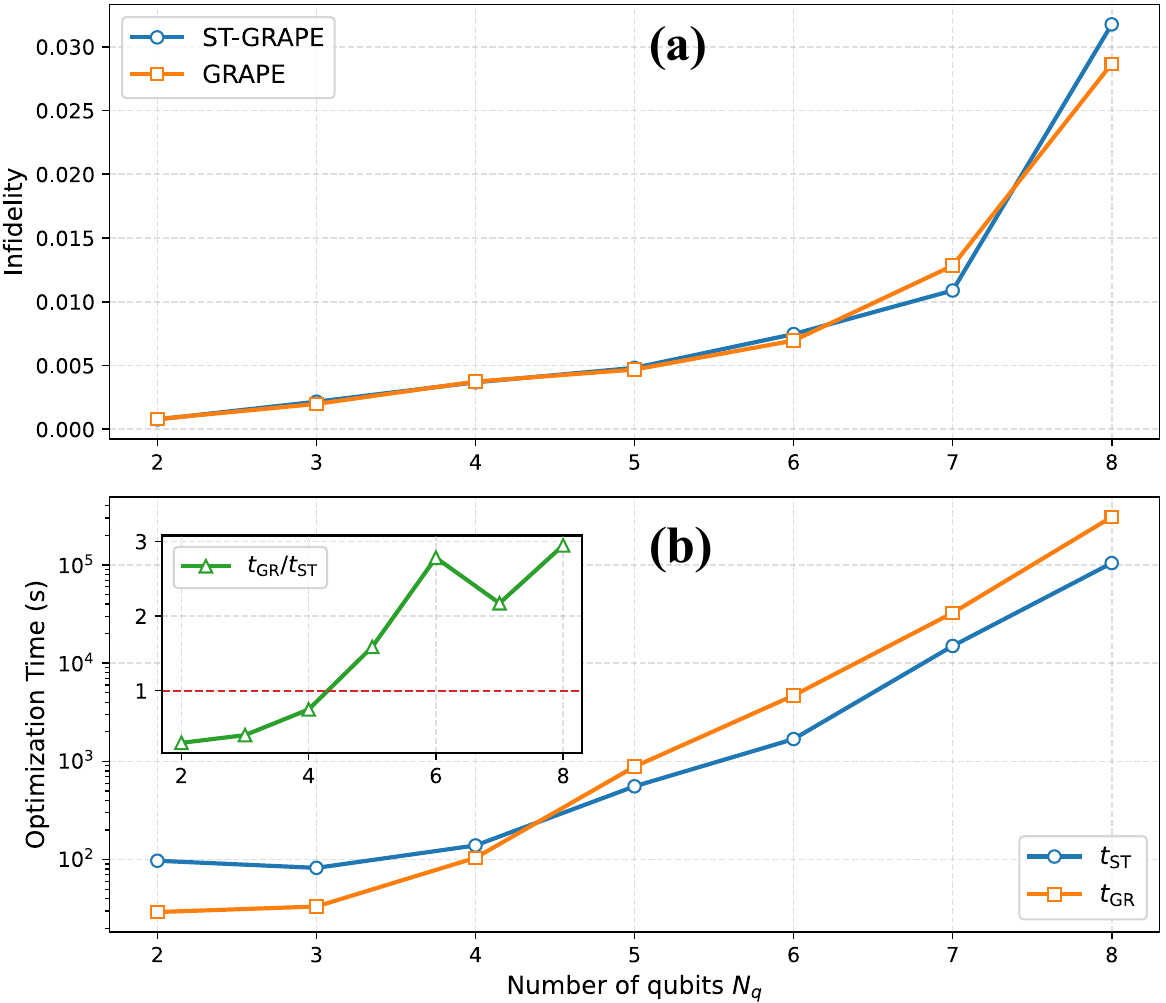}
		\caption{Scalability of the GRAPE and ST-GRAPE algorithms. (a) Infidelity $1-J$ of the final optimized solutions obtained by both algorithms. (b) Optimization time required for convergence. Here, $t_{\rm ST}$ and $t_{\rm GR}$ correspond to the optimization times of ST-GRAPE and GRAPE, respectively. ST-GRAPE is faster for $N_q\geq 5$. The inset shows the ratio $t_{\rm GR}/t_{\rm ST}$, which quantifies the speedup. }
		\label{fig:scalability_grapes}
	\end{figure}

	To manifest the scalability of the two algorithms, we perform the Hadamard transform task for varying numbers of qubits $N_q=2,3,\cdots,8$. The control pulse time $T=40$~ns and the robustness order $n=1$ are kept constant across all experiments. The corresponding dimensions of the augmented systems range over $48\leq d_{\rm aug}< 2\times 10^5$. Figure~\ref{fig:scalability_grapes}(a) shows the final infidelity $1-J$ of the optimized solution as a function of $N_q$. The objective function values achieved by both algorithms are comparable in all tested cases. Figure~\ref{fig:scalability_grapes}(b) illustrates the required optimization time for convergence. ST-GRAPE is more efficient for $N_q\geq 5$, and becomes more than twice faster than GRAPE to achieve robust solutions for $N_q\geq 6$. This finding is consistent with the trends observed in Fig.~\ref{fig:complexity_d}, though the speedup is less pronounced because ST-GRAPE tends to entail more iterations to converge. One possible explanation is that the landscape of the Trotterized objective function $\hat{J}$ may be less favorable for optimization. Therefore, the trade-off between computational efficiency—achieved through approximating real quantum dynamics—and solution quality demands a careful evaluation to ensure it brings substantial computational advantages in quantum control design. When it comes to enhancing the robustness, optimizing the Trotterized objective function $\hat{J}$ proves to be a practical and viable alternative, particularly when time consumption is of critical concern. The performance gap between the two algorithms could be narrowed with a finer time step $\Delta t$, as long as it is realizable by pulse generation devices~\cite{li2024realization, acharya2024quantum}. 
	
	\subsection{Example 2: Multi-qubit gates}
    Now consider a more challenging task for the synthesis of a robust entangling Toffoli gate (CCNOT gate) on $N_q=3$ qubits. In addition to $E_1$ and $E_2$ defined in Eq.~(\ref{eq:uncertain_E12}), we introduce two additional uncertainties: 
	\begin{equation}\label{eq:uncertain_E34}
		E_3 = \sigma_1^x\sigma_2^x + \sigma_1^y\sigma_2^y,~~E_4 = \sigma_2^x\sigma_3^x + \sigma_2^y\sigma_3^y,
	\end{equation}
    which account for drifts in the qubit-qubit coupling strengths. The nine basis density matrices $\{\vec{\rho}^{(i)}\}$ selected to evaluate the objective function in Eq.~(\ref{eq:robust_gate_synthesis}) can be found in Appendix~\ref{app:density_matrix}.
	
	\begin{figure}[htbp]
		\centering
		\includegraphics[width=\columnwidth]{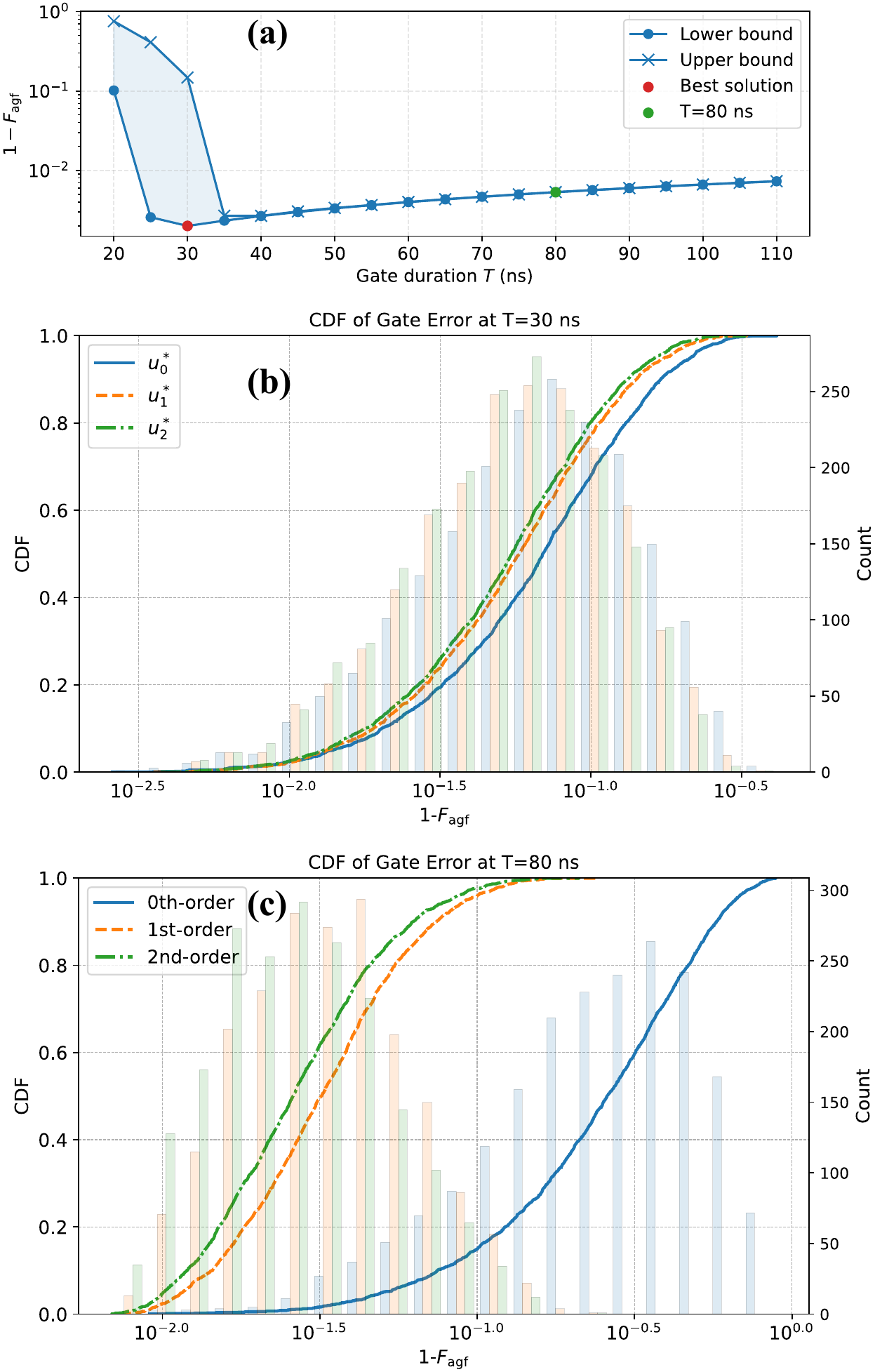}
		\caption{Robust control optimization for Toffoli gate synthesis. (a) Gate errors versus gate durations. For each gate duration $T = 20, 25, \dots, 110$~ns, ten random controls are initialized and then optimized using GRAPE under noiseless Lindblad evolution. The gate errors $1 - F_{\rm agf}$ of the final solutions are computed, with the upper and lower bounds of their distributions shown. For $T \leq 35$~ns, the $F_{\rm agf}$ values exhibit a wide distribution, as some optimization seeds may get trapped in false optima due to restricted controllability.  This issue disappears for $T\geq 40$~ns, where the performance is potentially limited by the finite coherence time. The best solution $u_0^*$, highlighted in red, achieves $F_{\rm agf}=0.9980$. A solution at $T=80$~ns, marked in green, is further investigated in panel~(c). (b) CDFs of gate errors under robust control solutions at $T=30$~ns. These distributions are calculated from $2~000$ noise samples $(\epsilon_1,\epsilon_2,\epsilon_3,\epsilon_4)$ drawn from the normal distribution $\mathcal{N}(0,\sigma^2I_4)$ with $\sigma/2\pi=2$~MHz. The first- and second-order robust solutions, denoted by $u_1^*$ and $u_2^*$, are obtained by applying ST-GRAPE starting from the initial seed that produces $u_0^*$. (c) CDFs of gate errors under robust control solutions at $T=80$~ns. The curves are derived from the same $2~000$ noise samples used to generate panel~(b). The background histograms in both panels~(b) and (c) count the number of samples within different error intervals.}
		\label{fig:duration_30ns}
	\end{figure}
	
	The gate duration $T$ plays a critical role in determining the performance limits of control solutions in open quantum environments~\cite{schulte2011optimal}. Specifically, a shorter gate duration may restrict the controllability of the quantum system, whereas a longer gate duration exacerbates decoherence effects. To identify an appropriate balance, multiple gate durations $T = 20, 25, \dots, 110$~ns are evaluated. For each duration, we perform standard GRAPE optimization under the ideal Lindblad dynamics with the robustness order $n=0$, starting from ten random initial seeds. We then assess the gate error $1-F_{\rm agf}$ in the absence of uncertainty, i.e., $\epsilon_i=0$ are set for $i=1,\cdots,4$. Here, $F_{\rm agf}$ denotes the average gate fidelity defined as~\cite{nielsen2002simple,pedersen2007fidelity}:
    \begin{equation}
        F_{\rm agf} = \int \bra{\psi} U_{\rm targ}^\dagger \mathcal{U}(\ket{\psi}\bra{\psi}) U_{\rm targ}\ket{\psi} d\psi,
    \end{equation}
    where $\mathcal{U}(\ket{\psi}\bra{\psi})$ represents the resulting density matrix evolved from the pure state $\rho(0)=\ket{\psi}\bra{\psi}$ under the Lindblad master equation~(\ref{eq:open_system_dynamics}), and the integral is taken over the uniform Haar measure on the space of pure states. This fidelity serves as an informative measure of how closely the implemented operation approximates the target unitary across the entire state space, and it can be efficiently estimated in experiments using randomized benchmarking~\cite{magesan2011scalable, proctor2017randomized}. The results are shown in Fig.~\ref{fig:duration_30ns}(a), where the best solution $u_0^{*}$ yielding $F_{\rm agf}=0.9980$ is achieved at $T=30$~ns. 
    
    Starting from the initial solution that generates $u_0^{*}$, we employ ST-GRAPE to find first-order and second-order robust control solutions, denoted by $u_1^*$ and $u_2^*$, respectively. Parallel computing is applied for accelerating the optimization~\cite{lambert2024qutip}. To assess the improved robustness, we randomly sample $2~000$ sets of uncertain parameters $(\epsilon_1,\epsilon_2,\epsilon_3,\epsilon_4)$ from a normal distribution $\mathcal{N}(0,\sigma^2I_4)$, with $\sigma/2\pi=2$~MHz. The gate errors under $u_0^*$, $u_1^*$, and $u_2^*$, averaged over the 2000 samples, are 0.0848, 0.0706, and 0.0663, respectively. The cumulative density functions (CDF) of the gate error are demonstrated in Fig.~\ref{fig:duration_30ns}(b). The value of the CDF curve at a prescribed error threshold indicates the probability that the gate error does not exceed the threshold. It can be observed that the CDF curves of the higher-order robust control solutions are always above those of lower-order ones, demonstrating stronger robustness in that the gate error is suppressed over a broader domain.

    It is also noted from Fig.~\ref{fig:duration_30ns}(b) that the robustness is not significantly enhanced. This is because the relatively fast gate operation makes it less susceptible to the accumulation of noise-induced corruption, thus leaves little room for the improvement of robustness. In practical quantum circuits, besides reducing the gate errors, the gate duration $T$ should also align with the overall timing logic of the circuit. Specifically, quantum gates within the same layer need to be synchronized to ensure consistency with the overall temporal structure of the circuit, maintaining both efficiency and accuracy. If a longer duration $T$ is needed in practice, the robust control design becomes more valuable. As an example, we select a non-robust control solution with $T=80$~ns, which is highlighted in Fig.~\ref{fig:duration_30ns}(a), and use ST-GRAPE to derive robust control solutions from its initial seed. The CDFs of these controls, evaluated under the same set of 2,000 noise samples, are shown in Fig.~\ref{fig:duration_30ns}(c). The results clearly demonstrate an evident performance improvement from the non-robust to the robust solutions. The average gate errors under the non-robust, the first-order, and the second-order robust solutions are 0.2978, 0.0392, and 0.0328, respectively. Thus, the robustness of quantum gates can be greatly enhanced by ST-GRAPE.
	
    We also simulate the case with $N_q=4$, where the target gate is the CCCNOT gate. This gate flips the fourth qubit if and only if the first three controlled qubits are all in the $\ket{1}$ state. The uncertainty terms remain to be the four specified in Eqs.~(\ref{eq:uncertain_E12}) and (\ref{eq:uncertain_E34}). The gate duration is set to $T=100$~ns. The augmented system has an high dimensionality $d_{\rm aug} = \binom{4+2}{2}\times (2^4)^2=3840$. At each optimization iteration, $d+1=17$ state preparation subroutines are executed. For this challenging task, the ST-GRAPE algorithm successfully obtains the first- and second-order robust control solutions, denoted as $u_1^*$ and $u_2^*$, respectively, with a little abuse of notations. 
    
\begin{figure*}
		\centering
		\includegraphics[width=0.95\textwidth]{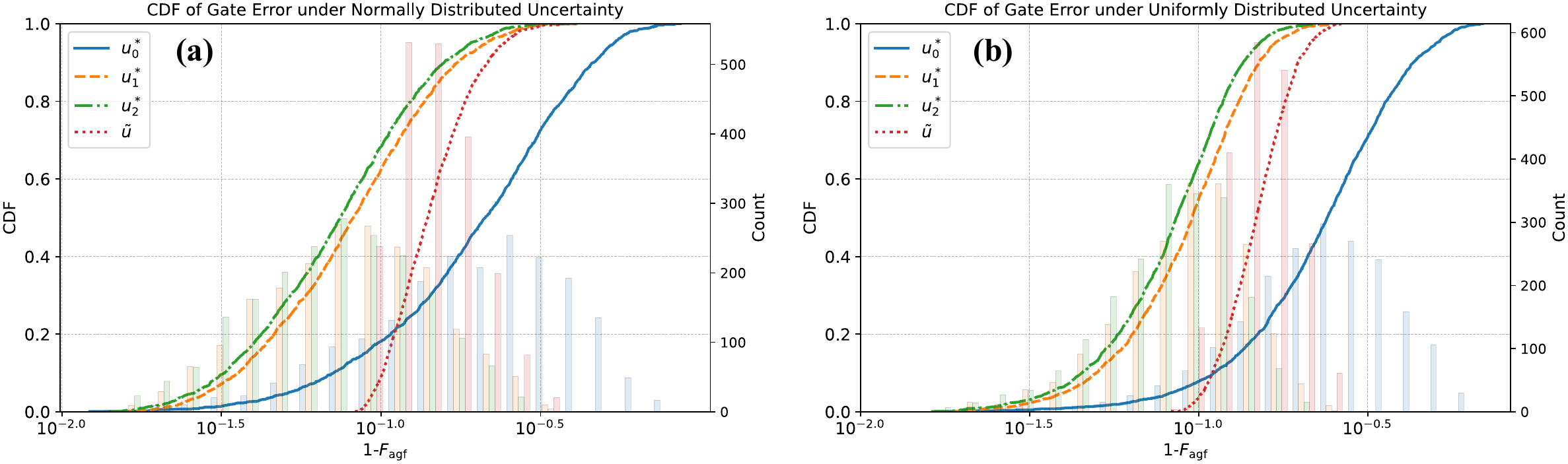}
		\caption{CDFs of gate errors under robust control solutions in CCCNOT gate synthesis. The non-robust ($u_0^*$), the first-order robust ($u_1^*$) and the second-order robust ($u_2^*$) control solutions are optimized from a random seed at $T=100$~ns by GRAPE and ST-GRAPE, respectively. The red line benchmarks the robust solution $\tilde{u}$, which is optimized using GRAPE under the alternative objective function $\tilde{J}$ defined in Eq.~(\ref{eq:average_control_objective}). Panel (a) shows results for $2~000$ uncertainty samples $(\epsilon_1,\cdots,\epsilon_4)$ drawn from a normal distribution $\mathcal{N}(0, \sigma^2 I_4)$ with $\sigma/2\pi=2$~MHz, yielding average errors of 0.2430, 0.0955, 0.0866, and 0.1500 for $u_0^*$, $u_1^*$, $u_2^*$, and $\tilde{u}$, respectively. Panel (b) presents the  case with another 2~000 samples, where each uncertainty $\epsilon_i$ is uniformly sampled from $[-2\sqrt{3}, 2\sqrt{3}]$~MHz, which exhibits a variance of $2$~MHz. The average gate errors are 0.2584, 0.0988, 0.0891, and 0.1516 for $u_0^*$, $u_1^*$, $u_2^*$, and $\tilde{u}$, respectively.} 
		\label{fig:gs_uniform_sampling}
	\end{figure*}

    The uncertain parameters are assumed to follow the same normal distribution as in the Toffoli gate example above, where the standard deviation $\sigma_i/2\pi = 2$~MHz. We then employ GRAPE to obtain another robust solution $\tilde{u}$ by optimizing the objective function $\tilde{J}$ defined in Eq.~(\ref{eq:average_control_objective}) as a comparison. Figure~\ref{fig:gs_uniform_sampling}(a) illustrates the CDFs of gate errors under these robust control solutions. It is evident that both $u_1^*$ and $u_2^*$ are more robust compared to $\tilde{u}$. This conclusion is further supported by the test using uniformly distributed noise, as illustrated in Fig.~\ref{fig:gs_uniform_sampling}(b), which yields consistent results. Therefore, our proposed algorithm is capable of achieving solutions that prioritize robustness to a greater extent. 
	
	\section{Conclusions}
	\label{sec:conclude_discuss}
	In conclusion, we proposed ST-GRAPE, an efficient algorithm for discovering robust control solutions for open quantum systems with parametric uncertainties. The process of robust control design involves computing the dynamics of a high-dimensional augmented system, and ST-GRAPE substantially accelerates this computation by utilizing the Suzuki-Trotter expansion method. The numerical tests on the robust state preparation show that ST-GRAPE outperforms the standard GRAPE algorithm in terms of computation time when the Hilbert dimension is modestly large. Furthermore, the successful synthesis of robust multi-qubit gates highlights the superior performance of ST-GRAPE in enhancing robustness. These numerical results validate the effectiveness of the proposed algorithm.

	The present work is readily extended to include the following potential improvements: (i) Mitigating more types of uncertainty. While this study focuses on additive noise, robustness against other types of noise, e.g., multiplicative noise~\cite{hocker2014characterization}, could be explored following the current methodology. (ii) Leveraging graphical processing units (GPUs). GPUs are increasingly utilized in the quantum control community to accelerate the solution of quantum system dynamics and exploit auto-differentiation for control optimization~\cite{leung2017speedup,abdelhafez2019gradient,puzzuoli2023qiskit,lambert2024qutip}. ST-GRAPE is well-suited for GPU hardware, as it reduces the complexity of matrix multiplication—an operation highly amenable to parallelization on GPUs. This capability would facilitate the efficient design of robust control solutions for larger quantum systems. These possibilities will be investigated in future studies.

\acknowledgements
We acknowledge helpful discussions with Dr. Zijie Chen and Prof. Chang-Ling Zou. The author RBW acknowledges support from Innovation Program for Quantum Science and
Technology (No.2021ZD0300200) and NSFC grant 62173201.

    \appendix
    \section{Implementations of superoperators}
	\label{app:superoperator}
	Based on the identity~\cite{goerz2014optimal}:
	\begin{equation}\label{eq:vectorization_technique}
		\op{vec}(AXB) = (B^T\otimes A)\op{vec}(X),
	\end{equation}
    we can derive the matrix representations of the superoperators $\mathcal{L}(t)$ and $\mathcal{E}_j$ defined in Eq.~(\ref{eq:quantum_channel}) as follows:
	\begin{equation}
		\begin{aligned}
			\op{mat}(\mathcal{L}(t)) & = -i(I_d\otimes H_S(t)-H_S^T(t)\otimes I_d) \\
			&\quad + \sum_{i=1}^{n_d}(\bar{c}_i\otimes c_i - \frac{1}{2}I_d\otimes c_i^\dagger c_i-\frac{1}{2}c_i^T\bar{c}_i\otimes I_d),\\
			\op{mat}(\mathcal{E}_j) & = -i(I_d\otimes E_j-E_j^T\otimes I_d),
		\end{aligned}
	\end{equation}
	where $\bar{c}_i$ denotes the complex conjugate of $c_i$. The matrix representations of $\mathbf{L}(t)$ and $\mathbf{E}_j$ can be obtained by substituting the corresponding superoperators $\mathcal{L}(t)$ and $\mathcal{E}_j$ in Eq.~(\ref{eq:L_and_E}) with their matrix forms derived above. The evolution of the augmented system specified in Eq.~(\ref{eq:augment_sys}) can then be computed by solving the corresponding ordinary differential equation.
	
	The adjoint of an operation $\mathcal{A}$ on the Liouville space is defined by the equivalence~\cite{goerz2014optimal}
	\begin{equation}
		\langle \rho_1,\mathcal{A}\rho_2\rangle = \langle \mathcal{A}^\dagger\rho_1,\rho_2\rangle,
	\end{equation}
	where the Hilbert-Schmidt inner product is given by $\langle\rho_1,\rho_2\rangle=\tr(\rho_1^\dagger\rho_2)$. The adjoint operators $\mathbf{L}(t)^\dagger$ and $\mathbf{E}_j^\dagger$ can be conveniently represented by taking the conjugate transpose of $\op{mat}(\mathbf{L}(t))$ and $\op{mat}(\mathbf{E}_j)$, respectively. These matrix representations are readily utilized in GRAPE for computing the backward propagation described in Eq.~(\ref{eq:backpropagate}). The implementation of ST-GRAPE needs the adjoint of $\hat{\mathbf{S}}_k$. It follows from the properties
	\begin{equation}
		(\mathcal{A}_1\mathcal{A}_2)^\dagger = \mathcal{A}_2^\dagger \mathcal{A}_1^\dagger,~~(e^{\mathcal{A}})^\dagger = e^{\mathcal{A}^\dagger}
	\end{equation}
	that the adjoint of $\hat{\mathbf{S}}_k$ is given by:
	\begin{equation}\label{eq:dagger_S}
		\begin{aligned}
			\hat{\mathbf{S}}^\dagger_k &= 
			\left( \prod_{j=m}^{1} e^{\mathbf{E}^\dagger_j\frac{\Delta t}{2}} \right) 
			e^{\mathbf{C}^\dagger\frac{\Delta t}{2}} 
			e^{u_c(t_k)\mathbf{H}^\dagger_{c}\frac{\Delta t}{2}} \\
			&\quad \times e^{\mathbf{H}^\dagger_{\rm eff}\Delta t} 
			e^{u_c(t_k)\mathbf{H}^\dagger_{c}\frac{\Delta t}{2}} 
			e^{\mathbf{C}^\dagger\frac{\Delta t}{2}} 
			\left( \prod_{j=1}^{m} e^{\mathbf{E}^\dagger_j\frac{\Delta t}{2}} \right).
		\end{aligned}
	\end{equation}
	
	The expression for $e^{\mathbf{H}^\dagger_{\rm eff}\Delta t}$ can be constructed from the following equivalences:
	\begin{equation}
		\begin{aligned}
			&\langle \vec{\rho}_1,e^{\mathbf{H}_{\rm eff}\Delta t}\vec{\rho}_2\rangle = \langle \vec{\rho}_1,(\mathbb{I}_N\otimes U_{\rm eff})\vec{\rho}_2 (\mathbb{I}_N\otimes U_{\rm eff}^\dagger)\rangle \\
			&  = \tr\left[\vec{\rho}^\dagger_1(\mathbb{I}_N\otimes U_{\rm eff})\vec{\rho}_2(\mathbb{I}_N\otimes U_{\rm eff}^\dagger)\right]\\
			& = \tr\left[(\mathbb{I}_N\otimes U_{\rm eff}^\dagger)\vec{\rho}^\dagger_1(\mathbb{I}_N\otimes U_{\rm eff})\vec{\rho}_2\right]\\
			& = \tr\left\{\left[(\mathbb{I}_N\otimes U_{\rm eff}^\dagger)\vec{\rho}_1(\mathbb{I}_N\otimes U_{\rm eff})\right]^\dagger\vec{\rho}_2\right\} \\
			& = \langle e^{\mathbf{H}^\dagger_{\rm eff}\Delta t} \vec{\rho}_1,\vec{\rho}_2\rangle,
		\end{aligned}
	\end{equation}
	which leads to
	\begin{equation}
		e^{\mathbf{H}^\dagger_{\rm eff}\Delta t} \vec{\rho} = (\mathbb{I}_N\otimes U_{\rm eff}^\dagger)\vec{\rho}(\mathbb{I}_N\otimes U_{\rm eff}).
	\end{equation}
	Similarly, the adjoint expressions for the remaining terms in Eq.~(\ref{eq:dagger_S}) can be derived as follows:
	\begin{equation}
		\begin{aligned}
			& e^{u_c(t_k)\mathbf{H}^\dagger_{c}\frac{\Delta t}{2}} \vec{\rho} 
			 = \left[\mathbb{I}_N\otimes U_c^\dagger(t_k)\right]\vec{\rho} \left[\mathbb{I}_N\otimes U_c(t_k)\right],\\
			& e^{\mathbf{E}^\dagger_j\frac{\Delta t}{2}}=I+\frac{\Delta t}{2}\mathbf{E}_j^\dagger+\cdots + \frac{1}{n!}\left(\frac{\Delta t}{2}\mathbf{E}_j^\dagger\right)^{n},\\
			& e^{\mathbf{C}^\dagger\frac{\Delta t}{2}} \approx I + \frac{\Delta t}{2}\mathbf{C}^\dagger + \frac{\Delta t^2}{8}{\mathbf{C}^\dagger}^2.
		\end{aligned}
	\end{equation}
	Here, $\mathbf{E}_j^\dagger = \mathbb{R}_j^\dagger\otimes \mathcal{E}_j^\dagger$ with $\mathcal{E}_j^\dagger\rho = i[E_j, \rho]$, and $\mathbf{C}^\dagger = \mathbb{I}_N\otimes \mathcal{C}^\dagger$ with $\mathcal{C}^\dagger\rho = \sum_{i=1}^{n_d}\gamma_ic_i^\dagger\rho c_i$. With these operations, the backward propagated state $\hat{\vec{O}}(t_{k+1}) = \hat{\mathbf{S}}^\dagger_{k+1}\cdots \hat{\mathbf{S}}^\dagger_{N_T-2}\hat{\mathbf{S}}^\dagger_{N_T-1}\vec{O}(T)$ in ST-GRAPE can be efficiently computed.

    Evaluating each element of $\mathbf{C}\vec{\rho}$, as defined in Eq.~(\ref{eq:collapse}) by $\sum_{i=1}^{n_d} c_i\rho c_i^\dagger$, involves $2n_d$ matrix multiplications. This computation can become time-consuming when $n_d$ is large.  In this work, we assume that the decoherence channels of each qubit are independent, resulting in $n_d$ being proportional to the number of qubits, which can become significant as the quantum system scales up. An alternative approach utilizes the vectorization technique~(\ref{eq:vectorization_technique}) as follows:
    \begin{equation}\label{eq:vec_crhoc}
        \op{vec}\left(\sum_{i=1}^{n_d} c_i\rho c_i^\dagger\right) = \left(\sum_{i=1}^{n_d} \bar{c}_i \otimes c_i\right)\op{vec}(\rho).
    \end{equation}
    The matrix $\sum_{i=1}^{n_d} \bar{c}_i \otimes c_i$ can be precomputed, so that the operation~(\ref{eq:vec_crhoc}) requires only one matrix multiplication. Although the theoretical computational complexity increases to $\mathcal{O}(d^4)$ due to the enlarged dimensionality, we observe that this method can be more efficient in practice, potentially due to the sparsity of $\sum_{i=1}^{n_d} \bar{c}_i \otimes c_i$. Therefore, it is recommended to evaluate the performance of both approaches and select the more efficient one for the specific system under consideration.
    
    \section{Multiple control Hamiltonians}
        \label{app:multi_control}
    In Sec.~\ref{subsec:suzuki_trotter}, we analyze the case where there only exists a single, diagonal control Hamiltonian. Here, we summarize the results of Ref.~\cite{jensen2021approximate} for a general setting involving multiple, non-diagonal Hamiltonians $\{H_c\}_{c=1}^{n_c}$.	
	Sort the control Hamiltonians into $Q$ sets of mutually commuting Hamiltonians as:
	\begin{equation}
		\{H_c\}_{c=1}^{n_c}=\cup_{q=1}^Q C_q,
	\end{equation}
    Each set $C_q$ consists of control Hamiltonians that commute with one another and can therefore be simultaneously diagonalized by a common unitary transformation $R_q$, i.e., $R_q^\dagger H_cR_q$ is diagonal for all $H_c\in C_q$. In the S-T expansion of $\hat{\mathbf{S}}_k$, the term $e^{u_c(t_k)\mathbf{H}_c\frac{\Delta t}{2}}$ generalizes to $e^{\sum_{c=1}^{n_c}u_c(t_k)\mathbf{H}_c\frac{\Delta t}{2}}$, which is further Trotterized as:
	\begin{equation}\label{eq:multicon_trotter}
		\begin{aligned}
			& \exp\left[\sum_{c=1}^{n_c}u_c(t_k)\mathbf{H}_c\frac{\Delta t}{2}\right] =\exp\left[\sum_{q=1}^{Q}\sum_{H_c\in C_q}u_c(t_k)\mathbf{H}_c\frac{\Delta t}{2}\right]\\
		& \approx \prod_{q=1}^Q \exp \left[\sum_{H_c\in C_{q}} u_c(t_k)\mathbf{H}_c\frac{\Delta t}{2}\right]
		    % = \prod_{q=1}^Q \mathbf{R}_q\exp \left[\left(\sum_{H_c\in C_{q}} u_c(t_k)\mathbf{R}_q^\dagger\mathbf{H}_c\mathbf{R}_q\right)\frac{\Delta t}{2}\right]\mathbf{R}_q^\dagger,
		\end{aligned}
	\end{equation}
    The superoperator $\mathbf{U}_q:=\exp\left[\sum_{H_c\in C_{q}} u_c(t_k)\mathbf{H}_c\frac{\Delta t}{2}\right]$ acts on each sub-block $\rho_{[l]}$ of $\vec{\rho}$ for $l=1,\cdots,N$ via the following transformation:
    \begin{equation}\label{eq:big_U_q}
        \rho_{[l]} \xrightarrow{\mathbf{U}_q} U_q (t_k) \rho_{[l]} U_q^\dagger (t_k),
    \end{equation}
    where 
    \begin{equation}
        \begin{aligned}
            U_q(t_k) & = \exp\left[-i\sum_{H_c\in C_{q}} u_c(t_k)H_c\frac{\Delta t}{2}\right] \\
            & = R_q\exp\left[-i\sum_{H_c\in C_{q}} u_c(t_k)R_q^\dagger H_c R_q\frac{\Delta t}{2}\right] R_q^\dagger\\
            & = R_q \left\{\prod_{H_c\in C_q}\exp\left[-i u_c(t_k)R_q^\dagger H_c R_q\frac{\Delta t}{2}\right] \right\} R_q^\dagger.
        \end{aligned}
    \end{equation}
    Since $R_q^\dagger H_c R_q$ is diagonal, both $U_q(t_k)$ and its gradient with respect to $u_c(t_k)$ can be computed efficiently. The transformation matrices $R_q$ and the diagonalized forms $R_q^\dagger H_c R_q$ can be precomputed once and reused throughout the optimization. The computational complexity of the transformation in Eq.~(\ref{eq:big_U_q}) still scales as $\mathcal{O}(Nd^3)$. The computational cost can be further reduced by representing the quantum system in a proper basis so that all control Hamiltonians in a group $C_q$ are diagonal. In this case, the corresponding $U_q(t_k)$ can be evaluated with a reduced complexity of $\mathcal{O}(d)$.

    Consider the spin-chain model simulated in Sec.~\ref{sec:simulation} as an example. The control Hamiltonians are sorted into two sets:
	\begin{equation}
		C_1 = \{\sigma_i^x\}|_{i=1,\cdots,N_q},~~C_2 = \{\sigma_i^y\}|_{i=1,\cdots,N_q}.
	\end{equation}
	Since $R_x^\dagger \sigma_x R_x = \sigma_z$, $R_y^\dagger \sigma_y R_y = \sigma_z$, where
	\begin{equation}
		R_x = \frac{1}{\sqrt{2}}\begin{bmatrix}
			1 & 1 \\
			1 & -1
		\end{bmatrix},~~R_y = \frac{1}{\sqrt{2}}\begin{bmatrix}
			1 & 1 \\
			i & -i
		\end{bmatrix},
	\end{equation}
	the unitary transformations can be chosen as: $R_1 = R_x\otimes R_x\cdots \otimes R_x$ and $R_2 = R_y\otimes R_y\cdots \otimes R_y$. The control Hamiltonians are diagonalized as:
	\begin{equation}
		R_1^\dagger \sigma_x^i R_1 = \sigma_z^i,~~R_2^\dagger \sigma_y^i R_2 = \sigma_z^i.
	\end{equation}
	\section{Properties of $\mathbb{R}_j$}
	\label{app:nilpotency}
	We first prove that $\mathbb{R}_j$ is nilpotent. The square of the matrix $\mathbb{R}_j$ defined in Eq.~(\ref{eq:L_and_E}) can be computed as:
	\begin{equation}
		(\mathbb{R}_j^2)_{kl} = \sum_{r}(\mathbb{R}_j)_{kr}(\mathbb{R}_j)_{rl}.
	\end{equation}
	We denote the order of $\rho[k]$ as $(p_1,\cdots,p_j,\cdots,p_m)$. Based on the structure of the matrix $\mathbb{R}_j$, each row of $\mathbb{R}_j$ has at most one non-zero element $(\mathbb{R}_j)_{kr}=1$, when the order of $\rho[r]$ is $(p_1,\cdots,p_j-1,\cdots,p_m)$. In that case, we find that
	\begin{equation}
		(\mathbb{R}_j^2)_{kl}=(\mathbb{R}_j)_{kr}(\mathbb{R}_j)_{rl}=1,
	\end{equation}
	when the order of $\rho[l]$ is $(p_1,\cdots,p_j-2,\cdots,p_m)$. All other terms are zero. It can be derived by induction that, for $\alpha>1$,  $(\mathbb{R}_j^\alpha)_{kl}=1$ when the order of $\rho[l]$ is $(p_1,\cdots,p_j-\alpha,\cdots,p_m)$.	For the $n$th-order robust control, the maximum allowed $\alpha$ that yields non-zero matrix elements is $\alpha=n$, where $(\mathbb{R}_j^{n})_{kl}=1$ when $\rho_{[k]}$ has the order $(0,\cdots,n,\cdots,0)$ and $\rho_{[l]}$ has the order $(0, \cdots, 0)$ (i.e., $\rho_{[l]}=\rho_0$, $l=N$).	For $\alpha>n$, $\mathbb{R}_j^{\alpha}=0$, indicating the nilpotency of $\mathbb{R}_j$.

    Following the expression in Eq.~(\ref{eq:expE}), $e^{\mathbf{E}_j\frac{\Delta t}{2}}\vec{\rho}$ is given by:
    \begin{equation}
        \begin{aligned}
            & e^{\mathbf{E}_j\frac{\Delta t}{2}}\vec{\rho} \\ 
            & = \vec{\rho}+\frac{\Delta t}{2}\mathbf{E}_j\vec{\rho}+\frac{1}{2!}\left(\frac{\Delta t}{2}\mathbf{E}_j\right)^{2}\vec{\rho}+\cdots + \frac{1}{n!}\left(\frac{\Delta t}{2}\mathbf{E}_j\right)^{n}\vec{\rho}\\
            & = \vec{\rho} + \frac{\Delta t}{2}\mathbf{E}_j\Bigg(\vec{\rho} + \frac{1}{2} \frac{\Delta t}{2}\mathbf{E}_j\Big(\vec{\rho} + \frac{1}{3} \frac{\Delta t}{2}\mathbf{E}_j \big(\cdots(\vec{\rho} \\
            & \quad + \frac{1}{n}\frac{\Delta t}{2}\mathbf{E}_j \vec{\rho})\big)\Big)\Bigg)
        \end{aligned}
    \end{equation}
    This nested formulation leads to the following algorithm for evaluating $e^{\mathbf{E}_j\frac{\Delta t}{2}}\vec{\rho}$:
    \begin{itemize}
        \item \textbf{Initialization:} $\vec{\rho}_{\rm nest} \gets \vec{\rho}$
        \item For $\ell = 0$ to $n-1$:
        \begin{itemize}
            \item $\vec{\rho}_{\rm nest} \gets \vec{\rho} + \frac{1}{n-l}\frac{\Delta t}{2}\mathbf{E}_j \vec{\rho}_{\rm nest}$ 
        \end{itemize}
        \item \textbf{Return:} $\vec{\rho}_{\rm nest}$
    \end{itemize}
    
    Recall that $\mathbf{E}_j=\mathbb{R}_j\otimes \mathcal{E}_j$, where the definition of $\mathcal{E}_j$ is given in Eq.~(\ref{eq:quantum_channel}). Since evaluating $\mathcal{E}_j \rho$ incurs a computational cost of $\mathcal{O}(d^3)$, the complexity of computing $e^{\mathbf{E}_j\frac{\Delta t}{2}}\vec{\rho}$ iteratively, following the nested algorithm above, is proportional to the number of times $\mathcal{E}_j(\cdot)$ is applied to some sub-blocks of $\vec{\rho}$ during the process. Based on the above derivation, the number of non-zero elements in $\mathbb{R}_j$ (i.e., $(\mathbb{R}_j)_{kl}=1$) equals the number of sub-blocks in $\vec{\rho}$ with an order satisfying $p_j\geq 1$. This number is given by
	\begin{equation}
		\begin{aligned}
			N_j & =\binom{n+m}{n}-\binom{n+m-1}{n}\\
			& =\frac{(m+n)!}{m!n!}-\frac{(m+n-1)!}{n!(m-1)!}=\frac{n}{m+n}N,
		\end{aligned}
	\end{equation}
	where $\binom{n+m-1}{n}$ corresponds to the number of terms with an order $p_j=0$. Consequently, each application of $\mathbf{E}_j$ requires $N_j$ evaluations of $\mathcal{E}_j(\cdot)$. As the algorithm involves $n$ such applications, and there are $m$ uncertain factors $\mathbf{E}_j$ in total ($j=1,\cdots,m$), the overall computational complexity is $\mathcal{O}(mnN_jd^3)=\mathcal{O}(\frac{mn^2}{m+n}Nd^3)$. Although robust control of any order is solvable, the robustness order of $n=1$ or $2$ is usually sufficient for practical design. In such cases, the factor $\frac{mn^2}{m+n}\approx n^2$ becomes constant for large $m$, and the computational complexity scales as $\mathcal{O}(Nd^3)$ with respect to $N$ and $d$.

	\section{Initialization of density matrices}
	\label{app:density_matrix}
	The evolution of a $d$-dimensional open quantum system, governed by the Lindblad master equation~(\ref{eq:open_system_dynamics}), is a completely positive and trace-preserving map, denoted as $\mathcal{U}(\cdot)$. It has been proven in Refs.~\cite{reich2013minimum,goerz2014optimal} that the target unitary transformation $U_{\rm targ}$ is achieved, i.e., $\mathcal{U}(\cdot)=U_{\rm targ}$, if and only if 
	\begin{equation}\label{eq:grk}
		\mathcal{U}\left[\rho^{(i)}(0)\right]=U_{\rm targ}\rho^{(i)}(0)U_{\rm targ}^\dagger,~~i=1,2,3,
	\end{equation}
	where the initial density matrices are chosen as
	\begin{equation}
		\begin{aligned}
			\left[\rho^{(1)}(0)\right]_{kl}&=\frac{2(d-k+1)}{d(d+1)} \delta_{kl},~~ \left[\rho^{(2)}(0)\right]_{kl}=\frac{1}{d},\\
			\rho^{(3)}(0)&=\mathbb{I}_d,
		\end{aligned}
	\end{equation}
    for $1\leq k,l\leq d$. The choice for the three states for the condition~(\ref{eq:grk}) is not unique. While three states are sufficient, a more stable reduction of gate error during optimization needs $d+1$ states~\cite{goerz2014optimal}:
	\begin{equation}
		\rho^{(i)}(0)=\ketbra{\psi_i}{\psi_i},~~ \left[\rho^{(d+1)}(0)\right]_{kl}=\frac{1}{d}
	\end{equation}
    for $i=1,\cdots,d$ and $1\leq k,l\leq d$, where $\{\ket{\psi_i}\}_{i=1,\cdots,d}$ form an arbitrary complete orthonormal basis of the Hilbert space. This approach is numerically more efficient than using a full set of $d^2$ states spanning the Liouville space~\cite{schulte2011optimal,floether2012robust}.
	
    \section{Average performance}
	\label{app:average_gate_performance}
    According to the Taylor series expansion defined in Eq.~(\ref{eq:taylor}), the second-order approximation to the final state $\rho(T)$ is given by:
    \begin{equation}
        \rho(T) = \rho_0(T)+\sum_{i=1}^m\epsilon_i\rho_i(T) + \sum_{i=1}^m\sum_{j=1}^m \epsilon_i\epsilon_j\rho_{ij}(T) + \mathcal{O}(\epsilon^3),
    \end{equation}
    where $\rho_i$ and $\rho_{ij}$ corresponds to the coefficients of the parameters $\epsilon_i$ and $\epsilon_i\epsilon_j$ in the series defined as 
    \begin{equation}
        \begin{aligned}
            \rho_i(T) :=\rho_{p_i=1}(T),\quad \rho_{ij}(T) :=\rho_{p_i=1,p_j=1}(T).
        \end{aligned}
    \end{equation}
	Under the assumption that the uncertain parameters are zero-mean and mutually independent, their statistics satisfy:
	\begin{equation}
		\langle \epsilon_i\rangle = 0,\quad \langle \epsilon_i\epsilon_j\rangle = 0,\quad\langle \epsilon_i^2\rangle = \sigma_i^2,
	\end{equation}
	for $1\leq i,j\leq m$, $i\neq j$. The average performance $F_{\rm avg}$ can be expanded as
	\begin{equation}\label{eq:agf}
		\begin{aligned}
			F_{\rm avg} & = \tr\left[\rho_{\rm targ} \rho_0(T)\right] + \sum_{i=1}^m\langle \epsilon_i\rangle\tr\left[\rho_{\rm targ} \rho_i(T)\right] \\
			&\quad + \sum_{i=1}^m\sum_{j=1}^m\langle \epsilon_i\epsilon_j\rangle\tr\left[\rho_{\rm targ} \rho_{ij}(T)\right] +  \mathcal{O}(\epsilon^3) \\
			&= \tr\left[\rho_{\rm targ} \rho_0(T)\right] + \sum_{i=1}^m\sigma_i^2 \tr\left[\rho_{\rm targ} \rho_{ii}(T)\right] +  \mathcal{O}(\epsilon^3).
		\end{aligned}
	\end{equation}
    Consequently, the control objective $\tilde{J}$ defined in Eq.~(\ref{eq:average_control_objective}) is a second-order approximation to $F_{\rm avg}$. To optimize $\tilde{J}$ using GRAPE, we need to compute the co-state:
	\begin{equation}
		\tilde{O}_{N_T} := \nabla_{\vec{\rho}(T)}\tilde{J},
	\end{equation}
	as outlined in Step 2 of the algorithm. Taking $\vec{\rho}(t)$ described by Eq.~(\ref{eq:m_2_n_2}) as an example, the corresponding co-state $\tilde{O}_{N_T}$ is derived as
	\begin{equation}
		\tilde{O}_{N_T} = \begin{bmatrix}
		    \sigma_1^2\rho_{\rm targ}\\
                0\\
                0\\
                \sigma_2^2\rho_{\rm targ}\\
                0\\
                \rho_{\rm targ}
		\end{bmatrix}.
	\end{equation}

\bibliography{reference}

\begin{thebibliography}{84}%
\makeatletter
\providecommand \@ifxundefined [1]{%
 \@ifx{#1\undefined}
}%
\providecommand \@ifnum [1]{%
 \ifnum #1\expandafter \@firstoftwo
 \else \expandafter \@secondoftwo
 \fi
}%
\providecommand \@ifx [1]{%
 \ifx #1\expandafter \@firstoftwo
 \else \expandafter \@secondoftwo
 \fi
}%
\providecommand \natexlab [1]{#1}%
\providecommand \enquote  [1]{``#1''}%
\providecommand \bibnamefont  [1]{#1}%
\providecommand \bibfnamefont [1]{#1}%
\providecommand \citenamefont [1]{#1}%
\providecommand \href@noop [0]{\@secondoftwo}%
\providecommand \href [0]{\begingroup \@sanitize@url \@href}%
\providecommand \@href[1]{\@@startlink{#1}\@@href}%
\providecommand \@@href[1]{\endgroup#1\@@endlink}%
\providecommand \@sanitize@url [0]{\catcode `\\12\catcode `\$12\catcode
  `\&12\catcode `\#12\catcode `\^12\catcode `\_12\catcode `\%12\relax}%
\providecommand \@@startlink[1]{}%
\providecommand \@@endlink[0]{}%
\providecommand \url  [0]{\begingroup\@sanitize@url \@url }%
\providecommand \@url [1]{\endgroup\@href {#1}{\urlprefix }}%
\providecommand \urlprefix  [0]{URL }%
\providecommand \Eprint [0]{\href }%
\providecommand \doibase [0]{https://doi.org/}%
\providecommand \selectlanguage [0]{\@gobble}%
\providecommand \bibinfo  [0]{\@secondoftwo}%
\providecommand \bibfield  [0]{\@secondoftwo}%
\providecommand \translation [1]{[#1]}%
\providecommand \BibitemOpen [0]{}%
\providecommand \bibitemStop [0]{}%
\providecommand \bibitemNoStop [0]{.\EOS\space}%
\providecommand \EOS [0]{\spacefactor3000\relax}%
\providecommand \BibitemShut  [1]{\csname bibitem#1\endcsname}%
\let\auto@bib@innerbib\@empty
%</preamble>
\bibitem [{\citenamefont {Krantz}\ \emph {et~al.}(2019)\citenamefont {Krantz},
  \citenamefont {Kjaergaard}, \citenamefont {Yan}, \citenamefont {Orlando},
  \citenamefont {Gustavsson},\ and\ \citenamefont
  {Oliver}}]{krantz2019quantum}%
  \BibitemOpen
  \bibfield  {author} {\bibinfo {author} {\bibfnamefont {P.}~\bibnamefont
  {Krantz}}, \bibinfo {author} {\bibfnamefont {M.}~\bibnamefont {Kjaergaard}},
  \bibinfo {author} {\bibfnamefont {F.}~\bibnamefont {Yan}}, \bibinfo {author}
  {\bibfnamefont {T.~P.}\ \bibnamefont {Orlando}}, \bibinfo {author}
  {\bibfnamefont {S.}~\bibnamefont {Gustavsson}},\ and\ \bibinfo {author}
  {\bibfnamefont {W.~D.}\ \bibnamefont {Oliver}},\ }\bibfield  {title}
  {\bibinfo {title} {A quantum engineer's guide to superconducting qubits},\
  }\href@noop {} {\bibfield  {journal} {\bibinfo  {journal} {Applied physics
  reviews}\ }\textbf {\bibinfo {volume} {6}} (\bibinfo {year}
  {2019})}\BibitemShut {NoStop}%
\bibitem [{\citenamefont {Preskill}(2018)}]{preskill2018quantum}%
  \BibitemOpen
  \bibfield  {author} {\bibinfo {author} {\bibfnamefont {J.}~\bibnamefont
  {Preskill}},\ }\bibfield  {title} {\bibinfo {title} {Quantum computing in the
  nisq era and beyond},\ }\href@noop {} {\bibfield  {journal} {\bibinfo
  {journal} {Quantum}\ }\textbf {\bibinfo {volume} {2}},\ \bibinfo {pages} {79}
  (\bibinfo {year} {2018})}\BibitemShut {NoStop}%
\bibitem [{\citenamefont {Nielsen}\ and\ \citenamefont
  {Chuang}(2010)}]{nielsen2010quantum}%
  \BibitemOpen
  \bibfield  {author} {\bibinfo {author} {\bibfnamefont {M.~A.}\ \bibnamefont
  {Nielsen}}\ and\ \bibinfo {author} {\bibfnamefont {I.~L.}\ \bibnamefont
  {Chuang}},\ }\href@noop {} {\emph {\bibinfo {title} {Quantum computation and
  quantum information}}}\ (\bibinfo  {publisher} {Cambridge university press},\
  \bibinfo {year} {2010})\BibitemShut {NoStop}%
\bibitem [{\citenamefont {Khaneja}\ \emph {et~al.}(2005)\citenamefont
  {Khaneja}, \citenamefont {Reiss}, \citenamefont {Kehlet}, \citenamefont
  {Schulte-Herbr{\"u}ggen},\ and\ \citenamefont {Glaser}}]{khaneja2005optimal}%
  \BibitemOpen
  \bibfield  {author} {\bibinfo {author} {\bibfnamefont {N.}~\bibnamefont
  {Khaneja}}, \bibinfo {author} {\bibfnamefont {T.}~\bibnamefont {Reiss}},
  \bibinfo {author} {\bibfnamefont {C.}~\bibnamefont {Kehlet}}, \bibinfo
  {author} {\bibfnamefont {T.}~\bibnamefont {Schulte-Herbr{\"u}ggen}},\ and\
  \bibinfo {author} {\bibfnamefont {S.~J.}\ \bibnamefont {Glaser}},\ }\bibfield
   {title} {\bibinfo {title} {Optimal control of coupled spin dynamics: design
  of nmr pulse sequences by gradient ascent algorithms},\ }\href@noop {}
  {\bibfield  {journal} {\bibinfo  {journal} {Journal of magnetic resonance}\
  }\textbf {\bibinfo {volume} {172}},\ \bibinfo {pages} {296} (\bibinfo {year}
  {2005})}\BibitemShut {NoStop}%
\bibitem [{\citenamefont {Koswara}\ \emph
  {et~al.}(2021{\natexlab{a}})\citenamefont {Koswara}, \citenamefont
  {Bhutoria},\ and\ \citenamefont {Chakrabarti}}]{koswara2021quantum}%
  \BibitemOpen
  \bibfield  {author} {\bibinfo {author} {\bibfnamefont {A.}~\bibnamefont
  {Koswara}}, \bibinfo {author} {\bibfnamefont {V.}~\bibnamefont {Bhutoria}},\
  and\ \bibinfo {author} {\bibfnamefont {R.}~\bibnamefont {Chakrabarti}},\
  }\bibfield  {title} {\bibinfo {title} {Quantum robust control theory for
  hamiltonian and control field uncertainty},\ }\href@noop {} {\bibfield
  {journal} {\bibinfo  {journal} {New Journal of Physics}\ }\textbf {\bibinfo
  {volume} {23}},\ \bibinfo {pages} {063046} (\bibinfo {year}
  {2021}{\natexlab{a}})}\BibitemShut {NoStop}%
\bibitem [{\citenamefont {Wu}\ \emph {et~al.}(2018)\citenamefont {Wu},
  \citenamefont {Chu}, \citenamefont {Owens},\ and\ \citenamefont
  {Rabitz}}]{wu2018data}%
  \BibitemOpen
  \bibfield  {author} {\bibinfo {author} {\bibfnamefont {R.-B.}\ \bibnamefont
  {Wu}}, \bibinfo {author} {\bibfnamefont {B.}~\bibnamefont {Chu}}, \bibinfo
  {author} {\bibfnamefont {D.~H.}\ \bibnamefont {Owens}},\ and\ \bibinfo
  {author} {\bibfnamefont {H.}~\bibnamefont {Rabitz}},\ }\bibfield  {title}
  {\bibinfo {title} {Data-driven gradient algorithm for high-precision quantum
  control},\ }\href@noop {} {\bibfield  {journal} {\bibinfo  {journal}
  {Physical Review A}\ }\textbf {\bibinfo {volume} {97}},\ \bibinfo {pages}
  {042122} (\bibinfo {year} {2018})}\BibitemShut {NoStop}%
\bibitem [{\citenamefont {Rol}\ \emph {et~al.}(2020)\citenamefont {Rol},
  \citenamefont {Ciorciaro}, \citenamefont {Malinowski}, \citenamefont
  {Tarasinski}, \citenamefont {Sagastizabal}, \citenamefont {Bultink},
  \citenamefont {Salathe}, \citenamefont {Haandb{\ae}k}, \citenamefont
  {Sedivy},\ and\ \citenamefont {DiCarlo}}]{rol2020time}%
  \BibitemOpen
  \bibfield  {author} {\bibinfo {author} {\bibfnamefont {M.~A.}\ \bibnamefont
  {Rol}}, \bibinfo {author} {\bibfnamefont {L.}~\bibnamefont {Ciorciaro}},
  \bibinfo {author} {\bibfnamefont {F.~K.}\ \bibnamefont {Malinowski}},
  \bibinfo {author} {\bibfnamefont {B.~M.}\ \bibnamefont {Tarasinski}},
  \bibinfo {author} {\bibfnamefont {R.~E.}\ \bibnamefont {Sagastizabal}},
  \bibinfo {author} {\bibfnamefont {C.~C.}\ \bibnamefont {Bultink}}, \bibinfo
  {author} {\bibfnamefont {Y.}~\bibnamefont {Salathe}}, \bibinfo {author}
  {\bibfnamefont {N.}~\bibnamefont {Haandb{\ae}k}}, \bibinfo {author}
  {\bibfnamefont {J.}~\bibnamefont {Sedivy}},\ and\ \bibinfo {author}
  {\bibfnamefont {L.}~\bibnamefont {DiCarlo}},\ }\bibfield  {title} {\bibinfo
  {title} {Time-domain characterization and correction of on-chip distortion of
  control pulses in a quantum processor},\ }\href@noop {} {\bibfield  {journal}
  {\bibinfo  {journal} {Applied Physics Letters}\ }\textbf {\bibinfo {volume}
  {116}} (\bibinfo {year} {2020})}\BibitemShut {NoStop}%
\bibitem [{\citenamefont {Zhao}\ \emph {et~al.}(2022)\citenamefont {Zhao},
  \citenamefont {Linghu}, \citenamefont {Li}, \citenamefont {Xu}, \citenamefont
  {Wang}, \citenamefont {Xue}, \citenamefont {Jin},\ and\ \citenamefont
  {Yu}}]{zhao2022quantum}%
  \BibitemOpen
  \bibfield  {author} {\bibinfo {author} {\bibfnamefont {P.}~\bibnamefont
  {Zhao}}, \bibinfo {author} {\bibfnamefont {K.}~\bibnamefont {Linghu}},
  \bibinfo {author} {\bibfnamefont {Z.}~\bibnamefont {Li}}, \bibinfo {author}
  {\bibfnamefont {P.}~\bibnamefont {Xu}}, \bibinfo {author} {\bibfnamefont
  {R.}~\bibnamefont {Wang}}, \bibinfo {author} {\bibfnamefont {G.}~\bibnamefont
  {Xue}}, \bibinfo {author} {\bibfnamefont {Y.}~\bibnamefont {Jin}},\ and\
  \bibinfo {author} {\bibfnamefont {H.}~\bibnamefont {Yu}},\ }\bibfield
  {title} {\bibinfo {title} {Quantum crosstalk analysis for simultaneous gate
  operations on superconducting qubits},\ }\href@noop {} {\bibfield  {journal}
  {\bibinfo  {journal} {PRX quantum}\ }\textbf {\bibinfo {volume} {3}},\
  \bibinfo {pages} {020301} (\bibinfo {year} {2022})}\BibitemShut {NoStop}%
\bibitem [{\citenamefont {Weidner}\ \emph {et~al.}(2024)\citenamefont
  {Weidner}, \citenamefont {Reed}, \citenamefont {Monroe}, \citenamefont
  {Sheller}, \citenamefont {O'Neil}, \citenamefont {Maas}, \citenamefont
  {Jonckheere}, \citenamefont {Langbein},\ and\ \citenamefont
  {Schirmer}}]{weidner2024robust}%
  \BibitemOpen
  \bibfield  {author} {\bibinfo {author} {\bibfnamefont {C.~A.}\ \bibnamefont
  {Weidner}}, \bibinfo {author} {\bibfnamefont {E.~A.}\ \bibnamefont {Reed}},
  \bibinfo {author} {\bibfnamefont {J.}~\bibnamefont {Monroe}}, \bibinfo
  {author} {\bibfnamefont {B.}~\bibnamefont {Sheller}}, \bibinfo {author}
  {\bibfnamefont {S.}~\bibnamefont {O'Neil}}, \bibinfo {author} {\bibfnamefont
  {E.}~\bibnamefont {Maas}}, \bibinfo {author} {\bibfnamefont {E.~A.}\
  \bibnamefont {Jonckheere}}, \bibinfo {author} {\bibfnamefont {F.~C.}\
  \bibnamefont {Langbein}},\ and\ \bibinfo {author} {\bibfnamefont {S.~G.}\
  \bibnamefont {Schirmer}},\ }\href {https://arxiv.org/abs/2401.00294}
  {\bibinfo {title} {Robust quantum control in closed and open systems: Theory
  and practice}} (\bibinfo {year} {2024}),\ \Eprint
  {https://arxiv.org/abs/2401.00294} {arXiv:2401.00294 [quant-ph]} \BibitemShut
  {NoStop}%
\bibitem [{\citenamefont {Chen}\ \emph {et~al.}(2014)\citenamefont {Chen},
  \citenamefont {Dong}, \citenamefont {Long}, \citenamefont {Petersen},\ and\
  \citenamefont {Rabitz}}]{chen2014sampling}%
  \BibitemOpen
  \bibfield  {author} {\bibinfo {author} {\bibfnamefont {C.}~\bibnamefont
  {Chen}}, \bibinfo {author} {\bibfnamefont {D.}~\bibnamefont {Dong}}, \bibinfo
  {author} {\bibfnamefont {R.}~\bibnamefont {Long}}, \bibinfo {author}
  {\bibfnamefont {I.~R.}\ \bibnamefont {Petersen}},\ and\ \bibinfo {author}
  {\bibfnamefont {H.~A.}\ \bibnamefont {Rabitz}},\ }\bibfield  {title}
  {\bibinfo {title} {Sampling-based learning control of inhomogeneous quantum
  ensembles},\ }\href@noop {} {\bibfield  {journal} {\bibinfo  {journal}
  {Physical Review A}\ }\textbf {\bibinfo {volume} {89}},\ \bibinfo {pages}
  {023402} (\bibinfo {year} {2014})}\BibitemShut {NoStop}%
\bibitem [{\citenamefont {Wu}\ \emph {et~al.}(2019)\citenamefont {Wu},
  \citenamefont {Ding}, \citenamefont {Dong},\ and\ \citenamefont
  {Wang}}]{wu2019learning}%
  \BibitemOpen
  \bibfield  {author} {\bibinfo {author} {\bibfnamefont {R.-B.}\ \bibnamefont
  {Wu}}, \bibinfo {author} {\bibfnamefont {H.}~\bibnamefont {Ding}}, \bibinfo
  {author} {\bibfnamefont {D.}~\bibnamefont {Dong}},\ and\ \bibinfo {author}
  {\bibfnamefont {X.}~\bibnamefont {Wang}},\ }\bibfield  {title} {\bibinfo
  {title} {Learning robust and high-precision quantum controls},\ }\href@noop
  {} {\bibfield  {journal} {\bibinfo  {journal} {Physical Review A}\ }\textbf
  {\bibinfo {volume} {99}},\ \bibinfo {pages} {042327} (\bibinfo {year}
  {2019})}\BibitemShut {NoStop}%
\bibitem [{\citenamefont {Turinici}(2019)}]{turinici2019stochasticgradient}%
  \BibitemOpen
  \bibfield  {author} {\bibinfo {author} {\bibfnamefont {G.}~\bibnamefont
  {Turinici}},\ }\bibfield  {title} {\bibinfo {title} {Stochastic learning
  control of inhomogeneous quantum ensembles},\ }\href
  {https://doi.org/10.1103/PhysRevA.100.053403} {\bibfield  {journal} {\bibinfo
   {journal} {Phys. Rev. A}\ }\textbf {\bibinfo {volume} {100}},\ \bibinfo
  {pages} {053403} (\bibinfo {year} {2019})}\BibitemShut {NoStop}%
\bibitem [{\citenamefont {Ge}\ \emph {et~al.}(2020)\citenamefont {Ge},
  \citenamefont {Ding}, \citenamefont {Rabitz},\ and\ \citenamefont
  {Wu}}]{ge2020robust}%
  \BibitemOpen
  \bibfield  {author} {\bibinfo {author} {\bibfnamefont {X.}~\bibnamefont
  {Ge}}, \bibinfo {author} {\bibfnamefont {H.}~\bibnamefont {Ding}}, \bibinfo
  {author} {\bibfnamefont {H.}~\bibnamefont {Rabitz}},\ and\ \bibinfo {author}
  {\bibfnamefont {R.-B.}\ \bibnamefont {Wu}},\ }\bibfield  {title} {\bibinfo
  {title} {Robust quantum control in games: An adversarial learning approach},\
  }\href@noop {} {\bibfield  {journal} {\bibinfo  {journal} {Physical Review
  A}\ }\textbf {\bibinfo {volume} {101}},\ \bibinfo {pages} {052317} (\bibinfo
  {year} {2020})}\BibitemShut {NoStop}%
\bibitem [{\citenamefont {Li}\ and\ \citenamefont
  {Khaneja}(2006)}]{li2006control}%
  \BibitemOpen
  \bibfield  {author} {\bibinfo {author} {\bibfnamefont {J.-S.}\ \bibnamefont
  {Li}}\ and\ \bibinfo {author} {\bibfnamefont {N.}~\bibnamefont {Khaneja}},\
  }\bibfield  {title} {\bibinfo {title} {Control of inhomogeneous quantum
  ensembles},\ }\href@noop {} {\bibfield  {journal} {\bibinfo  {journal}
  {Physical Review A—Atomic, Molecular, and Optical Physics}\ }\textbf
  {\bibinfo {volume} {73}},\ \bibinfo {pages} {030302} (\bibinfo {year}
  {2006})}\BibitemShut {NoStop}%
\bibitem [{\citenamefont {Van~Damme}\ \emph {et~al.}(2017)\citenamefont
  {Van~Damme}, \citenamefont {Ansel}, \citenamefont {Glaser},\ and\
  \citenamefont {Sugny}}]{van2017robust}%
  \BibitemOpen
  \bibfield  {author} {\bibinfo {author} {\bibfnamefont {L.}~\bibnamefont
  {Van~Damme}}, \bibinfo {author} {\bibfnamefont {Q.}~\bibnamefont {Ansel}},
  \bibinfo {author} {\bibfnamefont {S.}~\bibnamefont {Glaser}},\ and\ \bibinfo
  {author} {\bibfnamefont {D.}~\bibnamefont {Sugny}},\ }\bibfield  {title}
  {\bibinfo {title} {Robust optimal control of two-level quantum systems},\
  }\href@noop {} {\bibfield  {journal} {\bibinfo  {journal} {Physical Review
  A}\ }\textbf {\bibinfo {volume} {95}},\ \bibinfo {pages} {063403} (\bibinfo
  {year} {2017})}\BibitemShut {NoStop}%
\bibitem [{\citenamefont {Li}\ \emph {et~al.}(2022)\citenamefont {Li},
  \citenamefont {Zhang},\ and\ \citenamefont {Kuan}}]{li2022moment}%
  \BibitemOpen
  \bibfield  {author} {\bibinfo {author} {\bibfnamefont {J.-S.}\ \bibnamefont
  {Li}}, \bibinfo {author} {\bibfnamefont {W.}~\bibnamefont {Zhang}},\ and\
  \bibinfo {author} {\bibfnamefont {Y.-H.}\ \bibnamefont {Kuan}},\ }\bibfield
  {title} {\bibinfo {title} {Moment quantization of inhomogeneous spin
  ensembles},\ }\href@noop {} {\bibfield  {journal} {\bibinfo  {journal}
  {Annual Reviews in Control}\ }\textbf {\bibinfo {volume} {54}},\ \bibinfo
  {pages} {305} (\bibinfo {year} {2022})}\BibitemShut {NoStop}%
\bibitem [{\citenamefont {Cao}\ \emph {et~al.}(2024)\citenamefont {Cao},
  \citenamefont {Cui}, \citenamefont {Yung},\ and\ \citenamefont
  {Wu}}]{cao2024robust}%
  \BibitemOpen
  \bibfield  {author} {\bibinfo {author} {\bibfnamefont {X.}~\bibnamefont
  {Cao}}, \bibinfo {author} {\bibfnamefont {J.}~\bibnamefont {Cui}}, \bibinfo
  {author} {\bibfnamefont {M.~H.}\ \bibnamefont {Yung}},\ and\ \bibinfo
  {author} {\bibfnamefont {R.-B.}\ \bibnamefont {Wu}},\ }\bibfield  {title}
  {\bibinfo {title} {Robust control of single-qubit gates at the quantum speed
  limit},\ }\href@noop {} {\bibfield  {journal} {\bibinfo  {journal} {Physical
  Review A}\ }\textbf {\bibinfo {volume} {110}},\ \bibinfo {pages} {022603}
  (\bibinfo {year} {2024})}\BibitemShut {NoStop}%
\bibitem [{\citenamefont {Poggi}\ \emph {et~al.}(2024)\citenamefont {Poggi},
  \citenamefont {De~Chiara}, \citenamefont {Campbell},\ and\ \citenamefont
  {Kiely}}]{poggi2024universally}%
  \BibitemOpen
  \bibfield  {author} {\bibinfo {author} {\bibfnamefont {P.~M.}\ \bibnamefont
  {Poggi}}, \bibinfo {author} {\bibfnamefont {G.}~\bibnamefont {De~Chiara}},
  \bibinfo {author} {\bibfnamefont {S.}~\bibnamefont {Campbell}},\ and\
  \bibinfo {author} {\bibfnamefont {A.}~\bibnamefont {Kiely}},\ }\bibfield
  {title} {\bibinfo {title} {Universally robust quantum control},\ }\href@noop
  {} {\bibfield  {journal} {\bibinfo  {journal} {Physical review letters}\
  }\textbf {\bibinfo {volume} {132}},\ \bibinfo {pages} {193801} (\bibinfo
  {year} {2024})}\BibitemShut {NoStop}%
\bibitem [{\citenamefont {Propson}\ \emph {et~al.}(2022)\citenamefont
  {Propson}, \citenamefont {Jackson}, \citenamefont {Koch}, \citenamefont
  {Manchester},\ and\ \citenamefont {Schuster}}]{propson2022robust}%
  \BibitemOpen
  \bibfield  {author} {\bibinfo {author} {\bibfnamefont {T.}~\bibnamefont
  {Propson}}, \bibinfo {author} {\bibfnamefont {B.~E.}\ \bibnamefont
  {Jackson}}, \bibinfo {author} {\bibfnamefont {J.}~\bibnamefont {Koch}},
  \bibinfo {author} {\bibfnamefont {Z.}~\bibnamefont {Manchester}},\ and\
  \bibinfo {author} {\bibfnamefont {D.~I.}\ \bibnamefont {Schuster}},\
  }\bibfield  {title} {\bibinfo {title} {Robust quantum optimal control with
  trajectory optimization},\ }\href@noop {} {\bibfield  {journal} {\bibinfo
  {journal} {Physical Review Applied}\ }\textbf {\bibinfo {volume} {17}},\
  \bibinfo {pages} {014036} (\bibinfo {year} {2022})}\BibitemShut {NoStop}%
\bibitem [{\citenamefont {Liu}\ and\ \citenamefont
  {Nocedal}(1989)}]{liu1989limited}%
  \BibitemOpen
  \bibfield  {author} {\bibinfo {author} {\bibfnamefont {D.~C.}\ \bibnamefont
  {Liu}}\ and\ \bibinfo {author} {\bibfnamefont {J.}~\bibnamefont {Nocedal}},\
  }\bibfield  {title} {\bibinfo {title} {On the limited memory bfgs method for
  large scale optimization},\ }\href@noop {} {\bibfield  {journal} {\bibinfo
  {journal} {Mathematical programming}\ }\textbf {\bibinfo {volume} {45}},\
  \bibinfo {pages} {503} (\bibinfo {year} {1989})}\BibitemShut {NoStop}%
\bibitem [{\citenamefont {Byrd}\ \emph {et~al.}(1995)\citenamefont {Byrd},
  \citenamefont {Lu}, \citenamefont {Nocedal},\ and\ \citenamefont
  {Zhu}}]{byrd1995limited}%
  \BibitemOpen
  \bibfield  {author} {\bibinfo {author} {\bibfnamefont {R.~H.}\ \bibnamefont
  {Byrd}}, \bibinfo {author} {\bibfnamefont {P.}~\bibnamefont {Lu}}, \bibinfo
  {author} {\bibfnamefont {J.}~\bibnamefont {Nocedal}},\ and\ \bibinfo {author}
  {\bibfnamefont {C.}~\bibnamefont {Zhu}},\ }\bibfield  {title} {\bibinfo
  {title} {A limited memory algorithm for bound constrained optimization},\
  }\href@noop {} {\bibfield  {journal} {\bibinfo  {journal} {SIAM Journal on
  scientific computing}\ }\textbf {\bibinfo {volume} {16}},\ \bibinfo {pages}
  {1190} (\bibinfo {year} {1995})}\BibitemShut {NoStop}%
\bibitem [{\citenamefont {Green}\ \emph {et~al.}(2013)\citenamefont {Green},
  \citenamefont {Sastrawan}, \citenamefont {Uys},\ and\ \citenamefont
  {Biercuk}}]{green2013arbitrary}%
  \BibitemOpen
  \bibfield  {author} {\bibinfo {author} {\bibfnamefont {T.~J.}\ \bibnamefont
  {Green}}, \bibinfo {author} {\bibfnamefont {J.}~\bibnamefont {Sastrawan}},
  \bibinfo {author} {\bibfnamefont {H.}~\bibnamefont {Uys}},\ and\ \bibinfo
  {author} {\bibfnamefont {M.~J.}\ \bibnamefont {Biercuk}},\ }\bibfield
  {title} {\bibinfo {title} {Arbitrary quantum control of qubits in the
  presence of universal noise},\ }\href@noop {} {\bibfield  {journal} {\bibinfo
   {journal} {New Journal of Physics}\ }\textbf {\bibinfo {volume} {15}},\
  \bibinfo {pages} {095004} (\bibinfo {year} {2013})}\BibitemShut {NoStop}%
\bibitem [{\citenamefont {Koswara}\ and\ \citenamefont
  {Chakrabarti}(2014)}]{koswara2014robustness}%
  \BibitemOpen
  \bibfield  {author} {\bibinfo {author} {\bibfnamefont {A.}~\bibnamefont
  {Koswara}}\ and\ \bibinfo {author} {\bibfnamefont {R.}~\bibnamefont
  {Chakrabarti}},\ }\bibfield  {title} {\bibinfo {title} {Robustness of
  controlled quantum dynamics},\ }\href@noop {} {\bibfield  {journal} {\bibinfo
   {journal} {Physical Review A}\ }\textbf {\bibinfo {volume} {90}},\ \bibinfo
  {pages} {043414} (\bibinfo {year} {2014})}\BibitemShut {NoStop}%
\bibitem [{\citenamefont {Haas}\ \emph {et~al.}(2019)\citenamefont {Haas},
  \citenamefont {Puzzuoli}, \citenamefont {Zhang},\ and\ \citenamefont
  {Cory}}]{haas2019engineering}%
  \BibitemOpen
  \bibfield  {author} {\bibinfo {author} {\bibfnamefont {H.}~\bibnamefont
  {Haas}}, \bibinfo {author} {\bibfnamefont {D.}~\bibnamefont {Puzzuoli}},
  \bibinfo {author} {\bibfnamefont {F.}~\bibnamefont {Zhang}},\ and\ \bibinfo
  {author} {\bibfnamefont {D.~G.}\ \bibnamefont {Cory}},\ }\bibfield  {title}
  {\bibinfo {title} {Engineering effective hamiltonians},\ }\href@noop {}
  {\bibfield  {journal} {\bibinfo  {journal} {New Journal of Physics}\ }\textbf
  {\bibinfo {volume} {21}},\ \bibinfo {pages} {103011} (\bibinfo {year}
  {2019})}\BibitemShut {NoStop}%
\bibitem [{\citenamefont {Koswara}\ \emph
  {et~al.}(2021{\natexlab{b}})\citenamefont {Koswara}, \citenamefont
  {Bhutoria},\ and\ \citenamefont {Chakrabarti}}]{koswara2021robust}%
  \BibitemOpen
  \bibfield  {author} {\bibinfo {author} {\bibfnamefont {A.}~\bibnamefont
  {Koswara}}, \bibinfo {author} {\bibfnamefont {V.}~\bibnamefont {Bhutoria}},\
  and\ \bibinfo {author} {\bibfnamefont {R.}~\bibnamefont {Chakrabarti}},\
  }\bibfield  {title} {\bibinfo {title} {Robust control of quantum dynamics
  under input and parameter uncertainty},\ }\href@noop {} {\bibfield  {journal}
  {\bibinfo  {journal} {Physical Review A}\ }\textbf {\bibinfo {volume}
  {104}},\ \bibinfo {pages} {053118} (\bibinfo {year}
  {2021}{\natexlab{b}})}\BibitemShut {NoStop}%
\bibitem [{\citenamefont {Puzzuoli}\ \emph
  {et~al.}(2023{\natexlab{a}})\citenamefont {Puzzuoli}, \citenamefont {Lin},
  \citenamefont {Malekakhlagh}, \citenamefont {Pritchett}, \citenamefont
  {Rosand},\ and\ \citenamefont {Wood}}]{puzzuoli2023algorithms}%
  \BibitemOpen
  \bibfield  {author} {\bibinfo {author} {\bibfnamefont {D.}~\bibnamefont
  {Puzzuoli}}, \bibinfo {author} {\bibfnamefont {S.~F.}\ \bibnamefont {Lin}},
  \bibinfo {author} {\bibfnamefont {M.}~\bibnamefont {Malekakhlagh}}, \bibinfo
  {author} {\bibfnamefont {E.}~\bibnamefont {Pritchett}}, \bibinfo {author}
  {\bibfnamefont {B.}~\bibnamefont {Rosand}},\ and\ \bibinfo {author}
  {\bibfnamefont {C.~J.}\ \bibnamefont {Wood}},\ }\bibfield  {title} {\bibinfo
  {title} {Algorithms for perturbative analysis and simulation of quantum
  dynamics},\ }\href@noop {} {\bibfield  {journal} {\bibinfo  {journal}
  {Journal of Computational Physics}\ }\textbf {\bibinfo {volume} {489}},\
  \bibinfo {pages} {112262} (\bibinfo {year} {2023}{\natexlab{a}})}\BibitemShut
  {NoStop}%
\bibitem [{\citenamefont {Liu}\ \emph {et~al.}(2024)\citenamefont {Liu},
  \citenamefont {Yang},\ and\ \citenamefont {Li}}]{liu2024robust}%
  \BibitemOpen
  \bibfield  {author} {\bibinfo {author} {\bibfnamefont {R.}~\bibnamefont
  {Liu}}, \bibinfo {author} {\bibfnamefont {X.}~\bibnamefont {Yang}},\ and\
  \bibinfo {author} {\bibfnamefont {J.}~\bibnamefont {Li}},\ }\bibfield
  {title} {\bibinfo {title} {Robust quantum optimal control for markovian
  quantum systems},\ }\href@noop {} {\bibfield  {journal} {\bibinfo  {journal}
  {Physical Review A}\ }\textbf {\bibinfo {volume} {110}},\ \bibinfo {pages}
  {012402} (\bibinfo {year} {2024})}\BibitemShut {NoStop}%
\bibitem [{\citenamefont {Shao}\ \emph {et~al.}(2024)\citenamefont {Shao},
  \citenamefont {Yang}, \citenamefont {Liu}, \citenamefont {Zhai},
  \citenamefont {Lu}, \citenamefont {Xin},\ and\ \citenamefont
  {Li}}]{shao2024multiple}%
  \BibitemOpen
  \bibfield  {author} {\bibinfo {author} {\bibfnamefont {B.}~\bibnamefont
  {Shao}}, \bibinfo {author} {\bibfnamefont {X.}~\bibnamefont {Yang}}, \bibinfo
  {author} {\bibfnamefont {R.}~\bibnamefont {Liu}}, \bibinfo {author}
  {\bibfnamefont {Y.}~\bibnamefont {Zhai}}, \bibinfo {author} {\bibfnamefont
  {D.}~\bibnamefont {Lu}}, \bibinfo {author} {\bibfnamefont {T.}~\bibnamefont
  {Xin}},\ and\ \bibinfo {author} {\bibfnamefont {J.}~\bibnamefont {Li}},\
  }\bibfield  {title} {\bibinfo {title} {Multiple classical noise mitigation by
  multiobjective robust quantum optimal control},\ }\href@noop {} {\bibfield
  {journal} {\bibinfo  {journal} {Physical Review Applied}\ }\textbf {\bibinfo
  {volume} {21}},\ \bibinfo {pages} {034042} (\bibinfo {year}
  {2024})}\BibitemShut {NoStop}%
\bibitem [{\citenamefont {Chen}\ \emph {et~al.}(2025)\citenamefont {Chen},
  \citenamefont {Huang}, \citenamefont {Sun}, \citenamefont {Jie},
  \citenamefont {Zhou}, \citenamefont {Hua}, \citenamefont {Xu}, \citenamefont
  {Wang}, \citenamefont {Guo}, \citenamefont {Zou} \emph
  {et~al.}}]{chen2025robust}%
  \BibitemOpen
  \bibfield  {author} {\bibinfo {author} {\bibfnamefont {Z.-J.}\ \bibnamefont
  {Chen}}, \bibinfo {author} {\bibfnamefont {H.}~\bibnamefont {Huang}},
  \bibinfo {author} {\bibfnamefont {L.}~\bibnamefont {Sun}}, \bibinfo {author}
  {\bibfnamefont {Q.-X.}\ \bibnamefont {Jie}}, \bibinfo {author} {\bibfnamefont
  {J.}~\bibnamefont {Zhou}}, \bibinfo {author} {\bibfnamefont {Z.}~\bibnamefont
  {Hua}}, \bibinfo {author} {\bibfnamefont {Y.}~\bibnamefont {Xu}}, \bibinfo
  {author} {\bibfnamefont {W.}~\bibnamefont {Wang}}, \bibinfo {author}
  {\bibfnamefont {G.-C.}\ \bibnamefont {Guo}}, \bibinfo {author} {\bibfnamefont
  {C.-L.}\ \bibnamefont {Zou}}, \emph {et~al.},\ }\bibfield  {title} {\bibinfo
  {title} {Robust and optimal control of open quantum systems},\ }\href@noop {}
  {\bibfield  {journal} {\bibinfo  {journal} {Science Advances}\ }\textbf
  {\bibinfo {volume} {11}},\ \bibinfo {pages} {eadr0875} (\bibinfo {year}
  {2025})}\BibitemShut {NoStop}%
\bibitem [{\citenamefont {Schulte-Herbr{\"u}ggen}\ \emph
  {et~al.}(2011)\citenamefont {Schulte-Herbr{\"u}ggen}, \citenamefont
  {Sp{\"o}rl}, \citenamefont {Khaneja},\ and\ \citenamefont
  {Glaser}}]{schulte2011optimal}%
  \BibitemOpen
  \bibfield  {author} {\bibinfo {author} {\bibfnamefont {T.}~\bibnamefont
  {Schulte-Herbr{\"u}ggen}}, \bibinfo {author} {\bibfnamefont {A.}~\bibnamefont
  {Sp{\"o}rl}}, \bibinfo {author} {\bibfnamefont {N.}~\bibnamefont {Khaneja}},\
  and\ \bibinfo {author} {\bibfnamefont {S.}~\bibnamefont {Glaser}},\
  }\bibfield  {title} {\bibinfo {title} {Optimal control for generating quantum
  gates in open dissipative systems},\ }\href@noop {} {\bibfield  {journal}
  {\bibinfo  {journal} {Journal of Physics B: Atomic, Molecular and Optical
  Physics}\ }\textbf {\bibinfo {volume} {44}},\ \bibinfo {pages} {154013}
  (\bibinfo {year} {2011})}\BibitemShut {NoStop}%
\bibitem [{\citenamefont {Floether}\ \emph {et~al.}(2012)\citenamefont
  {Floether}, \citenamefont {De~Fouquieres},\ and\ \citenamefont
  {Schirmer}}]{floether2012robust}%
  \BibitemOpen
  \bibfield  {author} {\bibinfo {author} {\bibfnamefont {F.~F.}\ \bibnamefont
  {Floether}}, \bibinfo {author} {\bibfnamefont {P.}~\bibnamefont
  {De~Fouquieres}},\ and\ \bibinfo {author} {\bibfnamefont {S.~G.}\
  \bibnamefont {Schirmer}},\ }\bibfield  {title} {\bibinfo {title} {Robust
  quantum gates for open systems via optimal control: Markovian versus
  non-markovian dynamics},\ }\href@noop {} {\bibfield  {journal} {\bibinfo
  {journal} {New Journal of Physics}\ }\textbf {\bibinfo {volume} {14}},\
  \bibinfo {pages} {073023} (\bibinfo {year} {2012})}\BibitemShut {NoStop}%
\bibitem [{\citenamefont {Goerz}\ \emph {et~al.}(2014)\citenamefont {Goerz},
  \citenamefont {Reich},\ and\ \citenamefont {Koch}}]{goerz2014optimal}%
  \BibitemOpen
  \bibfield  {author} {\bibinfo {author} {\bibfnamefont {M.~H.}\ \bibnamefont
  {Goerz}}, \bibinfo {author} {\bibfnamefont {D.~M.}\ \bibnamefont {Reich}},\
  and\ \bibinfo {author} {\bibfnamefont {C.~P.}\ \bibnamefont {Koch}},\
  }\bibfield  {title} {\bibinfo {title} {Optimal control theory for a unitary
  operation under dissipative evolution},\ }\href@noop {} {\bibfield  {journal}
  {\bibinfo  {journal} {New Journal of Physics}\ }\textbf {\bibinfo {volume}
  {16}},\ \bibinfo {pages} {055012} (\bibinfo {year} {2014})}\BibitemShut
  {NoStop}%
\bibitem [{\citenamefont {Koch}(2016)}]{koch2016controlling}%
  \BibitemOpen
  \bibfield  {author} {\bibinfo {author} {\bibfnamefont {C.~P.}\ \bibnamefont
  {Koch}},\ }\bibfield  {title} {\bibinfo {title} {Controlling open quantum
  systems: tools, achievements, and limitations},\ }\href@noop {} {\bibfield
  {journal} {\bibinfo  {journal} {Journal of Physics: Condensed Matter}\
  }\textbf {\bibinfo {volume} {28}},\ \bibinfo {pages} {213001} (\bibinfo
  {year} {2016})}\BibitemShut {NoStop}%
\bibitem [{\citenamefont {Boutin}\ \emph {et~al.}(2017)\citenamefont {Boutin},
  \citenamefont {Andersen}, \citenamefont {Venkatraman}, \citenamefont
  {Ferris},\ and\ \citenamefont {Blais}}]{boutin2017resonator}%
  \BibitemOpen
  \bibfield  {author} {\bibinfo {author} {\bibfnamefont {S.}~\bibnamefont
  {Boutin}}, \bibinfo {author} {\bibfnamefont {C.~K.}\ \bibnamefont
  {Andersen}}, \bibinfo {author} {\bibfnamefont {J.}~\bibnamefont
  {Venkatraman}}, \bibinfo {author} {\bibfnamefont {A.~J.}\ \bibnamefont
  {Ferris}},\ and\ \bibinfo {author} {\bibfnamefont {A.}~\bibnamefont
  {Blais}},\ }\bibfield  {title} {\bibinfo {title} {Resonator reset in circuit
  {QED} by optimal control for large open quantum systems},\ }\href@noop {}
  {\bibfield  {journal} {\bibinfo  {journal} {Physical Review A}\ }\textbf
  {\bibinfo {volume} {96}},\ \bibinfo {pages} {042315} (\bibinfo {year}
  {2017})}\BibitemShut {NoStop}%
\bibitem [{\citenamefont {Gautier}\ \emph {et~al.}(2025)\citenamefont
  {Gautier}, \citenamefont {Genois},\ and\ \citenamefont
  {Blais}}]{gautier2025optimal}%
  \BibitemOpen
  \bibfield  {author} {\bibinfo {author} {\bibfnamefont {R.}~\bibnamefont
  {Gautier}}, \bibinfo {author} {\bibfnamefont {{\'E}.}~\bibnamefont
  {Genois}},\ and\ \bibinfo {author} {\bibfnamefont {A.}~\bibnamefont
  {Blais}},\ }\bibfield  {title} {\bibinfo {title} {Optimal control in large
  open quantum systems: the case of transmon readout and reset},\ }\href@noop
  {} {\bibfield  {journal} {\bibinfo  {journal} {Physical Review Letters}\
  }\textbf {\bibinfo {volume} {134}},\ \bibinfo {pages} {070802} (\bibinfo
  {year} {2025})}\BibitemShut {NoStop}%
\bibitem [{\citenamefont {Breuer}\ and\ \citenamefont
  {Petruccione}(2002)}]{breuer2002theory}%
  \BibitemOpen
  \bibfield  {author} {\bibinfo {author} {\bibfnamefont {H.-P.}\ \bibnamefont
  {Breuer}}\ and\ \bibinfo {author} {\bibfnamefont {F.}~\bibnamefont
  {Petruccione}},\ }\href@noop {} {\emph {\bibinfo {title} {The theory of open
  quantum systems}}}\ (\bibinfo  {publisher} {Oxford University Press, USA},\
  \bibinfo {year} {2002})\BibitemShut {NoStop}%
\bibitem [{\citenamefont {Jing}\ and\ \citenamefont {Wu}(2022)}]{jing2022one}%
  \BibitemOpen
  \bibfield  {author} {\bibinfo {author} {\bibfnamefont {J.}~\bibnamefont
  {Jing}}\ and\ \bibinfo {author} {\bibfnamefont {L.-A.}\ \bibnamefont {Wu}},\
  }\bibfield  {title} {\bibinfo {title} {One-component quantum mechanics and
  dynamical leakage-free paths},\ }\href@noop {} {\bibfield  {journal}
  {\bibinfo  {journal} {Scientific reports}\ }\textbf {\bibinfo {volume}
  {12}},\ \bibinfo {pages} {9247} (\bibinfo {year} {2022})}\BibitemShut
  {NoStop}%
\bibitem [{\citenamefont {Abdelhafez}\ \emph {et~al.}(2019)\citenamefont
  {Abdelhafez}, \citenamefont {Schuster},\ and\ \citenamefont
  {Koch}}]{abdelhafez2019gradient}%
  \BibitemOpen
  \bibfield  {author} {\bibinfo {author} {\bibfnamefont {M.}~\bibnamefont
  {Abdelhafez}}, \bibinfo {author} {\bibfnamefont {D.~I.}\ \bibnamefont
  {Schuster}},\ and\ \bibinfo {author} {\bibfnamefont {J.}~\bibnamefont
  {Koch}},\ }\bibfield  {title} {\bibinfo {title} {Gradient-based optimal
  control of open quantum systems using quantum trajectories and automatic
  differentiation},\ }\href@noop {} {\bibfield  {journal} {\bibinfo  {journal}
  {Physical Review A}\ }\textbf {\bibinfo {volume} {99}},\ \bibinfo {pages}
  {052327} (\bibinfo {year} {2019})}\BibitemShut {NoStop}%
\bibitem [{\citenamefont {Dum}\ \emph {et~al.}(1992)\citenamefont {Dum},
  \citenamefont {Parkins}, \citenamefont {Zoller},\ and\ \citenamefont
  {Gardiner}}]{dum1992monte}%
  \BibitemOpen
  \bibfield  {author} {\bibinfo {author} {\bibfnamefont {R.}~\bibnamefont
  {Dum}}, \bibinfo {author} {\bibfnamefont {A.}~\bibnamefont {Parkins}},
  \bibinfo {author} {\bibfnamefont {P.}~\bibnamefont {Zoller}},\ and\ \bibinfo
  {author} {\bibfnamefont {C.}~\bibnamefont {Gardiner}},\ }\bibfield  {title}
  {\bibinfo {title} {Monte carlo simulation of master equations in quantum
  optics for vacuum, thermal, and squeezed reservoirs},\ }\href@noop {}
  {\bibfield  {journal} {\bibinfo  {journal} {Physical Review A}\ }\textbf
  {\bibinfo {volume} {46}},\ \bibinfo {pages} {4382} (\bibinfo {year}
  {1992})}\BibitemShut {NoStop}%
\bibitem [{\citenamefont {Daley}(2014)}]{daley2014quantum}%
  \BibitemOpen
  \bibfield  {author} {\bibinfo {author} {\bibfnamefont {A.~J.}\ \bibnamefont
  {Daley}},\ }\bibfield  {title} {\bibinfo {title} {Quantum trajectories and
  open many-body quantum systems},\ }\href@noop {} {\bibfield  {journal}
  {\bibinfo  {journal} {Advances in Physics}\ }\textbf {\bibinfo {volume}
  {63}},\ \bibinfo {pages} {77} (\bibinfo {year} {2014})}\BibitemShut {NoStop}%
\bibitem [{\citenamefont {Dalgaard}\ \emph {et~al.}(2022)\citenamefont
  {Dalgaard}, \citenamefont {Weidner},\ and\ \citenamefont
  {Motzoi}}]{dalgaard2022dynamical}%
  \BibitemOpen
  \bibfield  {author} {\bibinfo {author} {\bibfnamefont {M.}~\bibnamefont
  {Dalgaard}}, \bibinfo {author} {\bibfnamefont {C.~A.}\ \bibnamefont
  {Weidner}},\ and\ \bibinfo {author} {\bibfnamefont {F.}~\bibnamefont
  {Motzoi}},\ }\bibfield  {title} {\bibinfo {title} {Dynamical uncertainty
  propagation with noisy quantum parameters},\ }\href@noop {} {\bibfield
  {journal} {\bibinfo  {journal} {Physical Review Letters}\ }\textbf {\bibinfo
  {volume} {128}},\ \bibinfo {pages} {150503} (\bibinfo {year}
  {2022})}\BibitemShut {NoStop}%
\bibitem [{\citenamefont {Bhole}\ and\ \citenamefont
  {Jones}(2018)}]{bhole2018practical}%
  \BibitemOpen
  \bibfield  {author} {\bibinfo {author} {\bibfnamefont {G.}~\bibnamefont
  {Bhole}}\ and\ \bibinfo {author} {\bibfnamefont {J.~A.}\ \bibnamefont
  {Jones}},\ }\bibfield  {title} {\bibinfo {title} {Practical pulse
  engineering: Gradient ascent without matrix exponentiation},\ }\href@noop {}
  {\bibfield  {journal} {\bibinfo  {journal} {Frontiers of Physics}\ }\textbf
  {\bibinfo {volume} {13}},\ \bibinfo {pages} {1} (\bibinfo {year}
  {2018})}\BibitemShut {NoStop}%
\bibitem [{\citenamefont {Jensen}\ \emph
  {et~al.}(2021{\natexlab{a}})\citenamefont {Jensen}, \citenamefont
  {M{\o}ller}, \citenamefont {S{\o}rensen},\ and\ \citenamefont
  {Sherson}}]{jensen2021approximate}%
  \BibitemOpen
  \bibfield  {author} {\bibinfo {author} {\bibfnamefont {J.~H.~M.}\
  \bibnamefont {Jensen}}, \bibinfo {author} {\bibfnamefont {F.~S.}\
  \bibnamefont {M{\o}ller}}, \bibinfo {author} {\bibfnamefont {J.~J.}\
  \bibnamefont {S{\o}rensen}},\ and\ \bibinfo {author} {\bibfnamefont {J.~F.}\
  \bibnamefont {Sherson}},\ }\bibfield  {title} {\bibinfo {title} {Approximate
  dynamics leading to more optimal control: Efficient exact derivatives},\
  }\href@noop {} {\bibfield  {journal} {\bibinfo  {journal} {Physical Review
  A}\ }\textbf {\bibinfo {volume} {103}},\ \bibinfo {pages} {062612} (\bibinfo
  {year} {2021}{\natexlab{a}})}\BibitemShut {NoStop}%
\bibitem [{\citenamefont {Jensen}\ \emph
  {et~al.}(2021{\natexlab{b}})\citenamefont {Jensen}, \citenamefont
  {M{\o}ller}, \citenamefont {S{\o}rensen},\ and\ \citenamefont
  {Sherson}}]{jensen2021achieving}%
  \BibitemOpen
  \bibfield  {author} {\bibinfo {author} {\bibfnamefont {J.~H.~M.}\
  \bibnamefont {Jensen}}, \bibinfo {author} {\bibfnamefont {F.~S.}\
  \bibnamefont {M{\o}ller}}, \bibinfo {author} {\bibfnamefont {J.~J.}\
  \bibnamefont {S{\o}rensen}},\ and\ \bibinfo {author} {\bibfnamefont {J.~F.}\
  \bibnamefont {Sherson}},\ }\bibfield  {title} {\bibinfo {title} {Achieving
  fast high-fidelity optimal control of many-body quantum dynamics},\
  }\href@noop {} {\bibfield  {journal} {\bibinfo  {journal} {Physical Review
  A}\ }\textbf {\bibinfo {volume} {104}},\ \bibinfo {pages} {052210} (\bibinfo
  {year} {2021}{\natexlab{b}})}\BibitemShut {NoStop}%
\bibitem [{\citenamefont {Fowler}\ \emph {et~al.}(2012)\citenamefont {Fowler},
  \citenamefont {Mariantoni}, \citenamefont {Martinis},\ and\ \citenamefont
  {Cleland}}]{fowler2012surface}%
  \BibitemOpen
  \bibfield  {author} {\bibinfo {author} {\bibfnamefont {A.~G.}\ \bibnamefont
  {Fowler}}, \bibinfo {author} {\bibfnamefont {M.}~\bibnamefont {Mariantoni}},
  \bibinfo {author} {\bibfnamefont {J.~M.}\ \bibnamefont {Martinis}},\ and\
  \bibinfo {author} {\bibfnamefont {A.~N.}\ \bibnamefont {Cleland}},\
  }\bibfield  {title} {\bibinfo {title} {Surface codes: Towards practical
  large-scale quantum computation},\ }\href@noop {} {\bibfield  {journal}
  {\bibinfo  {journal} {Physical Review A—Atomic, Molecular, and Optical
  Physics}\ }\textbf {\bibinfo {volume} {86}},\ \bibinfo {pages} {032324}
  (\bibinfo {year} {2012})}\BibitemShut {NoStop}%
\bibitem [{\citenamefont {Michael}\ \emph {et~al.}(2016)\citenamefont
  {Michael}, \citenamefont {Silveri}, \citenamefont {Brierley}, \citenamefont
  {Albert}, \citenamefont {Salmilehto}, \citenamefont {Jiang},\ and\
  \citenamefont {Girvin}}]{michael2016new}%
  \BibitemOpen
  \bibfield  {author} {\bibinfo {author} {\bibfnamefont {M.~H.}\ \bibnamefont
  {Michael}}, \bibinfo {author} {\bibfnamefont {M.}~\bibnamefont {Silveri}},
  \bibinfo {author} {\bibfnamefont {R.}~\bibnamefont {Brierley}}, \bibinfo
  {author} {\bibfnamefont {V.~V.}\ \bibnamefont {Albert}}, \bibinfo {author}
  {\bibfnamefont {J.}~\bibnamefont {Salmilehto}}, \bibinfo {author}
  {\bibfnamefont {L.}~\bibnamefont {Jiang}},\ and\ \bibinfo {author}
  {\bibfnamefont {S.~M.}\ \bibnamefont {Girvin}},\ }\bibfield  {title}
  {\bibinfo {title} {New class of quantum error-correcting codes for a bosonic
  mode},\ }\href@noop {} {\bibfield  {journal} {\bibinfo  {journal} {Physical
  Review X}\ }\textbf {\bibinfo {volume} {6}},\ \bibinfo {pages} {031006}
  (\bibinfo {year} {2016})}\BibitemShut {NoStop}%
\bibitem [{\citenamefont {Acharya}\ \emph {et~al.}(2024)\citenamefont
  {Acharya}, \citenamefont {Aghababaie-Beni}, \citenamefont {Aleiner},
  \citenamefont {Andersen}, \citenamefont {Ansmann}, \citenamefont {Arute},
  \citenamefont {Arya}, \citenamefont {Asfaw}, \citenamefont {Astrakhantsev},
  \citenamefont {Atalaya} \emph {et~al.}}]{acharya2024quantum}%
  \BibitemOpen
  \bibfield  {author} {\bibinfo {author} {\bibfnamefont {R.}~\bibnamefont
  {Acharya}}, \bibinfo {author} {\bibfnamefont {L.}~\bibnamefont
  {Aghababaie-Beni}}, \bibinfo {author} {\bibfnamefont {I.}~\bibnamefont
  {Aleiner}}, \bibinfo {author} {\bibfnamefont {T.~I.}\ \bibnamefont
  {Andersen}}, \bibinfo {author} {\bibfnamefont {M.}~\bibnamefont {Ansmann}},
  \bibinfo {author} {\bibfnamefont {F.}~\bibnamefont {Arute}}, \bibinfo
  {author} {\bibfnamefont {K.}~\bibnamefont {Arya}}, \bibinfo {author}
  {\bibfnamefont {A.}~\bibnamefont {Asfaw}}, \bibinfo {author} {\bibfnamefont
  {N.}~\bibnamefont {Astrakhantsev}}, \bibinfo {author} {\bibfnamefont
  {J.}~\bibnamefont {Atalaya}}, \emph {et~al.},\ }\bibfield  {title} {\bibinfo
  {title} {Quantum error correction below the surface code threshold},\
  }\href@noop {} {\bibfield  {journal} {\bibinfo  {journal} {arXiv preprint
  arXiv:2408.13687}\ } (\bibinfo {year} {2024})}\BibitemShut {NoStop}%
\bibitem [{\citenamefont {Leung}\ \emph {et~al.}(2017)\citenamefont {Leung},
  \citenamefont {Abdelhafez}, \citenamefont {Koch},\ and\ \citenamefont
  {Schuster}}]{leung2017speedup}%
  \BibitemOpen
  \bibfield  {author} {\bibinfo {author} {\bibfnamefont {N.}~\bibnamefont
  {Leung}}, \bibinfo {author} {\bibfnamefont {M.}~\bibnamefont {Abdelhafez}},
  \bibinfo {author} {\bibfnamefont {J.}~\bibnamefont {Koch}},\ and\ \bibinfo
  {author} {\bibfnamefont {D.}~\bibnamefont {Schuster}},\ }\bibfield  {title}
  {\bibinfo {title} {Speedup for quantum optimal control from automatic
  differentiation based on graphics processing units},\ }\href@noop {}
  {\bibfield  {journal} {\bibinfo  {journal} {Physical Review A}\ }\textbf
  {\bibinfo {volume} {95}},\ \bibinfo {pages} {042318} (\bibinfo {year}
  {2017})}\BibitemShut {NoStop}%
\bibitem [{\citenamefont {Shi}\ \emph {et~al.}(2019)\citenamefont {Shi},
  \citenamefont {Leung}, \citenamefont {Gokhale}, \citenamefont {Rossi},
  \citenamefont {Schuster}, \citenamefont {Hoffmann},\ and\ \citenamefont
  {Chong}}]{shi2019optimized}%
  \BibitemOpen
  \bibfield  {author} {\bibinfo {author} {\bibfnamefont {Y.}~\bibnamefont
  {Shi}}, \bibinfo {author} {\bibfnamefont {N.}~\bibnamefont {Leung}}, \bibinfo
  {author} {\bibfnamefont {P.}~\bibnamefont {Gokhale}}, \bibinfo {author}
  {\bibfnamefont {Z.}~\bibnamefont {Rossi}}, \bibinfo {author} {\bibfnamefont
  {D.~I.}\ \bibnamefont {Schuster}}, \bibinfo {author} {\bibfnamefont
  {H.}~\bibnamefont {Hoffmann}},\ and\ \bibinfo {author} {\bibfnamefont
  {F.~T.}\ \bibnamefont {Chong}},\ }\bibfield  {title} {\bibinfo {title}
  {Optimized compilation of aggregated instructions for realistic quantum
  computers},\ }in\ \href@noop {} {\emph {\bibinfo {booktitle} {Proceedings of
  the Twenty-Fourth International Conference on Architectural Support for
  Programming Languages and Operating Systems}}}\ (\bibinfo {year} {2019})\
  pp.\ \bibinfo {pages} {1031--1044}\BibitemShut {NoStop}%
\bibitem [{\citenamefont {Klimov}\ \emph {et~al.}(2020)\citenamefont {Klimov},
  \citenamefont {Kelly}, \citenamefont {Martinis},\ and\ \citenamefont
  {Neven}}]{klimov2020snake}%
  \BibitemOpen
  \bibfield  {author} {\bibinfo {author} {\bibfnamefont {P.~V.}\ \bibnamefont
  {Klimov}}, \bibinfo {author} {\bibfnamefont {J.}~\bibnamefont {Kelly}},
  \bibinfo {author} {\bibfnamefont {J.~M.}\ \bibnamefont {Martinis}},\ and\
  \bibinfo {author} {\bibfnamefont {H.}~\bibnamefont {Neven}},\ }\bibfield
  {title} {\bibinfo {title} {The snake optimizer for learning quantum processor
  control parameters},\ }\href@noop {} {\bibfield  {journal} {\bibinfo
  {journal} {arXiv preprint arXiv:2006.04594}\ } (\bibinfo {year}
  {2020})}\BibitemShut {NoStop}%
\bibitem [{\citenamefont {Piveteau}\ and\ \citenamefont
  {Sutter}(2023)}]{piveteau2023circuit}%
  \BibitemOpen
  \bibfield  {author} {\bibinfo {author} {\bibfnamefont {C.}~\bibnamefont
  {Piveteau}}\ and\ \bibinfo {author} {\bibfnamefont {D.}~\bibnamefont
  {Sutter}},\ }\bibfield  {title} {\bibinfo {title} {Circuit knitting with
  classical communication},\ }\href@noop {} {\bibfield  {journal} {\bibinfo
  {journal} {IEEE Transactions on Information Theory}\ } (\bibinfo {year}
  {2023})}\BibitemShut {NoStop}%
\bibitem [{\citenamefont {Heeres}\ \emph {et~al.}(2017)\citenamefont {Heeres},
  \citenamefont {Reinhold}, \citenamefont {Ofek}, \citenamefont {Frunzio},
  \citenamefont {Jiang}, \citenamefont {Devoret},\ and\ \citenamefont
  {Schoelkopf}}]{heeres2017implementing}%
  \BibitemOpen
  \bibfield  {author} {\bibinfo {author} {\bibfnamefont {R.~W.}\ \bibnamefont
  {Heeres}}, \bibinfo {author} {\bibfnamefont {P.}~\bibnamefont {Reinhold}},
  \bibinfo {author} {\bibfnamefont {N.}~\bibnamefont {Ofek}}, \bibinfo {author}
  {\bibfnamefont {L.}~\bibnamefont {Frunzio}}, \bibinfo {author} {\bibfnamefont
  {L.}~\bibnamefont {Jiang}}, \bibinfo {author} {\bibfnamefont {M.~H.}\
  \bibnamefont {Devoret}},\ and\ \bibinfo {author} {\bibfnamefont {R.~J.}\
  \bibnamefont {Schoelkopf}},\ }\bibfield  {title} {\bibinfo {title}
  {Implementing a universal gate set on a logical qubit encoded in an
  oscillator},\ }\href@noop {} {\bibfield  {journal} {\bibinfo  {journal}
  {Nature communications}\ }\textbf {\bibinfo {volume} {8}},\ \bibinfo {pages}
  {94} (\bibinfo {year} {2017})}\BibitemShut {NoStop}%
\bibitem [{\citenamefont {Hu}\ \emph {et~al.}(2019)\citenamefont {Hu},
  \citenamefont {Ma}, \citenamefont {Cai}, \citenamefont {Mu}, \citenamefont
  {Xu}, \citenamefont {Wang}, \citenamefont {Wu}, \citenamefont {Wang},
  \citenamefont {Song}, \citenamefont {Zou} \emph {et~al.}}]{hu2019quantum}%
  \BibitemOpen
  \bibfield  {author} {\bibinfo {author} {\bibfnamefont {L.}~\bibnamefont
  {Hu}}, \bibinfo {author} {\bibfnamefont {Y.}~\bibnamefont {Ma}}, \bibinfo
  {author} {\bibfnamefont {W.}~\bibnamefont {Cai}}, \bibinfo {author}
  {\bibfnamefont {X.}~\bibnamefont {Mu}}, \bibinfo {author} {\bibfnamefont
  {Y.}~\bibnamefont {Xu}}, \bibinfo {author} {\bibfnamefont {W.}~\bibnamefont
  {Wang}}, \bibinfo {author} {\bibfnamefont {Y.}~\bibnamefont {Wu}}, \bibinfo
  {author} {\bibfnamefont {H.}~\bibnamefont {Wang}}, \bibinfo {author}
  {\bibfnamefont {Y.}~\bibnamefont {Song}}, \bibinfo {author} {\bibfnamefont
  {C.-L.}\ \bibnamefont {Zou}}, \emph {et~al.},\ }\bibfield  {title} {\bibinfo
  {title} {Quantum error correction and universal gate set operation on a
  binomial bosonic logical qubit},\ }\href@noop {} {\bibfield  {journal}
  {\bibinfo  {journal} {Nature Physics}\ }\textbf {\bibinfo {volume} {15}},\
  \bibinfo {pages} {503} (\bibinfo {year} {2019})}\BibitemShut {NoStop}%
\bibitem [{\citenamefont {Ni}\ \emph {et~al.}(2023)\citenamefont {Ni},
  \citenamefont {Li}, \citenamefont {Deng}, \citenamefont {Cai}, \citenamefont
  {Zhang}, \citenamefont {Wang}, \citenamefont {Yang}, \citenamefont {Yu},
  \citenamefont {Yan}, \citenamefont {Liu} \emph {et~al.}}]{ni2023beating}%
  \BibitemOpen
  \bibfield  {author} {\bibinfo {author} {\bibfnamefont {Z.}~\bibnamefont
  {Ni}}, \bibinfo {author} {\bibfnamefont {S.}~\bibnamefont {Li}}, \bibinfo
  {author} {\bibfnamefont {X.}~\bibnamefont {Deng}}, \bibinfo {author}
  {\bibfnamefont {Y.}~\bibnamefont {Cai}}, \bibinfo {author} {\bibfnamefont
  {L.}~\bibnamefont {Zhang}}, \bibinfo {author} {\bibfnamefont
  {W.}~\bibnamefont {Wang}}, \bibinfo {author} {\bibfnamefont {Z.-B.}\
  \bibnamefont {Yang}}, \bibinfo {author} {\bibfnamefont {H.}~\bibnamefont
  {Yu}}, \bibinfo {author} {\bibfnamefont {F.}~\bibnamefont {Yan}}, \bibinfo
  {author} {\bibfnamefont {S.}~\bibnamefont {Liu}}, \emph {et~al.},\ }\bibfield
   {title} {\bibinfo {title} {Beating the break-even point with a
  discrete-variable-encoded logical qubit},\ }\href@noop {} {\bibfield
  {journal} {\bibinfo  {journal} {Nature}\ }\textbf {\bibinfo {volume} {616}},\
  \bibinfo {pages} {56} (\bibinfo {year} {2023})}\BibitemShut {NoStop}%
\bibitem [{\citenamefont {Van~Loan}(1978)}]{van1978computing}%
  \BibitemOpen
  \bibfield  {author} {\bibinfo {author} {\bibfnamefont {C.}~\bibnamefont
  {Van~Loan}},\ }\bibfield  {title} {\bibinfo {title} {Computing integrals
  involving the matrix exponential},\ }\href@noop {} {\bibfield  {journal}
  {\bibinfo  {journal} {IEEE transactions on automatic control}\ }\textbf
  {\bibinfo {volume} {23}},\ \bibinfo {pages} {395} (\bibinfo {year}
  {1978})}\BibitemShut {NoStop}%
\bibitem [{\citenamefont {Carbonell}\ \emph {et~al.}(2008)\citenamefont
  {Carbonell}, \citenamefont {Jimenez},\ and\ \citenamefont
  {Pedroso}}]{carbonell2008computing}%
  \BibitemOpen
  \bibfield  {author} {\bibinfo {author} {\bibfnamefont {F.}~\bibnamefont
  {Carbonell}}, \bibinfo {author} {\bibfnamefont {J.}~\bibnamefont {Jimenez}},\
  and\ \bibinfo {author} {\bibfnamefont {L.}~\bibnamefont {Pedroso}},\
  }\bibfield  {title} {\bibinfo {title} {Computing multiple integrals involving
  matrix exponentials},\ }\href@noop {} {\bibfield  {journal} {\bibinfo
  {journal} {Journal of Computational and Applied Mathematics}\ }\textbf
  {\bibinfo {volume} {213}},\ \bibinfo {pages} {300} (\bibinfo {year}
  {2008})}\BibitemShut {NoStop}%
\bibitem [{\citenamefont {Goodwin}\ and\ \citenamefont
  {Kuprov}(2015)}]{goodwin2015auxiliary}%
  \BibitemOpen
  \bibfield  {author} {\bibinfo {author} {\bibfnamefont {D.~L.}\ \bibnamefont
  {Goodwin}}\ and\ \bibinfo {author} {\bibfnamefont {I.}~\bibnamefont
  {Kuprov}},\ }\bibfield  {title} {\bibinfo {title} {Auxiliary matrix formalism
  for interaction representation transformations, optimal control, and spin
  relaxation theories},\ }\href@noop {} {\bibfield  {journal} {\bibinfo
  {journal} {The Journal of chemical physics}\ }\textbf {\bibinfo {volume}
  {143}} (\bibinfo {year} {2015})}\BibitemShut {NoStop}%
\bibitem [{joh(2013)}]{johansson2013qutip}%
  \BibitemOpen
  \bibfield  {title} {\bibinfo {title} {Qutip 2: A python framework for the
  dynamics of open quantum systems},\ }\href
  {https://doi.org/https://doi.org/10.1016/j.cpc.2012.11.019} {\bibfield
  {journal} {\bibinfo  {journal} {Computer Physics Communications}\ }\textbf
  {\bibinfo {volume} {184}},\ \bibinfo {pages} {1234} (\bibinfo {year}
  {2013})}\BibitemShut {NoStop}%
\bibitem [{\citenamefont {Havel}(2003)}]{havel2003robust}%
  \BibitemOpen
  \bibfield  {author} {\bibinfo {author} {\bibfnamefont {T.~F.}\ \bibnamefont
  {Havel}},\ }\bibfield  {title} {\bibinfo {title} {Robust procedures for
  converting among lindblad, kraus and matrix representations of quantum
  dynamical semigroups},\ }\href@noop {} {\bibfield  {journal} {\bibinfo
  {journal} {Journal of Mathematical Physics}\ }\textbf {\bibinfo {volume}
  {44}},\ \bibinfo {pages} {534} (\bibinfo {year} {2003})}\BibitemShut
  {NoStop}%
\bibitem [{\citenamefont {Higham}(2005)}]{higham2005scaling}%
  \BibitemOpen
  \bibfield  {author} {\bibinfo {author} {\bibfnamefont {N.~J.}\ \bibnamefont
  {Higham}},\ }\bibfield  {title} {\bibinfo {title} {The scaling and squaring
  method for the matrix exponential revisited},\ }\href@noop {} {\bibfield
  {journal} {\bibinfo  {journal} {SIAM Journal on Matrix Analysis and
  Applications}\ }\textbf {\bibinfo {volume} {26}},\ \bibinfo {pages} {1179}
  (\bibinfo {year} {2005})}\BibitemShut {NoStop}%
\bibitem [{\citenamefont {Al-Mohy}\ and\ \citenamefont
  {Higham}(2010)}]{al2010new}%
  \BibitemOpen
  \bibfield  {author} {\bibinfo {author} {\bibfnamefont {A.~H.}\ \bibnamefont
  {Al-Mohy}}\ and\ \bibinfo {author} {\bibfnamefont {N.~J.}\ \bibnamefont
  {Higham}},\ }\bibfield  {title} {\bibinfo {title} {A new scaling and squaring
  algorithm for the matrix exponential},\ }\href@noop {} {\bibfield  {journal}
  {\bibinfo  {journal} {SIAM Journal on Matrix Analysis and Applications}\
  }\textbf {\bibinfo {volume} {31}},\ \bibinfo {pages} {970} (\bibinfo {year}
  {2010})}\BibitemShut {NoStop}%
\bibitem [{\citenamefont {Virtanen}\ \emph {et~al.}(2020)\citenamefont
  {Virtanen}, \citenamefont {Gommers}, \citenamefont {Oliphant}, \citenamefont
  {Haberland}, \citenamefont {Reddy}, \citenamefont {Cournapeau}, \citenamefont
  {Burovski}, \citenamefont {Peterson}, \citenamefont {Weckesser},
  \citenamefont {Bright} \emph {et~al.}}]{virtanen2020scipy}%
  \BibitemOpen
  \bibfield  {author} {\bibinfo {author} {\bibfnamefont {P.}~\bibnamefont
  {Virtanen}}, \bibinfo {author} {\bibfnamefont {R.}~\bibnamefont {Gommers}},
  \bibinfo {author} {\bibfnamefont {T.~E.}\ \bibnamefont {Oliphant}}, \bibinfo
  {author} {\bibfnamefont {M.}~\bibnamefont {Haberland}}, \bibinfo {author}
  {\bibfnamefont {T.}~\bibnamefont {Reddy}}, \bibinfo {author} {\bibfnamefont
  {D.}~\bibnamefont {Cournapeau}}, \bibinfo {author} {\bibfnamefont
  {E.}~\bibnamefont {Burovski}}, \bibinfo {author} {\bibfnamefont
  {P.}~\bibnamefont {Peterson}}, \bibinfo {author} {\bibfnamefont
  {W.}~\bibnamefont {Weckesser}}, \bibinfo {author} {\bibfnamefont
  {J.}~\bibnamefont {Bright}}, \emph {et~al.},\ }\bibfield  {title} {\bibinfo
  {title} {Scipy 1.0: fundamental algorithms for scientific computing in
  python},\ }\href@noop {} {\bibfield  {journal} {\bibinfo  {journal} {Nature
  methods}\ }\textbf {\bibinfo {volume} {17}},\ \bibinfo {pages} {261}
  (\bibinfo {year} {2020})}\BibitemShut {NoStop}%
\bibitem [{\citenamefont {Kr{\"a}mer}\ \emph {et~al.}(2018)\citenamefont
  {Kr{\"a}mer}, \citenamefont {Plankensteiner}, \citenamefont {Ostermann},\
  and\ \citenamefont {Ritsch}}]{kramer2018quantumoptics}%
  \BibitemOpen
  \bibfield  {author} {\bibinfo {author} {\bibfnamefont {S.}~\bibnamefont
  {Kr{\"a}mer}}, \bibinfo {author} {\bibfnamefont {D.}~\bibnamefont
  {Plankensteiner}}, \bibinfo {author} {\bibfnamefont {L.}~\bibnamefont
  {Ostermann}},\ and\ \bibinfo {author} {\bibfnamefont {H.}~\bibnamefont
  {Ritsch}},\ }\bibfield  {title} {\bibinfo {title} {Quantumoptics. jl: A julia
  framework for simulating open quantum systems},\ }\href@noop {} {\bibfield
  {journal} {\bibinfo  {journal} {Computer Physics Communications}\ }\textbf
  {\bibinfo {volume} {227}},\ \bibinfo {pages} {109} (\bibinfo {year}
  {2018})}\BibitemShut {NoStop}%
\bibitem [{\citenamefont {Puzzuoli}\ \emph
  {et~al.}(2023{\natexlab{b}})\citenamefont {Puzzuoli}, \citenamefont {Wood},
  \citenamefont {Egger}, \citenamefont {Rosand},\ and\ \citenamefont
  {Ueda}}]{puzzuoli2023qiskit}%
  \BibitemOpen
  \bibfield  {author} {\bibinfo {author} {\bibfnamefont {D.}~\bibnamefont
  {Puzzuoli}}, \bibinfo {author} {\bibfnamefont {C.~J.}\ \bibnamefont {Wood}},
  \bibinfo {author} {\bibfnamefont {D.~J.}\ \bibnamefont {Egger}}, \bibinfo
  {author} {\bibfnamefont {B.}~\bibnamefont {Rosand}},\ and\ \bibinfo {author}
  {\bibfnamefont {K.}~\bibnamefont {Ueda}},\ }\bibfield  {title} {\bibinfo
  {title} {Qiskit dynamics: A python package for simulating the time dynamics
  of quantum systems},\ }\href@noop {} {\bibfield  {journal} {\bibinfo
  {journal} {Journal of Open Source Software}\ }\textbf {\bibinfo {volume}
  {8}},\ \bibinfo {pages} {5853} (\bibinfo {year}
  {2023}{\natexlab{b}})}\BibitemShut {NoStop}%
\bibitem [{\citenamefont {Hindmarsh}(1983)}]{hindmarsh1983odepack}%
  \BibitemOpen
  \bibfield  {author} {\bibinfo {author} {\bibfnamefont {A.~C.}\ \bibnamefont
  {Hindmarsh}},\ }\bibfield  {title} {\bibinfo {title} {{ODEPACK}, a systemized
  collection of ode solvers},\ }\href@noop {} {\bibfield  {journal} {\bibinfo
  {journal} {Scientific computing}\ } (\bibinfo {year} {1983})}\BibitemShut
  {NoStop}%
\bibitem [{\citenamefont {Hairer}\ \emph {et~al.}(1993)\citenamefont {Hairer},
  \citenamefont {Wanner},\ and\ \citenamefont {Nørsett}}]{Hairer1993}%
  \BibitemOpen
  \bibfield  {author} {\bibinfo {author} {\bibfnamefont {E.}~\bibnamefont
  {Hairer}}, \bibinfo {author} {\bibfnamefont {G.}~\bibnamefont {Wanner}},\
  and\ \bibinfo {author} {\bibfnamefont {S.~P.}\ \bibnamefont {Nørsett}},\
  }\href {https://doi.org/10.1007/978-3-540-78862-1} {\emph {\bibinfo {title}
  {Solving Ordinary Differential Equations I: Nonstiff Problems}}},\ \bibinfo
  {series} {Springer Series in Computational Mathematics}, Vol.~\bibinfo
  {volume} {8}\ (\bibinfo  {publisher} {Springer-Verlag},\ \bibinfo {address}
  {Berlin, Heidelberg},\ \bibinfo {year} {1993})\BibitemShut {NoStop}%
\bibitem [{\citenamefont {Suzuki}(1991)}]{suzuki1991general}%
  \BibitemOpen
  \bibfield  {author} {\bibinfo {author} {\bibfnamefont {M.}~\bibnamefont
  {Suzuki}},\ }\bibfield  {title} {\bibinfo {title} {General theory of fractal
  path integrals with applications to many-body theories and statistical
  physics},\ }\href@noop {} {\bibfield  {journal} {\bibinfo  {journal} {Journal
  of mathematical physics}\ }\textbf {\bibinfo {volume} {32}},\ \bibinfo
  {pages} {400} (\bibinfo {year} {1991})}\BibitemShut {NoStop}%
\bibitem [{\citenamefont {Childs}\ \emph {et~al.}(2021)\citenamefont {Childs},
  \citenamefont {Su}, \citenamefont {Tran}, \citenamefont {Wiebe},\ and\
  \citenamefont {Zhu}}]{childs2021theory}%
  \BibitemOpen
  \bibfield  {author} {\bibinfo {author} {\bibfnamefont {A.~M.}\ \bibnamefont
  {Childs}}, \bibinfo {author} {\bibfnamefont {Y.}~\bibnamefont {Su}}, \bibinfo
  {author} {\bibfnamefont {M.~C.}\ \bibnamefont {Tran}}, \bibinfo {author}
  {\bibfnamefont {N.}~\bibnamefont {Wiebe}},\ and\ \bibinfo {author}
  {\bibfnamefont {S.}~\bibnamefont {Zhu}},\ }\bibfield  {title} {\bibinfo
  {title} {Theory of trotter error with commutator scaling},\ }\href@noop {}
  {\bibfield  {journal} {\bibinfo  {journal} {Physical Review X}\ }\textbf
  {\bibinfo {volume} {11}},\ \bibinfo {pages} {011020} (\bibinfo {year}
  {2021})}\BibitemShut {NoStop}%
\bibitem [{\citenamefont {Han}\ \emph {et~al.}(2021)\citenamefont {Han},
  \citenamefont {Cai}, \citenamefont {Hu}, \citenamefont {Mu}, \citenamefont
  {Ma}, \citenamefont {Xu}, \citenamefont {Wang}, \citenamefont {Wang},
  \citenamefont {Song}, \citenamefont {Zou} \emph
  {et~al.}}]{han2021experimental}%
  \BibitemOpen
  \bibfield  {author} {\bibinfo {author} {\bibfnamefont {J.}~\bibnamefont
  {Han}}, \bibinfo {author} {\bibfnamefont {W.}~\bibnamefont {Cai}}, \bibinfo
  {author} {\bibfnamefont {L.}~\bibnamefont {Hu}}, \bibinfo {author}
  {\bibfnamefont {X.}~\bibnamefont {Mu}}, \bibinfo {author} {\bibfnamefont
  {Y.}~\bibnamefont {Ma}}, \bibinfo {author} {\bibfnamefont {Y.}~\bibnamefont
  {Xu}}, \bibinfo {author} {\bibfnamefont {W.}~\bibnamefont {Wang}}, \bibinfo
  {author} {\bibfnamefont {H.}~\bibnamefont {Wang}}, \bibinfo {author}
  {\bibfnamefont {Y.}~\bibnamefont {Song}}, \bibinfo {author} {\bibfnamefont
  {C.-L.}\ \bibnamefont {Zou}}, \emph {et~al.},\ }\bibfield  {title} {\bibinfo
  {title} {Experimental simulation of open quantum system dynamics via
  trotterization},\ }\href@noop {} {\bibfield  {journal} {\bibinfo  {journal}
  {Physical Review Letters}\ }\textbf {\bibinfo {volume} {127}},\ \bibinfo
  {pages} {020504} (\bibinfo {year} {2021})}\BibitemShut {NoStop}%
\bibitem [{\citenamefont {Ge}\ and\ \citenamefont {Wu}(2021)}]{ge2021risk}%
  \BibitemOpen
  \bibfield  {author} {\bibinfo {author} {\bibfnamefont {X.}~\bibnamefont
  {Ge}}\ and\ \bibinfo {author} {\bibfnamefont {R.-B.}\ \bibnamefont {Wu}},\
  }\bibfield  {title} {\bibinfo {title} {Risk-sensitive optimization for robust
  quantum controls},\ }\href@noop {} {\bibfield  {journal} {\bibinfo  {journal}
  {Physical Review A}\ }\textbf {\bibinfo {volume} {104}},\ \bibinfo {pages}
  {012422} (\bibinfo {year} {2021})}\BibitemShut {NoStop}%
\bibitem [{\citenamefont {Barends}\ \emph {et~al.}(2014)\citenamefont
  {Barends}, \citenamefont {Kelly}, \citenamefont {Megrant}, \citenamefont
  {Veitia}, \citenamefont {Sank}, \citenamefont {Jeffrey}, \citenamefont
  {White}, \citenamefont {Mutus}, \citenamefont {Fowler}, \citenamefont
  {Campbell} \emph {et~al.}}]{barends2014superconducting}%
  \BibitemOpen
  \bibfield  {author} {\bibinfo {author} {\bibfnamefont {R.}~\bibnamefont
  {Barends}}, \bibinfo {author} {\bibfnamefont {J.}~\bibnamefont {Kelly}},
  \bibinfo {author} {\bibfnamefont {A.}~\bibnamefont {Megrant}}, \bibinfo
  {author} {\bibfnamefont {A.}~\bibnamefont {Veitia}}, \bibinfo {author}
  {\bibfnamefont {D.}~\bibnamefont {Sank}}, \bibinfo {author} {\bibfnamefont
  {E.}~\bibnamefont {Jeffrey}}, \bibinfo {author} {\bibfnamefont {T.~C.}\
  \bibnamefont {White}}, \bibinfo {author} {\bibfnamefont {J.}~\bibnamefont
  {Mutus}}, \bibinfo {author} {\bibfnamefont {A.~G.}\ \bibnamefont {Fowler}},
  \bibinfo {author} {\bibfnamefont {B.}~\bibnamefont {Campbell}}, \emph
  {et~al.},\ }\bibfield  {title} {\bibinfo {title} {Superconducting quantum
  circuits at the surface code threshold for fault tolerance},\ }\href@noop {}
  {\bibfield  {journal} {\bibinfo  {journal} {Nature}\ }\textbf {\bibinfo
  {volume} {508}},\ \bibinfo {pages} {500} (\bibinfo {year}
  {2014})}\BibitemShut {NoStop}%
\bibitem [{\citenamefont {Palao}\ and\ \citenamefont
  {Kosloff}(2003)}]{palao2003optimal}%
  \BibitemOpen
  \bibfield  {author} {\bibinfo {author} {\bibfnamefont {J.~P.}\ \bibnamefont
  {Palao}}\ and\ \bibinfo {author} {\bibfnamefont {R.}~\bibnamefont
  {Kosloff}},\ }\bibfield  {title} {\bibinfo {title} {Optimal control theory
  for unitary transformations},\ }\href@noop {} {\bibfield  {journal} {\bibinfo
   {journal} {Physical Review A}\ }\textbf {\bibinfo {volume} {68}},\ \bibinfo
  {pages} {062308} (\bibinfo {year} {2003})}\BibitemShut {NoStop}%
\bibitem [{\citenamefont {Reich}\ \emph {et~al.}(2013)\citenamefont {Reich},
  \citenamefont {Gualdi},\ and\ \citenamefont {Koch}}]{reich2013minimum}%
  \BibitemOpen
  \bibfield  {author} {\bibinfo {author} {\bibfnamefont {D.~M.}\ \bibnamefont
  {Reich}}, \bibinfo {author} {\bibfnamefont {G.}~\bibnamefont {Gualdi}},\ and\
  \bibinfo {author} {\bibfnamefont {C.~P.}\ \bibnamefont {Koch}},\ }\bibfield
  {title} {\bibinfo {title} {Minimum number of input states required for
  quantum gate characterization},\ }\href@noop {} {\bibfield  {journal}
  {\bibinfo  {journal} {Physical Review A—Atomic, Molecular, and Optical
  Physics}\ }\textbf {\bibinfo {volume} {88}},\ \bibinfo {pages} {042309}
  (\bibinfo {year} {2013})}\BibitemShut {NoStop}%
\bibitem [{\citenamefont {Rabitz}\ \emph {et~al.}(2004)\citenamefont {Rabitz},
  \citenamefont {Hsieh},\ and\ \citenamefont {Rosenthal}}]{rabitz2004quantum}%
  \BibitemOpen
  \bibfield  {author} {\bibinfo {author} {\bibfnamefont {H.~A.}\ \bibnamefont
  {Rabitz}}, \bibinfo {author} {\bibfnamefont {M.~M.}\ \bibnamefont {Hsieh}},\
  and\ \bibinfo {author} {\bibfnamefont {C.~M.}\ \bibnamefont {Rosenthal}},\
  }\bibfield  {title} {\bibinfo {title} {Quantum optimally controlled
  transition landscapes},\ }\href@noop {} {\bibfield  {journal} {\bibinfo
  {journal} {Science}\ }\textbf {\bibinfo {volume} {303}},\ \bibinfo {pages}
  {1998} (\bibinfo {year} {2004})}\BibitemShut {NoStop}%
\bibitem [{\citenamefont {Rabitz}\ \emph {et~al.}(2006)\citenamefont {Rabitz},
  \citenamefont {Hsieh},\ and\ \citenamefont {Rosenthal}}]{rabitz2006optimal}%
  \BibitemOpen
  \bibfield  {author} {\bibinfo {author} {\bibfnamefont {H.}~\bibnamefont
  {Rabitz}}, \bibinfo {author} {\bibfnamefont {M.}~\bibnamefont {Hsieh}},\ and\
  \bibinfo {author} {\bibfnamefont {C.}~\bibnamefont {Rosenthal}},\ }\bibfield
  {title} {\bibinfo {title} {Optimal control landscapes for quantum
  observables},\ }\href@noop {} {\bibfield  {journal} {\bibinfo  {journal} {The
  Journal of chemical physics}\ }\textbf {\bibinfo {volume} {124}} (\bibinfo
  {year} {2006})}\BibitemShut {NoStop}%
\bibitem [{\citenamefont {Rabitz}\ \emph {et~al.}(2005)\citenamefont {Rabitz},
  \citenamefont {Hsieh},\ and\ \citenamefont
  {Rosenthal}}]{rabitz2005landscape}%
  \BibitemOpen
  \bibfield  {author} {\bibinfo {author} {\bibfnamefont {H.}~\bibnamefont
  {Rabitz}}, \bibinfo {author} {\bibfnamefont {M.}~\bibnamefont {Hsieh}},\ and\
  \bibinfo {author} {\bibfnamefont {C.}~\bibnamefont {Rosenthal}},\ }\bibfield
  {title} {\bibinfo {title} {Landscape for optimal control of
  quantum-mechanical unitary transformations},\ }\href@noop {} {\bibfield
  {journal} {\bibinfo  {journal} {Physical Review A—Atomic, Molecular, and
  Optical Physics}\ }\textbf {\bibinfo {volume} {72}},\ \bibinfo {pages}
  {052337} (\bibinfo {year} {2005})}\BibitemShut {NoStop}%
\bibitem [{\citenamefont {Goerz}\ \emph {et~al.}(2022)\citenamefont {Goerz},
  \citenamefont {Carrasco},\ and\ \citenamefont
  {Malinovsky}}]{goerz2022quantum}%
  \BibitemOpen
  \bibfield  {author} {\bibinfo {author} {\bibfnamefont {M.~H.}\ \bibnamefont
  {Goerz}}, \bibinfo {author} {\bibfnamefont {S.~C.}\ \bibnamefont
  {Carrasco}},\ and\ \bibinfo {author} {\bibfnamefont {V.~S.}\ \bibnamefont
  {Malinovsky}},\ }\bibfield  {title} {\bibinfo {title} {Quantum optimal
  control via semi-automatic differentiation},\ }\href@noop {} {\bibfield
  {journal} {\bibinfo  {journal} {Quantum}\ }\textbf {\bibinfo {volume} {6}},\
  \bibinfo {pages} {871} (\bibinfo {year} {2022})}\BibitemShut {NoStop}%
\bibitem [{\citenamefont {Lambert}\ \emph {et~al.}(2024)\citenamefont
  {Lambert}, \citenamefont {Giguère}, \citenamefont {Menczel}, \citenamefont
  {Li}, \citenamefont {Hopf}, \citenamefont {Suárez}, \citenamefont {Gali},
  \citenamefont {Lishman}, \citenamefont {Gadhvi}, \citenamefont {Agarwal},
  \citenamefont {Galicia}, \citenamefont {Shammah}, \citenamefont {Nation},
  \citenamefont {Johansson}, \citenamefont {Ahmed}, \citenamefont {Cross},
  \citenamefont {Pitchford},\ and\ \citenamefont {Nori}}]{lambert2024qutip}%
  \BibitemOpen
  \bibfield  {author} {\bibinfo {author} {\bibfnamefont {N.}~\bibnamefont
  {Lambert}}, \bibinfo {author} {\bibfnamefont {E.}~\bibnamefont {Giguère}},
  \bibinfo {author} {\bibfnamefont {P.}~\bibnamefont {Menczel}}, \bibinfo
  {author} {\bibfnamefont {B.}~\bibnamefont {Li}}, \bibinfo {author}
  {\bibfnamefont {P.}~\bibnamefont {Hopf}}, \bibinfo {author} {\bibfnamefont
  {G.}~\bibnamefont {Suárez}}, \bibinfo {author} {\bibfnamefont
  {M.}~\bibnamefont {Gali}}, \bibinfo {author} {\bibfnamefont {J.}~\bibnamefont
  {Lishman}}, \bibinfo {author} {\bibfnamefont {R.}~\bibnamefont {Gadhvi}},
  \bibinfo {author} {\bibfnamefont {R.}~\bibnamefont {Agarwal}}, \bibinfo
  {author} {\bibfnamefont {A.}~\bibnamefont {Galicia}}, \bibinfo {author}
  {\bibfnamefont {N.}~\bibnamefont {Shammah}}, \bibinfo {author} {\bibfnamefont
  {P.}~\bibnamefont {Nation}}, \bibinfo {author} {\bibfnamefont {J.~R.}\
  \bibnamefont {Johansson}}, \bibinfo {author} {\bibfnamefont {S.}~\bibnamefont
  {Ahmed}}, \bibinfo {author} {\bibfnamefont {S.}~\bibnamefont {Cross}},
  \bibinfo {author} {\bibfnamefont {A.}~\bibnamefont {Pitchford}},\ and\
  \bibinfo {author} {\bibfnamefont {F.}~\bibnamefont {Nori}},\ }\href
  {https://arxiv.org/abs/2412.04705} {\bibinfo {title} {Qutip 5: The quantum
  toolbox in python}} (\bibinfo {year} {2024}),\ \Eprint
  {https://arxiv.org/abs/2412.04705} {arXiv:2412.04705 [quant-ph]} \BibitemShut
  {NoStop}%
\bibitem [{\citenamefont {Li}\ \emph {et~al.}(2024)\citenamefont {Li},
  \citenamefont {Kubo}, \citenamefont {Ho}, \citenamefont {Yan}, \citenamefont
  {Nakamura},\ and\ \citenamefont {Goto}}]{li2024realization}%
  \BibitemOpen
  \bibfield  {author} {\bibinfo {author} {\bibfnamefont {R.}~\bibnamefont
  {Li}}, \bibinfo {author} {\bibfnamefont {K.}~\bibnamefont {Kubo}}, \bibinfo
  {author} {\bibfnamefont {Y.}~\bibnamefont {Ho}}, \bibinfo {author}
  {\bibfnamefont {Z.}~\bibnamefont {Yan}}, \bibinfo {author} {\bibfnamefont
  {Y.}~\bibnamefont {Nakamura}},\ and\ \bibinfo {author} {\bibfnamefont
  {H.}~\bibnamefont {Goto}},\ }\bibfield  {title} {\bibinfo {title}
  {Realization of high-fidelity cz gate based on a double-transmon coupler},\
  }\href@noop {} {\bibfield  {journal} {\bibinfo  {journal} {Physical Review
  X}\ }\textbf {\bibinfo {volume} {14}},\ \bibinfo {pages} {041050} (\bibinfo
  {year} {2024})}\BibitemShut {NoStop}%
\bibitem [{\citenamefont {Nielsen}(2002)}]{nielsen2002simple}%
  \BibitemOpen
  \bibfield  {author} {\bibinfo {author} {\bibfnamefont {M.~A.}\ \bibnamefont
  {Nielsen}},\ }\bibfield  {title} {\bibinfo {title} {A simple formula for the
  average gate fidelity of a quantum dynamical operation},\ }\href@noop {}
  {\bibfield  {journal} {\bibinfo  {journal} {Physics Letters A}\ }\textbf
  {\bibinfo {volume} {303}},\ \bibinfo {pages} {249} (\bibinfo {year}
  {2002})}\BibitemShut {NoStop}%
\bibitem [{\citenamefont {Pedersen}\ \emph {et~al.}(2007)\citenamefont
  {Pedersen}, \citenamefont {M{\o}ller},\ and\ \citenamefont
  {M{\o}lmer}}]{pedersen2007fidelity}%
  \BibitemOpen
  \bibfield  {author} {\bibinfo {author} {\bibfnamefont {L.~H.}\ \bibnamefont
  {Pedersen}}, \bibinfo {author} {\bibfnamefont {N.~M.}\ \bibnamefont
  {M{\o}ller}},\ and\ \bibinfo {author} {\bibfnamefont {K.}~\bibnamefont
  {M{\o}lmer}},\ }\bibfield  {title} {\bibinfo {title} {Fidelity of quantum
  operations},\ }\href@noop {} {\bibfield  {journal} {\bibinfo  {journal}
  {Physics Letters A}\ }\textbf {\bibinfo {volume} {367}},\ \bibinfo {pages}
  {47} (\bibinfo {year} {2007})}\BibitemShut {NoStop}%
\bibitem [{\citenamefont {Magesan}\ \emph {et~al.}(2011)\citenamefont
  {Magesan}, \citenamefont {Gambetta},\ and\ \citenamefont
  {Emerson}}]{magesan2011scalable}%
  \BibitemOpen
  \bibfield  {author} {\bibinfo {author} {\bibfnamefont {E.}~\bibnamefont
  {Magesan}}, \bibinfo {author} {\bibfnamefont {J.~M.}\ \bibnamefont
  {Gambetta}},\ and\ \bibinfo {author} {\bibfnamefont {J.}~\bibnamefont
  {Emerson}},\ }\bibfield  {title} {\bibinfo {title} {Scalable and robust
  randomized benchmarking of quantum processes},\ }\href@noop {} {\bibfield
  {journal} {\bibinfo  {journal} {Physical review letters}\ }\textbf {\bibinfo
  {volume} {106}},\ \bibinfo {pages} {180504} (\bibinfo {year}
  {2011})}\BibitemShut {NoStop}%
\bibitem [{\citenamefont {Proctor}\ \emph {et~al.}(2017)\citenamefont
  {Proctor}, \citenamefont {Rudinger}, \citenamefont {Young}, \citenamefont
  {Sarovar},\ and\ \citenamefont {Blume-Kohout}}]{proctor2017randomized}%
  \BibitemOpen
  \bibfield  {author} {\bibinfo {author} {\bibfnamefont {T.}~\bibnamefont
  {Proctor}}, \bibinfo {author} {\bibfnamefont {K.}~\bibnamefont {Rudinger}},
  \bibinfo {author} {\bibfnamefont {K.}~\bibnamefont {Young}}, \bibinfo
  {author} {\bibfnamefont {M.}~\bibnamefont {Sarovar}},\ and\ \bibinfo {author}
  {\bibfnamefont {R.}~\bibnamefont {Blume-Kohout}},\ }\bibfield  {title}
  {\bibinfo {title} {What randomized benchmarking actually measures},\
  }\href@noop {} {\bibfield  {journal} {\bibinfo  {journal} {Physical review
  letters}\ }\textbf {\bibinfo {volume} {119}},\ \bibinfo {pages} {130502}
  (\bibinfo {year} {2017})}\BibitemShut {NoStop}%
\bibitem [{\citenamefont {Hocker}\ \emph {et~al.}(2014)\citenamefont {Hocker},
  \citenamefont {Brif}, \citenamefont {Grace}, \citenamefont {Donovan},
  \citenamefont {Ho}, \citenamefont {Tibbetts}, \citenamefont {Wu},\ and\
  \citenamefont {Rabitz}}]{hocker2014characterization}%
  \BibitemOpen
  \bibfield  {author} {\bibinfo {author} {\bibfnamefont {D.}~\bibnamefont
  {Hocker}}, \bibinfo {author} {\bibfnamefont {C.}~\bibnamefont {Brif}},
  \bibinfo {author} {\bibfnamefont {M.~D.}\ \bibnamefont {Grace}}, \bibinfo
  {author} {\bibfnamefont {A.}~\bibnamefont {Donovan}}, \bibinfo {author}
  {\bibfnamefont {T.-S.}\ \bibnamefont {Ho}}, \bibinfo {author} {\bibfnamefont
  {K.~M.}\ \bibnamefont {Tibbetts}}, \bibinfo {author} {\bibfnamefont
  {R.}~\bibnamefont {Wu}},\ and\ \bibinfo {author} {\bibfnamefont
  {H.}~\bibnamefont {Rabitz}},\ }\bibfield  {title} {\bibinfo {title}
  {Characterization of control noise effects in optimal quantum unitary
  dynamics},\ }\href@noop {} {\bibfield  {journal} {\bibinfo  {journal}
  {Physical Review A}\ }\textbf {\bibinfo {volume} {90}},\ \bibinfo {pages}
  {062309} (\bibinfo {year} {2014})}\BibitemShut {NoStop}%
\end{thebibliography}%
\end{document}